\documentclass[english,11pt]{article}
\pdfoutput=1
\usepackage{jcappub}
\usepackage{lmodern}
\usepackage{lmodern}
\usepackage{graphicx}
\usepackage{enumitem}
\usepackage{float}
\usepackage{amsmath}
\usepackage{amssymb}
\usepackage{mathtools}
\usepackage{hyperref}
\usepackage{xcolor}
\usepackage{subcaption}
\usepackage{cellspace}
\usepackage{widetext}
\usepackage{etex}
\usepackage{url}

\hypersetup{colorlinks = true, urlcolor = blue, linkcolor = blue, frenchlinks = true, pdfborder = {0 0 0}}

\usepackage{romannum}

\begin{document}

\title{Machian Gravity:  A mathematical formulation for Mach's Principle}

\author{Santanu Das}%\email{sanjone@gmail.com}
%\affil{\orgname{Raman Research Institute}, \state{Bangalore}, \country{India}}
\affiliation{Raman Research Institute, Bangalore, India}
\emailAdd{santanu@cosmocommunity.in}

\abstract{The general theory of relativity (GR) was proposed with an aim of incorporating Mach's principle mathematically. Despite early hopes, it became evident that GR did not follow Mach's principle.  Over time, multiple researchers attempted to develop gravity theories aligned with Machian idea. Although these theories successfully explained various aspects of Mach's principle, each of these theories possessed its own strengths and weaknesses. In this paper, we discuss some of these theories and then try to combine these theories into a single framework that can fully embrace Mach's principle. This new theory, termed Machian Gravity (MG) is a metric-based theory, and can be derived from the action principle, ensuring compliance with all conservation laws.
\;\\

The theory converges to GR at solar system scales, but at larger scales, it diverges from GR and aligns with various modified gravity models proposed to explain dark sectors of the Universe. We have tested our theory against multiple observational data. It explains the galactic rotation curve without requiring additional dark matter (DM). The theory also resolves the discrepancy between dynamic mass and photometric mass in galaxy clusters without resorting to DM, but it introduces two additional parameters. It can also explain the expansion history of the Universe without requiring dark components. }

\maketitle

%\tableofcontents 
\section{Introduction}

Newtonian gravity can provide a very accurate description of gravity, provided the gravitational field is weak, not time-varying and the 
concerned velocities are much less than the speed of light. It can accurately describe the motions of planets and satellites in the solar system. Einstein formulated GR to provide a complete geometric approach to gravity. GR is designed to follow Newtonian gravity at a large scale. It can explain the perihelion precession of Mercury's orbit and the bending of light by the Sun, which were never realized before, using Newtonian mechanics. Over the years, numerous predictions of GR, such as the existence of black holes, gravitational waves, etc. have been observed. This makes GR one of the most well-accepted theories of gravity. 

However, the drawbacks of GR come to light when GR is applied on the galactic and cosmological scale. It fails to produce the galactic velocity profiles, provided that calculations are made just considering the visible matter in the galaxy. This led researchers to postulate a new form of weakly interacting matter named dark matter. Earlier it was commonly believed that dark matter (DM) is made up of particles predicted from supersymmetry theory~\cite{MARTIN_1998}. However, the lack of evidence of these particles from Large Hadron Collider (LHC) strengthens the proposition of other candidates, such as Axions, ultra-light scalar field dark matter, etc~\cite{PhysRevLett.40.279,PhysRevD.53.2236}. A further mysterious puzzle is the dark energy (DE) because that requires to produce a repulsive gravitation force. Cosmological constant or $\Lambda$-term provides an excellent solution for this. However, as the observations become more precise, multiple inconsistencies come to light~\cite{Di_Valentino_2021,Joudaki_2016,Hildebrandt_2016,Riess_2019,Ade2016pap16,Addison_2016,Kitching_2016,Couchot_2017}. 

There can be two ways to solve the dark sector of the Universe. 
Firstly, we can assume that there is in need some type of matter that
does not interact with standard-model particles and acts as dark matter, and we have some form of energy with a negative pressure and provide a dark-energy-like behavior. While this can, in need, be the case, the possibility that the GR fails to explain the true nature of gravity in kilo-parsec scale can also not be overlooked. In such a case, we need an alternate theory of gravity that can replicate GR on a relatively smaller scale while deviating from it on a galactic scale. 

Several theories have been proposed in the last decade to explain DM and DE. Empirical theories like Modified Newtonian Dynamics (MOND), proposed by \cite{Milgrim1983,Milgrim1983a,Milgrim1983b,Milgrom2011} can
explain the galactic velocity profiles extremely well but violates
momentum conservation principles. Therefore, if a mathematically sound theory is developed that can mimic the MOND empirically, then that can explain the dark matter. Bekenstein proposed A QUAdratic Lagrangian (AQUAL) \cite{Bekenstein1984,Bekenstein2009,Milgrom1986}
to provide a physical ground to MOND. Other theories, such as Modified gravity, Scalar-Tensor-Vector Gravity (STVG) \cite{Moffat2005,Brownstein2005,Brownstein2005a,Moffat2005a}, 
Tensor–Vector–Scalar gravity (TeVeS) \cite{Bekenstein2005}, Massive gravity \cite{Dam1970,Zakharov1970,Babichev2010,Babichev2013} etc. are also proposed to match the galactic velocity profiles without dark matter. 
Other higher dimensional theories such as induced matter theory \cite{Overduin1998,Ponce1993,Wesson1992,de2010schwarzschild,moraes2016cosmic} etc. are also proposed by researchers. However, all these theories came from the natural desire to explain the observational data and not build on a solid logical footing.

Now, let us shift our focus to another aspect of GR. In the early 20th century, Earnest Mach hypothesized that the inertial properties of matter must depend on the distant matters of the Universe. Einstein was intrigued by Mach's Principle and tried to provide a mathematical construct of it through the GR. He later realized that his field equations imply that a test particle in an otherwise empty Universe has inertial properties, which contradicts Mach's argument~\cite{Einstein}. However, intrigued by the overwhelming success of GR in explaining different observational data, he did not make any further attempt to explain Mach's principle.

In view of this, it is worthwhile searching for a theory that implies that matter has inertia only in the presence of other matter. Several theories that abide by Mach's principle have been postulated in the last century. Among these, the most prominent are Sciama's vector potential theory~\cite{sciama1953origin}, Brans Dicke (BD) theory or the scalar-tensor theory of gravity~\cite{Brans1961, fujii2003scalar, faraoni2004cosmology} and Hoyle Narlikar theory \cite{Hoyle1964,hoyle1964c, hoyle1964time, hoyle1964avoidance, hoyle1964gravitational} etc. Although each of these theories addresses certain aspects of Mach's principle as discussed in the respective articles, none offers a complete explanation. Thus, only a unified theory that combines these approaches could provide a comprehensive understanding of Mach's principle in its entirety. %Such a theory would also be able to replicate the behavior of these individual theories under the specific conditions for which they were developed.

In this article, we address all the issues described above and propose a theory of gravity based on Mach's principle. It is based on the following premises.

\begin{itemize}
\item Action principle: The theory should be derived
from an action principle to guarantee that the theory does not violate conservation laws.

\item Equivalence principle : 
Various research groups have tested the Weak Equivalence Principle (WEP) at an exquisite procession. Therefore, any theory must follow the weak equivalence principle\cite{PhysRevD.104.044001,Wagner:2012ui,rosi2017quantum,10.1093/mnrasl/slaa143,PhysRevLett.129.121102,huber2000precision,Yang:2019tzi}. However, the strong equivalence principle has not been tested on a large scale~\cite{Voisin:2020lqi}. If the ratio of the inertial mass and the gravitational mass changes over space-time (on a galactic scale or cosmological scale), then that does not violate results from our local measurements. In accordance with Mach's principle, the inertial properties of matter come from all the distant matter of the Universe. As the matter distribution at different parts of the Universe is different, the theory may not follow the strong equivalence principle.

\item Departure from GR: As GR provides an excellent result in the solar system scale, the proposed theory should follow GR on that scale, and it only deviates from GR at the galactic scale to mimic some of the modified gravity theories proposed by researchers to explain the dark sectors of the Universe. Along with this, the proposed theory should also be able to replicate the behavior of theories like Sciama's theory or BD theory under the specific circumstances for which they were proposed. 
\end{itemize}

The paper is organized as follows. In the second section I briefly discuss previous developments in gravity theory to explain Mach's principle. Most of the points covered in this section are generally known positions form various previous research. However, since Mach's principle is not a mainstream area of study, this discussion is necessary and important for understanding the new insights presented in this article. In some cases I interpret these established ideas from the perspective of this paper, that will help me to built the gravity theory in the later section. In the next section, we explain Mach's principle and discuss the mathematical tools used to formulate the theory. We present the source-free field equations for the theory in the same section. The static spherically symmetric solution for the theory in weak field approximation is presented in the fourth section. We show that the solution follows Newtonian gravity and GR at a smaller scale but deviates from it at a large scale. Section five presents examples of galactic rotation curves and galaxy cluster mass distributions, demonstrating that the theory yields results in close agreement with observations. The source term of the theory has been described in the 6th section. In the next section, we provide the cosmological solutions to the MG model. The final section is the conclusion and discussion section. We have also added five appendices where we describe the nitty-gritty of the calculations and add multiple illustrations. 

\section{Mach's Principle: Foundations, History, and Gravitational Theories}

\subsection{Concept of Mach's principle}

To explain Mach's principle, I start from the definition of mass itself. Although mass is a fundamental concept in physics, it lacks a well-defined definition. Yet, to do physics, we need a working notion of mass, which is defined in two ways: inertial mass, defined based on inertial properties of matter, and gravitational mass, which is defined based on the gravitational properties. Gravitational mass can be further split into active and passive types, but we will not go into that here~\cite{Jammer}. Several experiments have been conducted to measure the ratio between inertial and passive gravitational mass. However, they came out to be the same for all the materials. We term it as the weak equivalence principle.  

The inertial mass of a particle is measured based on its motion in an inertial coordinate frame. Therefore, determining the inertial coordinate system is important for measuring the inertial mass. However, it is difficult to determine a perfect inertial coordinate system because there is no external reference frame to measure its acceleration. Ernest Mach postulated that the inertial reference could be determined by measuring the motions of distant objects in the Universe. This implies that the distant objects in the Universe actually determine the inertial properties of matter, which is the famous Mach’s principle. Therefore, two identical objects depending on their background mass distribution may show different the inertial properties. We elaborate this concept below.

\subsubsection{Relativity of motions}
\label{sec2}
The velocity or acceleration of a particle are relative quantities, i.e. 
they are always measured with respect to some reference frame. While measuring the velocity or the acceleration of a running train they are measured with respect to the surface of the Earth. However, the Earth is orbiting the sun, which is again circling our galaxy. The galaxy also has some random motion in the galaxy cluster, and so on.
Therefore, if the origin of the coordinate system is chosen to be at
the center of the galaxy, then the velocity and acceleration of the train will be completely different~\cite{Hoyle1980}. 

As the acceleration of a particle is related to the force exerted on the particle, the force is also associated with the coordinate system. Let us consider that a stone is tied with a string and whirled around in a circle. We define two reference frames, one with the origin at the center of the circle, which is fixed with respect to us, and the other is fixed at the stone. In the reference frame that is fixed at the center of the circle, we can analyze the forces on the stone using Newton's law. If $m_0$ is the inertial mass, $v$ is the velocity of the stone, $r$ is the radius of the circle, and $T$ is the tension on the string, then using Newton's law, we can write 

\begin{equation}
m_0\frac{v^{2}}{r}=T\,.\label{eq:stone}
\end{equation}

On the other hand, in the reference frame fixed to the stone, the stone has no velocity, $v=0$. Therefore, the left-hand side of Eq.(\ref{eq:stone}) becomes zero. However, the right-hand side, i.e., the tension on the string towards the center, remains the same as $T$. Therefore, the equality of the Eq.(\ref{eq:stone}) does not hold in this frame. Newton's law is not applicable in this reference frame, and it's called a non-inertial reference frame. To balance the equation in such frames, we need additional fictitious forces, known as inertial forces. In this example, the fictitious force is the centrifugal force, and it is equal and opposite to the force $T$, i.e., $-T$. 

The source of these inertial forces is still unknown. Interestingly, in the second reference frame, the rest of the Universe, i.e. the distant starts, galaxies, etc, are rotating. Therefore, according to Mach's hypothesis these inertial forces are generated due to the rotation of all the distant stars in the non-inertial coordinate system~ \cite{Jammer,Weinberg:1972kfs}. 

Here some researchers may argue that velocities are relative. We can not construct a meaningful scalar quantity out of the four velocity, as $u^\mu u_\mu =1$. However, the proper acceleration of particle can be contracted with itself to provide a meaningful scalar, $a^\mu a_\mu = a^2$. As the scalar is independent of the reference frame, we should be able to measure the scalar quantity in any reference frame. If there is a closed lab floating in the Universe, an experimentalist inside the lab can measure its acceleration through experiments, whereas it's velocity cannot be determined through experiments within the lab.

Although this logic may appear to be correct at a first glance, my counter will be that there is no experiment that can distinguish between acceleration and gravity inside the lab. All the matter in the Universe creates a collective gravitational field at the location of the particle. If the motion of the particle causes all other objects in the Universe to create a gravitational field that mimics inertial forces, then no experiment inside the lab can determine whether it is experiencing acceleration or gravity. In other words, if the closed lab is placed in a Universe where there are no other matter particles, will it still measure acceleration? According to Mach, it will not, because accelerations will have no meaning in such Universe. 

In fact if we assume that there are two closed laboratories that are accelerating with respect to each other in an otherwise empty Universe, if we ignore the gravitational force between them, and if we detect absolute acceleration for both the frames to be different, then that should imply that the spacetime is an absolute quantity. Otherwise, both reference frames have identical external conditions but still observe different accelerations, which is illogical --- since the laws of physics must be of such a nature that they apply to systems of reference in any kind of motion.

\subsection{General theory of relativity}
GR was proposed to extend the above philosophy. The logic behind Einstein's GR can be described as follows. According to Newton's gravity, if we place a particle in a gravitational field, then 

\begin{equation}
\begin{array}{c}
\texttt{Inertial mass} \;\times\; \texttt{acceleration} \\  
= \\
\texttt{passive gravitational mass} \;\times\; \texttt{gravitational field}
\end{array}
\end{equation}
\newline

\noindent Therefore, if we consider the inertial mass is equal to the passive gravitational mass, then in a small region of spacetime, it is impossible to distinguish between the acceleration and the gravitational field. Then Einstein showed that spacetime gets curved in an accelerated reference frame for deriving GR. 

He used a simple thought experiment to do this. He considers two systems of coordinates, $K$ and $K'$, whose $z$ axes are aligned, where $K$ is an inertial frame and $K'$ is a noninertial frame that is rotating with constant angular velocity with respect to $K$ frame. Let us take a circle on the $x'y'$ plane of the $K'$ reference frame and fill one of its diameter and perimeter with small rigid rods of length $l$. If $U$ is the number of these rods along the 
periphery, $D$ the number along the diameter, then, if $K'$ does not rotate relatively to $K$, we shall have $U/D = \pi$.
But if $K'$ rotates then from the perspective of $K$, all the rods upon the periphery experience the Lorentz contraction,  but the rods upon the diameter do not experience this contraction and we have $U/D > \pi$. Therefore, the laws of configuration of rigid bodies do not follow the Euclidean geometry with respect to the $K'$ reference frame. He used Riemannian geometry to show that this curvature in space can be used to explain the acceleration~\cite{Einstein}. (note that a change in coordinate system does not change the Riemann curvature of the entire spacetime, i.e. the Riemann curvature of the spacetime will remain $0$. It is because the 4-D curvature is not measured using the meter sticks placed on the circumference of the circle, but with respect to the 4D line element.) As the acceleration and gravitational field cannot be distinguished within a small enough region of spacetime, in GR, the spacetime curvature is related to the stress-energy tensor to formulate the theory of gravitation.

\subsubsection{Issues between General Relativity and Mach's principle}

GR was proposed based on the philosophy that ``the laws of physics must be of such a nature that they apply to all the systems of reference"~\cite{Einstein:1916vd}. However, GR does not fully adhere to this principle and thereby does not follow Mach's principle as it was understood by Einstein himself and many others~\cite{Einstein,sciama1953origin,bondi1952cosmology}. 

To understand the issue more clearly, let us go back to the Einstein's thought experiment discussed in the above section. In that setup, the frame $K$ is assumed to be an inertial reference frame. An observer in $K$ would sees the background formed by distant stars and galaxies as static and non-rotating. Now, if $K'$ rotates with respect to $K$ at some angular velocity then from the perspective of $K'$, the distant stars and galaxies would also appear to rotate with the same angular velocity but in opposite direction. Since $K'$ is a non-inertial reference frame, any observer within it will experience inertial forces. Moreover, as demonstrated in the previous section, the laws governing the configuration of rigid bodies no longer follow Euclidean geometry in the $K'$ frame. 

We can extend the above thought experiment and consider that the entire Universe begins to rotate, such that the observer in the $K'$ reference frame perceives herself as stationary with respect to the distant stars. Of course, if we assume spacetime to be absolute, it might  violate fundamental physical laws. However, as long as we consider that the space is given by the distance between two particle it will not violate any laws of physics such as causality etc. because rotating the entire Universe will not change the motion between any two particles with respect to one another.  

The observer in the $K$ reference frame would now perceive the background stars as rotating. Thus, although neither the observer in $K$ nor the one in $K'$ has made any physical change, the very definition of an inertial reference frame appears to shift. Since, in the $K'$ frame, the distant stars now appear stationary, $K'$ should be treated as the inertial frame. Consequently, one would expect no inertial forces to appear in $K'$, and the geometry of space to be Euclidean. On the other hand, the $K$ frame, in which the distant stars appear to rotate, would require a non-Euclidean (Riemannian) geometric description of spacetime to account for the resulting inertial effects.

Now for deriving GR we assume that in a small enough region of spacetime it is impossible to distinguish between acceleration and gravity. Consider a central mass and assume that the spacetime around it is described by the Schwarzschild metric. At a radial distance $r$ from the central object any observer should experience an acceleration $\ddot{r} \approx -\frac{GM}{r^2}$, provided $r$ is not very close to the event horizon and the gravitational field is not time varying. The observer who is experiencing this acceleration is seeing that the distant stars are fixed in his coordinate frame. Now, suppose we rotate the entire Universe around the same central mass at a constant angular velocity , without changing the positions of the observer or the mass. In this new situation, the same observer should experience an acceleration  $\ddot{r} \approx -\frac{GM}{r^2} - \omega^2 r$. This indicates that the distant matter in the Universe is influencing the local inertial behavior of the observer. However, the Einstein field equations in GR do not contain a term that can account for this additional acceleration as GR has no term that comes from the background mass distribution. GR will still give me the same acceleration as before.

This issue arises because, in GR, the concept of inertia --- or the definition of inertial reference frames -- must be introduced \emph{a priori}.  To address this limitation, we require a theoretical framework in which inertia emerges naturally from the dynamics of the theory, rather than being imposed externally.

\subsection{\label{SciamaTheory}Sciama's attempt to incorporate Mach's principle in gravity}

In 1954, Sciama proposed a model of gravity where he postulated that ``In the rest-frame of any body the total gravitational field at the body arising from all the other matter in the Universe is $0$"~\cite{sciama1953origin,sciama1964physical,berman2008machian,Licata_2016}. He derived a vector potential from gravity and the velocity of a particle and showed that the Universe in a noninertial reference frame could provide inertial forces, such as the Coriolis force or the centrifugal force, etc. 

To give a brief overview of Scima's theory, he considers that the Universe creates a potential on the test particle. If the Universe has a uniform density $\rho$ with respect to any test particle, then the total potential of the Universe on that particle will be 
\begin{equation}
 \Phi=-G\int_V \frac{\rho}{r}dV.   
\end{equation}

\noindent Here, $G$ is the gravitational constant. $dV$ is the volume of the Universe within a distance $r$ and $r+dr$ from the test particle. If the particle moves with a velocity $-\vec{v}$ with respect to the smoothed-out Universe, then in the rest frame of the test particle, the Universe will move with a velocity $\vec{v}$. So the vector potential exerted by the Universe on the particle will be given by 

\begin{equation}
 \vec{\mathcal{A}}=-G\int_V \frac{\vec{v}\rho}{r}dV = -G\vec{v}\int_V \frac{\rho}{r}dV = \vec{v}\Phi.   
\end{equation}

\noindent Here, we take the velocity outside the integral because the velocity $\vec{v}$ is independent of $r$. We have also taken the speed of light to be $c=1$. The relativistic effects are not considered here, i.e. the $v^2\ll 1$ and hence the higher order  terms are neglected. In the relativistic limit, we need to use the four-velocity, which will give us the four-potential, exactly like electromagnetism. This actually solves the problem of inertia related to the Mach principle~\footnote{At  this point I would like to inform the readers that according to GR gravity is a spin $2$ field. However, under a weak gravitational field or an almost flat space-time we can we can well approximate the equations by equations that have the same form as in classical electromagnetism. Under such circumstances, the spacetime metric can be expanded as $g_{\mu\nu} = \eta_{\mu\nu} + h_{\mu\nu}$. Here, $\eta_{\mu\nu}$ is the Minkowski metric. Let us define $\bar{h}_{\mu\nu} = h_{\mu\nu} - \frac{1}{2}\eta_{\mu\nu} h$, where $h = \text{tr}(h_{\mu\nu})$. 
 We can define $\bar{h}_{00} = 4\Phi/c^2$ and $\bar{h}_{0i} = -2\vec{\mathcal{A}}_i/c^2$, where $i\in(1,2,3)$. As $\bar{h}_{ij}\sim \mathcal{O}(1/c^2)$ we ignore those terms. We can take $\Phi(t,x)$ as the gravitoelectric potential, which is nothing but the Newtonian potential, and $\mathcal{A}(t,x)$ as the gravitomagnetic vector potential. These quantities follow the classical electromagnetic equations~\cite{mashhoon2003gravitoelectromagnetism,
mashhoon2001gravitoelectromagnetism1,
mashhoon2001gravitomagnetism}. Sciama used the same gravito-magnetic vector potential in the equations.}. To understand that, we will consider two cases.

\subsubsection*{Linear motion}
First, let's consider linear motion. In the rest frame of the test particle, the Universe is moving with a linear velocity $v$. Therefore, the gravitational field the Universe will create due to the motion of the particle is given by 
\begin{equation}
\vec{E} = - \vec{\nabla} \Phi -\frac{\partial \vec{\mathcal{A}}}{\partial t} = -\Phi \frac{\partial \vec{v}}{\partial t}   
\end{equation}

\noindent Here the $\vec{\nabla} \Phi = 0$ as we are considering an isotropic Universe and also the variation of $\Phi$ with respect to $t$ is significantly small and hence $\frac{\partial \Phi}{\partial t} = 0$. So, in the rest frame of the particle, an observer will experience this kind of gravitation field, and if the passive gravitational mass of the test particle is $m_g$ then the particle will see a gravitational force $\vec{F}=m_g\vec{E}$ coming from the Universe. Here it's important to note that if we define the inertial mass $m_0=m_g\Phi$, then it will recover Newton's law. Therefore, we don't need to consider any additional pseudo forces to balance the equations. 

Exactly like electromagnetism, we can also introduce a gravitomagnetic field given by $\vec{B} = \vec{\nabla} \times \vec{\mathcal{A}}$. However, in the above case $\vec{\nabla} \times \vec{\mathcal{A}} = 0$.

\subsubsection*{Circular motion}

When the test particle is static with respect to the Universe, the gravitational potential on the test particle from the Universe is given by~\cite{sciama1953origin} 
\begin{equation}
\vec{\mathcal{A}} = 0 \qquad \text{and} \qquad\Phi = -G\int_V \frac{\rho}{r}dV \,.  
\end{equation}

\noindent If the test particle rotates in an orbit, then in the test particle's reference frame, the Universe should rotate. Therefore, the rotating Universe should create a potential on the test particle. If the particle rotates at an angular velocity $\omega$ in the $x-y$ plane, then the four potential can be written as 

\begin{align}
\mathcal{A}_x = \omega y I\,,\qquad \mathcal{A}_y = -\omega x I\,,\qquad \mathcal{A}_z = 0\,,\qquad 
\Phi = - [1+\omega^2 r^2]^\frac{1}{2} I\,.
\end{align}

\noindent where $r^2 = x^2 + y^2$ and $I = G\int_V \frac{\rho}{r}dV$.

As before, we can calculate the gravitational field of the Universe as 
\begin{equation}
\vec{E} = - \vec{\nabla} \Phi -\frac{\partial \vec{\mathcal{A}}}{\partial t} = \frac{\omega^2 r}{(1+\omega^2 r^2)^\frac{1}{2}} I \approx \omega^2 r I \;, \quad \text{for} \quad \omega r \ll 1\,.
\end{equation}

Therefore, in the rest frame of the Universe, the test particle will follow the standard Newtonian equations, and it will have the centripetal force. However, in the rest frame of the particle, the Universe is providing it a gravitation field $\vec{E}$, which has similar expressions as the centripetal force and will act as the centrifugal force but with a negative sign. Therefore, we don't require any pseudo forces. 

Here the gravitomagnetic field is also nonzero and is given by 

\begin{equation}
    \vec{B} = \vec{\nabla} \times \vec{\mathcal{A}} = 2\vec{\omega}I \,.
\end{equation}

If the test particle is not moving then it will not see this force. However, if a second test particle moves at a velocity $\vec{v}$ in this frame, it will experience a force due to this gravitomagnetic effect, and similar to the electromagnetism it will be given by 

\begin{equation}
    \vec{v}\times \vec{B} = 2\vec{v} \times \vec{\omega} I \,.
\end{equation}

\noindent This field corresponds to the Coriolis field in the Newtonian theory. Finally, we can also recover the Euler force in an accelerated rotation using straightforward calculations. 

At a given time, the value of $I$ should remain constant. However, as the Universe expands, the value of $I$ will change. Ideally, $I$ should be equal to $1$, assuming Sciama's theory is correct. Nonetheless, Newton's gravitational law is not applicable in an expanding Universe where the gravitational field varies with time. In Appendix~\ref{gravitypotentialuniverse}, we demonstrate that an approximate calculation of the Newtonian potential (assuming the Newtonian gravitational law holds in the  entire Universe) results in $I \sim \mathcal{O}(1)$. In fact, if we assume that the Universe has no dark energy content then $I$ will be  exactly $1$.  In Appendix~\ref{generalizedinertia}, we show how all inertial forces arise in a non-inertial reference frame in the non-relativistic and relativistic case.

\subsubsection{ Limitations of the theory}
\label{LimitationsSciama}
Even though Sciama's theory brilliantly incorporates Mach's principle, there are multiple limitations (most of which are pointed out by Sciama himself in the original article). 

\begin{enumerate}[wide, labelwidth=!, labelindent=0pt]
\item The theory has been derived for a vector potential from Newtonian mechanics. However, the field equations of gravity are given by the variation of $g_{\mu\nu}$. Therefore a vector field equation can not provide a complete theory of gravity, and we need a tensor field equation. Also, the theory has not addressed the effect of expanding Universe. 

\item In \cite{sciama1953origin}, it is assumed that the integral 
$I\rightarrow 1$ leads to the correct value of the inertial forces. However, the value will depend on the content of the Universe and the distribution of matter in the Universe. 
The gravitational force depends on the passive gravitational mass of the test particle and the inertial force depends on the inertial mass. Therefore, we need to make the integral unity to achieve equality between the passive gravitational mass and the inertial mass. Therefore, $\Omega_m$, $\Omega_r$, and $\Omega_\Lambda$ of the Universe need to be such that the integral is unity.

\item Most of the contribution of the inertia comes from the distant objects of the Universe. The local objects do not have that much influence on the inertia of the test particle because most of the contribution in the integral comes from the objects from far away. The integral, even at the galactic or cluster scale, is significantly small about ($\sim 10^{-7} - 10^{-9}$) of the full contribution. While this is not a drawback, this is also an important property of the theory. 
\end{enumerate}

\subsection{Brans Dicke theory and the variation of the  gravitation constant}
\label{sec:bd}

Sciama's theory provides an excellent description of inertial effects on a reference frame arising from its motion relative to the averaged-out mass distribution of the Universe. However, the mass distribution of the Universe is nonuniform, e.g., the density of the Universe at two different redshifts is different. Therefore, if we place two particles at those different redshifts and calculate their inertial responses, we should expect those responses to differ. 

Brans and Dicke in~\cite{Brans1961} addressed this particular aspect of Mach's principle by proposing that effect can be accounted for through a varying Newtonian gravitational constant. They presented an interesting interpretation of the idea that inertia arises from gravitational interactions with distant matter in the Universe. Their reasoning is as follows:

\noindent Consider a test body that is falling towards Sun. In a coordinate system, where the object is not accelerating, the Sun's gravitational pull may be considered balanced by another gravitational pull from the rest of the Universe, which we call an inertial reaction. Now if we double all the gravitational forces, the balance is not getting disturbed. Thus the acceleration is determined by the gravitational pull of the Universe but is independent of the strength of the gravitational interaction. If the active gravitational mass of Sun is $m_s$ and the distance of the test particle from the Sun is $r$, then the acceleration of the test particle due to the gravity of Sun will be $a=Gm_s/r^2$. As the acceleration is independent of the gravitational constant $G$ through the dimensional analysis, we can write $a\sim m_s R_Uc^2 / M_U r^2$, where $M_U$ can be considered as some effective mass of the visible Universe and $R_U$ is the Hubble radius near the test particle. These two equations can be combined to get 

\begin{equation}
    GM_U / R_Uc^2 \sim 1 \qquad\quad\implies\qquad\quad G^{-1} \sim M_U / R_Uc^2\,.
    \label{eqwave11}
\end{equation}

Now, either the $M_U/R_U$ should be fixed by theory, or $G$ must vary over space and time and will be determined by the mass distribution around a point where we want to measure $G$. 

As we are considering that the inertial reaction may be interpreted as a gravitational force due to distant accelerated matter, one may expect that the locally observed values of the inertial masses of particles would depend upon the distribution of matter about the point. In such cases there is no direct way one can compare the mass of a particle at two different places in the spacetime. If we define the $\hbar$ and $c$ constants, we can define a mass

\begin{equation}
    (\hbar c/G)^{\frac{1}{2}} = 2.16\times 10^{-5}{\rm g} \quad 
    \cdots \quad(\text{in our neighborhood})
\end{equation}
and this mass can be used as a characteristic mass based on which we can measure other particles in that part of the Universe. (Here it should be noted that $\hbar$ and $c$ are defined to be constant. The above-mentioned effect can be taken care of by varying $G$ or inertial or gravitational mass; or even by any of their combination to be varying. However, for simplicity, they, in their paper consider the masses to be constant and $G$ is varying.)

Brans and Dicke assumed that $G$ is a function of a scalar field $\phi$ and that $G^{-1} \sim \phi$. In such a case the a simple wave equation for $\phi$ with a scalar matter density as a source will give an equation roughly similar to Eq.~\ref{eqwave11}. This leads to the BD Action

\begin{eqnarray}
\mathcal{S}_{BD}&=&\frac{1}{16 \pi} \int d^4 x \sqrt{-g}\left(\phi R-\frac{w_{D}}{\phi} \partial_\alpha \phi \partial^\alpha \phi\right) %\nonumber \\ &&
+\int d^4 x \sqrt{-g} \mathcal{L}_{\mathrm{M}}\,,
\end{eqnarray}

\noindent where $w_D$  is the dimensionless constant known as Dicke coupling constant. $R$ is the Ricci Scalar and $\mathcal{L}_{\mathrm{M}}$ is the Lagrangian contribution from the matter density. This gives the field equations of the Brans-Dicke theory as

\begin{eqnarray}
G_{\alpha \beta} &=& \frac{8 \pi}{\phi} T_{\alpha \beta} + \frac{w_D}{\phi^2} \left(\partial_\alpha \phi \partial_\beta \phi - \frac{1}{2} g_{\alpha \beta} \partial_\gamma \phi \partial^\gamma \phi\right) %\nonumber\\  && 
+ \frac{1}{\phi}\left(\nabla_\alpha\nabla_\beta\phi - g_{\alpha \beta} \square\phi\right), \\
\square\phi &=& \frac{8 \pi}{3 + 2 w_D} T \,.
\label{Eq21_BD_Scalar}
\end{eqnarray}

\noindent where $g_{\alpha \beta}$ is the metric tensor, $G_{\alpha \beta}=R_{\alpha \beta}-\frac{1}{2} R g_{\alpha \beta}$ is the Einstein tensor, 
$T_{\alpha \beta}$ is the stress-energy tensor, $T=T_\alpha^\alpha$ is the trace of the stress-energy tensor;
$\phi$ is the scalar field.  $\square \phi=(\sqrt{-g})^{-1}(\sqrt{-g}g^{\alpha \beta}\partial_\beta \phi)_{;\alpha}$. (please check \cite{Brans1961} for detailed discussion)

Many researchers have measured the variation of the gravitational constant. \cite{Verbiest2008} measure orbital period rates of pulsars and set a limit of
$|\dot{G}/G| = 23 \times 10^{-12} {\rm yr^{-1}}$. Other such limits from pulser timings are researched by \cite{taylor1993particle,Nordtvedt1990,kaspi1994high}. From white dwarf cooling an upper bound $\dot{G}/G = -1.8 \times 10^{-12}{\rm yr^{-1}}$ is set~\cite{Garc_a_Berro_2011,garcia1995rate} and from  a white dwarf pulsation limit of $\dot{G}/G = -1.3 \times 10^{-10}{\rm yr^{-1}}$~\cite{corsico2013independent,Althaus_2011,garcia1995rate}. Variation of the gravitational constant has also been derived from Lunar laser ranging to be $\dot{G}/G = (4 \pm 9) \times 10^{-13} {\rm yr^{-1}}$~\cite{PhysRevLett.93.261101}. Current planetary radar experiments have measured a significant linear increase of $dAU/dt = 0.15 \pm 0.04 {\rm m yr^{-1}}$, which may imply $\dot{G}/G = (-10 \pm 3) \times 10^{-13} {\rm yr^{-1}}$\cite{krasinsky2004secular}.
Constrain has also been put forward by the supernova Sn1a.  An equivalent dimensionless limit is $-0.5 < \dot{G}/(G H_0) < 1$, where $H_0$ is the Hubble parameter at present.

Brans and Dicke through their logic showed that explaining Mach's principle requires the gravitational constant, ( or some equivalent quantity)  to vary as a function of space-time. However, in my opinion, introducing a scalar field for the gravitational constant and adding a kinematic term to the action feels somewhat \textit{ad hoc}. While in accordence with their logic the gravitational constant or something equivalent must vary across the spacetime, it is unclear why it should behave as a scalar field and what it physically means. It also destroys the beauty of general relativity that only originated from the geometry of space-time. Finally, the observational data shows that the Dicke parameter $w_D > 10^4$, in which case the field equation roughly tends to GR~\cite{Psaltis:2005ai,Avilez:2013dxa,Tan:2023fyl,Amirhashchi:2019jpf}. ( note that, it is often taught that the BD theory tends to GR for $w_D \xrightarrow{} \infty$~\cite{Weinberg:1972kfs,Billyard:1998kg,barrow1997behavior,mimoso1995anisotropic,barrow1994perfect}. However this assumption brakes down when the trace of the energy–momentum tensor vanishes, as pointed out in~\cite{Faraoni:1999yp}. )

\section{Developing the Mathematical Framework for Machian Gravity \label{Sec:Theory}}

GR explains gravity with remarkable precision. Although Einstein originally developed GR with the aim of incorporating Mach's principle, it soon became clear that GR does not fully adhere to it. GR's field equations require prior knowledge of inertia, whereas in a theory that truly satisfies Mach's principle, inertia should emerge from the theory itself, based on the matter distribution in the Universe.

Sciama attempted to construct such a theory. His work showed that the background mass distribution and the relative motion of distant stars and galaxies influence the inertia of a particle. The core idea is that when the smoothed-out Universe moves relative to a particle, it creates a Newtonian gravitational field that manifests as pseudo-forces, or inertial effects. While Sciama's theory provides a compelling explanation for the origin of inertia, it has certain limitations. Notably, it does not account for non-uniform mass distributions in the Universe.

Brans and Dicke, in their theory, referenced Sciama's work but focused on a different aspect of Mach’s principle—how a particle’s inertia is affected by changing mass distributions around it. For instance, in an expanding Universe, the mass distribution evolves over time, and Brans-Dicke theory addresses how this evolution impacts gravity.

If we look closely, we can see that while each of these theories explains different aspects of gravity and Mach’s principle, together they encompass Mach’s principle in its entirety. Mach's principle is a logical principle. Therefore, over the year different authors interpreted it in multiple ways. In \cite{Bondi:1996md}, Bondi and Samuel listed eleven distinct statements that researchers pass over the year under the disguise of “Mach’s principle” (It's not an exhaustive list, but it's comprehensive enough to cover most aspects of Mach's principle). Here I have listed 8 those (rest of three are ignored as they are based on spatially closed Universe):

\hspace{.5em}
\begin{itemize}
\setlength{\itemsep}{.5em}
    \item The Universe, as represented by the average motion of distant galaxies, does not appear to rotate relative to local inertial frames.
    
    \item Newton's gravitational constant $G$ is a dynamical field.
    \item An isolated body in otherwise empty space has no inertia.
    \item Local inertial frames are affected by the cosmic motion and distribution of matter.
    \item Inertial mass is affected by the global distribution of matter.
    \item If you take away all matter, there is no more space.
    \item $4\pi \rho G T^2$ is a definite number, of order unity, where $\rho$ is the mean density of matter in the Universe, and $T$ is the Hubble time.
    \item The theory contains no absolute elements.
\end{itemize}
\hspace{.5em}

\noindent Each of these distinct statements are satisfied by at least one of the above theories. Therefore, if we can construct a unified framework that synthesizes the insights of GR, Sciama’s theory, and Brans-Dicke theory, it may be possible to develop a comprehensive theory that fully realizes Mach’s principle in its entirety.

\subsection{Tensorial Reformulation of Sciama’s Theory}
According to GR, particles moves in a geodesic path in the spacetime. The equation for the geodesic path for the spacetime metric $g_{\beta\gamma}$ can be written as 

\begin{equation}
\frac{\mathrm{d}^2 x^\alpha}{\mathrm{d} \tau^2}+\Gamma_{\beta \gamma}^\alpha \frac{\mathrm{d} x^\beta}{\mathrm{d} \tau} \frac{\mathrm{d} x^\gamma}{\mathrm{d} \tau}=0\,.
\end{equation}

\noindent Here $\tau$ is some affine parameter. For massive particles it can be taken as proper time of the particle. The Christoffel symbols, $\Gamma_{\beta \gamma}^\alpha$ are given by the standard formula

\begin{equation}
\Gamma_{\beta \gamma}^\alpha=\frac{1}{2} g^{\alpha \rho}\left(\frac{\partial g_{\beta \rho}}{\partial x^\gamma}+\frac{\partial g_{\gamma \rho}}{\partial x^\beta}-\frac{\partial g_{\beta \gamma}}{\partial x^\rho}\right)\;.
\end{equation}

It is straight forward to show that the Hamiltonian that leads to the above geodesic equation is given by

\begin{equation}
H = \frac{1}{2m_0} g^{\alpha\beta}p_\alpha p_\beta \,.
\label{Eq:18}
\end{equation}

\noindent Here $p_\alpha$ are the conjugate momentum to the coordinate $x^\alpha$. Also we have assume that $m_0$ is the inertial mass of the particle. The Hamiltonian Equation of motion for this particle moving freely in the spacetime can be written as 
\begin{eqnarray}
\frac{\mathrm{d} x^\alpha}{\mathrm{d} \tau}=\frac{\partial H}{\partial p_\alpha}, \qquad\quad \frac{\mathrm{d} p_\alpha}{\mathrm{d} \tau}=-\frac{\partial H}{\partial x^\alpha}\;.
\label{Eq:19}
\end{eqnarray}

\noindent According to the Sciama's theory the entire Universe exerts a gravitational potential on an object. Thus, in the observer's reference frame, any motion of the background induces a vector potential, which is proportional to the observer's proper velocity. Sciama in the theory treated the gravitational potential from the background as a vector field as an extension to the Newtonian gravity. However, here for formulating the theory, we will assume that the background will provide a vector potential instead of going into details of why the field from the background will act as some vector field instead of a spin 2 field. 
For formulating a theory, based on this, let us first define the vector potential to be $\mathcal{A}_\alpha$ and the passive gravitational mass of the particle to be $m_g$. While building our theory we should follow the weak equivalence principle, meaning that the ratio of gravitational to inertial mass for a particle should remain constant at any given point in the Universe. However, we do not assume the strong equivalence principle, so this ratio may vary across different spacetime locations. Accordingly, in this formulation, we treat gravitational and inertial masses as independent quantities.

Now let us write the vector potential as $\mathcal{A}_\alpha = I A_\alpha$, where $I=G\int_V\frac{\rho}{r}dV$, i.e. some Newtonian type of potential from the entire Universe on the particle as described in the Sciama's theory. As discussed earlier, calculating this Newtonian-type potential for the whole Universe is challenging, as Newton’s law of gravitation may not hold at cosmological scales. However, for our purposes, the precise value of this quantity is not required, so we will simply denote it by $I$. Given that the Universe, in my reference frame, produces a vector potential $\mathcal{A}_\alpha$, the Hamiltonian for this setup, following the structure of electromagnetism, can be expressed as 

\begin{eqnarray}
H &=& \frac{1}{2 m_0} g^{\alpha \beta}\left(p_\alpha-m_g I  A_\alpha\right)\left(p_\beta- m_g I A_\beta\right) \nonumber \\
&=& \frac{1}{2 m_0}\left[g^{\alpha \beta} p_\alpha p_\beta-2 m_g I A^\alpha p_\alpha+\left(m_g I\right)^2 A^\alpha A_\alpha\right]\,.
\label{Eq:20}
\end{eqnarray}

Interestingly this Hamiltonian can also be written exactly in the same form as Eq.~\ref{Eq:18} but in 5 dimension, i.e. 
\begin{equation}
    H = \frac{1}{2m_0} \widetilde{g}^{AB} p_A p_B\,. 
    \label{Eq:27}
\end{equation}

\noindent The Latin indices $A,B,...$ runs from $0$ to $4$ and $\widetilde{g}^{AB}$ is the 5 dimensional metric. Separating out the fifth dimension, we can get

\begin{equation}
    H = \frac{1}{2m_0} \left(\widetilde{g}^{\alpha\beta} p_\alpha p_\beta + 2\widetilde{g}^{\alpha 4} p_\alpha p_4 + \widetilde{g}^{4 4} p_4 p_4 \right)\,. 
\label{Eq:22}
\end{equation}

\noindent However, the original Hamiltonian was not a function of $x^4$. Therefore, $p_4$ must be a constant of motion in accordance with the Hamiltonian Equation Eq.~\ref{Eq:19}. By comparing Eq.~\ref{Eq:20} and Eq.~\ref{Eq:22}, we can get

\begin{eqnarray}
\widetilde{g}^{\alpha \beta}=g^{\alpha \beta}, \qquad\quad \widetilde{g}^{\alpha 4} p_4=-m_g I  A^\alpha,  \qquad\quad %\nonumber \\ 
\quad \widetilde{g}^{44} p_4 p_4=\left(m_g I\right)^2 A^\alpha A_\alpha  \;.  
\end{eqnarray}

\noindent We know $p_4$ is a constant of motion. Therefore, let us assume that $p_4 = \left(\frac{m_g I }{\upsilon}\right) $, where $\upsilon$ is some other constant. So, in terms of $\upsilon$, the metric can be written as 

\begin{eqnarray}
\widetilde{g}^{\alpha \beta}=g^{\alpha \beta}, \qquad\qquad \widetilde{g}^{\alpha 4} =-\upsilon A^\alpha, \qquad\qquad %\nonumber \\ 
\quad \widetilde{g}^{44} =1 + \upsilon^2 A^\alpha A_\alpha  \;.  
\label{Eq:24}
\end{eqnarray}

\noindent Here in the last term we have arbitrarily added a constant term $1$, because we want the metric $g^{\alpha\beta}$ to be non-singular when $A^\alpha A_\alpha = 0$. The inclusion of this additional $1$, is equivalent to adding a constant term to the Hamiltonian in Eq.~\ref{Eq:22} which does not affect the equation of motion in anyway. 

We can easily calculate the inverse metric as

\begin{eqnarray}
\widetilde{g}_{\alpha \beta}=g_{\alpha \beta}+\upsilon^2 A_\alpha A_\beta, \quad \widetilde{g}_{\alpha 4}=\upsilon A_\alpha, \quad \widetilde{g}_{44}=1 \,.
\label{metric_lower}
\end{eqnarray}

We have taken the signature of the metric to be $(-1, 1, 1, 1, 1)$, which can be easily seen because the value of $\widetilde{g}_{44}$ is always positive.

\begin{eqnarray}
\widetilde{g}_{A B} \mathrm{~d} x^A \mathrm{~d} x^B &=& \widetilde{g}_{\alpha \beta} \mathrm{d} x^\alpha \mathrm{d} x^\beta+2 \widetilde{g}_{\alpha 4} \mathrm{~d} x^\alpha \mathrm{d} x^4+\widetilde{g}_{44} \mathrm{~d} x^4 \mathrm{~d} x^4 \nonumber\\
&=& g_{\alpha \beta} \mathrm{d} x^\alpha \mathrm{d} x^\beta+\left(\upsilon A_\alpha \mathrm{d} x^\alpha+\mathrm{d} x^4\right)^2    \,.
\end{eqnarray}

This is, of course, the standard Kaluza-Klein metric~\cite{delCastillo:2020wka}, but a more generalized form involve a the scalar field $\phi$. An interesting point to note is that under the transformation  $x^4 \rightarrow x^4 + \upsilon \lambda$, the field $A_\alpha$ transforms as $A_\alpha \rightarrow A_\alpha + \upsilon \partial_\alpha \lambda$.

It is also noteworthy that the five-dimensional metric inherently accounts for both the background and the motion of the particle. In contrast, special relativity does not incorporate the effect of the background, thus cannot account for pseudo-forces. Here, $I$ does not necessarily need to be unity, as required by Sciama’s theory. The additional constant $\upsilon$ can take care of this.

We can derive the equation of motion of the particle from this 5D metric as (check Appendix~\ref{KaluzaKleinEquationC2} for details)

\begin{equation}
\frac{d^2 x^\mu}{d\tau^2} +\Gamma^\mu_{\nu\lambda}\frac{d x^\nu}{d\tau}\frac{d x^\lambda}{d\tau} = \left(\frac{m_g I}{m_0}\right)g^{\mu\alpha}F_{\nu\alpha}\frac{dx^\nu}{d\tau} \,. 
\label{Eq:33}
\end{equation}

If we assume that the vector field is constant, i.e., there is no influence from the background, the motion of a free particle follows the standard four-dimensional geodesic, and the right-hand side of the equation is zero. However, if we consider that the background can influence the particle's motion, the vector field appears on the right-hand side. In an inertial frame, the vector field remains constant, leading to a zero contribution from the right-hand side. Conversely, in a non-inertial reference frame, the background velocity is dependent on the space-time. This 
generates a field tensor similar to an electromagnetic tensor. The right hand side of the equation gives all the pseudo forces.

\subsubsection{Pseudo forces from the vector field}
\label{sec322}

Sciama demonstrated that, in the non-relativistic limit, the vector potential arising from distant objects can produce pseudo-forces. However, in the relativistic case, the situation is more complex and requires detailed mathematical analysis. To address the relativistic case, we need to show that, locally any pseudo force or acceleration can be expressed as a pseudo-antisymmetric tensor multiplied with the four-velocity. Since introducing the vector field in our theory naturally yields an antisymmetric tensor, it follows that the vector field can provide the pseudo-forces in the relativistic regime. 

For our calculations, we will consider the $p_4$ to be constant and the metric to be independent on $x^4$ as discussed in the above section.
We can always define a flat spacetime in a sufficiently small region. Suppose that, in my reference frame, the particle has an acceleration given by $a^\mu = \frac{dU^\mu}{d\tau}$, where $U^\mu = \frac{dx^\mu}{d\tau}$ represents the four-velocity of the particle. In the locally flat spacetime, we can express $\eta_{\mu\nu} U^\mu U^\nu = -1$. By differentiating this with respect to $\tau$, we obtain~\cite{friedman2013covariant}

\begin{equation}
    \eta_{\mu\nu}U^\mu a^\nu = 0 \,.
    \label{eq37}
\end{equation}

\noindent We can also write the the four-velocity as 
\begin{equation}
    U^\mu = \gamma (1,\vec{v})\,,
\end{equation}

\noindent where $\vec{v} = \frac{dx^i}{dt}$ is the 3-velocity, and $\gamma$ is given by 
\begin{equation}
    \gamma = \frac{1}{\sqrt{1-\left|\vec{v}\right|^2}}\,.
\end{equation}

Now if in our reference there is some acceleration then we should see that there is some force which is acted on the particle. If the force is $\vec{F}$, then we can write the force as 

\begin{equation}
    \vec{F} = m_0 \left[\frac{d(\gamma\vec{v})}{dt} - \vec{\Omega}\times \vec{v}\right] \,.
\end{equation}

\noindent Some simple algebraic manipulations can show that 

\begin{equation}
    \frac{d\gamma}{d\tau} = \frac{\gamma \,\vec{v}\,. \vec{F}}{m_0} \,,\qquad\qquad \frac{d(\gamma \vec{v} )}{d\tau} = \frac{\gamma}{m_0}\vec{F} +  \vec{\Omega} \times (\gamma\vec{v})\,.
    \label{eqn32}
\end{equation}

\noindent where $\vec{\Omega}$ is a 3 vector related to the angular velocity. We can combine these and write these as

\begin{equation}
    \frac{dU^\mu}{d\tau} = D^\mu_\nu U^\nu \,.
    \label{eqn33}
\end{equation}

\noindent From the above equation, we can see that $D^\mu_\nu$ is a pseudo-antisymmetric tensor (i.e. $D_{\mu\nu}$ is an antisymmetric tensor). However, to show it explicitly, we can use Eq.~\ref{eq37}

\begin{equation}
    \eta_{\mu\nu}U^\mu a^\nu = \eta_{\mu\nu}U^\mu D^\nu_\lambda U^\lambda = D_{\mu\lambda}U^\mu U^\lambda = 0 \,.
\end{equation}

\noindent Therefore, it implies $D_{\mu\nu} = - D_{\nu\mu}$. From Eq.~\ref{eqn32} and Eq.~\ref{eqn33}, we can see that $D_{\mu\nu}$ is given by $D_{00} = 0$, $D_{i0} = -D_{0i} = F_i$ and $D_{ij} = \epsilon_{ijk} \Omega^k$, where $i,j,k\in(1,2,3)$ and $\epsilon_{ijk}$ is a Levi-Civita symbol. The antisymmetric tensor can be understood as follows. The Lorentz transformation produces a rotation in Minkowski spacetime. Similarly, four-acceleration can be interpreted as a rotation of the four-velocity, since the four-acceleration is orthogonal to the four-velocity.

From Eq.~\ref{Eq:33}, we observe that the pseudo-force originates from a vector field. Thus, in a local Minkowski spacetime, the motion of the background can always generate a vector potential capable of producing the required pseudo-force. For instance, consider a lab that is accelerating. An observer inside the lab would perceive all distant objects as having some acceleration. Locally, the spacetime remains Minkowski for her, but she experiences a force resembling an electromagnetic force, as if she possesses some kind of electric charge.

Note that the Hamiltonion in Eq.~\ref{Eq:20} essentially lead us to the  field equation Eq.~\ref{Eq:33}. If $(p_\mu + A_\mu)$ is fixed then the Hamiltonian will remain fixed and thereby the field equation will remain unchanged. So, given $p_\mu$ we can always calculate the vector potential $A_\mu$ in the reference frame that can keep the Hamiltonian fixed.

Here one should realize that $A_\mu$ depends on the reference frame. If we define different coordinate system then in a reference frame we can transform the values of $A_\mu$ from one to another coordinate by simple tensor transform. However, if we take two non-inertial reference frames which are accelerating with respect to one another, the vector of $A_\mu$ itself  in those two reference frames will be different. Therefore, its not mare a transformation of coordinates. In an inertial reference frame the values of $A_\mu$ will be constant, and $F_{\mu\nu} = 0$.

To better understand this, lets say there are two reference frames. $K$ is an inertial reference frame and $K'$ is an non-inertial reference frame. Suppose in $K$ frame an observer is observing a particle accelerating at $\frac{d^2 x^\mu}{d\tau^2} = 0$. In $K$ frame $F_{\mu\nu} =0$. If we do the coordinate transform and then we can find that an observer in $K'$ frame will see its acceleration as $\frac{d^2 x'^\alpha}{d\tau^2} = -\Gamma'^\alpha_{\beta\gamma}\frac{d x'^\beta}{d\tau}\frac{d x'^\gamma}{d\tau}$. As $F^{(K)}_{\mu\nu}$ were $0$, it will remain $0$ even after coordinate transform, i.e. $F'^{(K)}_{\alpha\beta}$ is $0$. So from the perspective of an observer sitting in $K$, an observer in $K'$ frame will detect and acceleration $\frac{d^2 x^{\prime \alpha}}{d\tau^2}$. However, for the observer who is sitting in $K'$ frame for him the local spacetime in Minkowskian and the Christoffel's symbols are $0$, but he sees pseudo forces. So those pseudo forces will come from the vector potential from the background. Therefore, $F^{(K')}_{\alpha\beta}\ne 0$. Or we can say that $F_{\alpha\beta}^{(K')} \ne F_{\alpha\beta}'^{(K)}$. In Appendix~\ref{rotating_frame}, I have provided an illustrative example that sheds more light on this.

\subsection{Adding the scalar field}

While deducing our theory based on the Sciama's vector potential theory, we have not considered the possibility of an expanding Universe. In an expanding Universe its not possible to have a background that has no acceleration. (The net acceleration of the entire background may be zero. However, since distant galaxies are receding from us, the matter distribution of the Universe cannot remain fixed with respect to any fixed point.) Therefore, we need further modification in the theory. We can modify Eq.~\ref{Eq:24}, where we initially added a $+1$ to the $g^{44}$ term. Instead of adding $+1$, we can introduce a scalar field $\phi$ and use $g^{44} = A^\beta A_\beta +\frac{1}{\phi^2}$. This adjustment will not alter the vector potential equations; the momentum will still be balanced by the vector potentials without any changes in that regard. However, this additional scalar field can counterbalance some of the derivatives of the $g^{\mu\nu}$ term. The equation of motion for this case can then be written as

\begin{eqnarray}
\frac{d^2 x^\mu}{d\tau^2} +\Gamma^\mu_{\nu\lambda}\frac{d x^\nu}{d\tau}\frac{d x^\lambda}{d\tau} &=& \left(\frac{m_g I}{m_0}\right)g^{\mu\alpha}F_{\nu\alpha}\frac{dx^\nu}{d\tau}  %\nonumber\\ && 
+ \left(\frac{m_g I}{m_0 \upsilon}\right)^2 g^{\mu\beta} \frac{\partial_\beta \phi}{\phi^3}\,.
\end{eqnarray}

\noindent Thus, the right-hand side term, which combines the derivative of a scalar field along with the vector field term, can counterbalance the term containing the Christoffel symbols in any non-inertial coordinate system. 

Interpreting these terms, especially the scalar field term is somewhat challenging. In an expanding Universe, any chosen reference frame is inherently non-inertial, as all distant objects (smoothed out Universe) will go away from it. As a result, the gravitational field acting on an object induces a pseudo-force. This pseudo-force can be accounted for by the scalar field. To better understand this, let us assume that the Newtonian gravitational constant varies in the Universe in a direction-dependent manner --- changing more rapidly in one direction of the Universe than in others. In such a scenario, the gravitational attraction on a particle from the matter distribution in one direction would increase or decrease more quickly than from other directions. As a result, the particle would experience an additional force originating from this uneven gravitational force. The term related to the scalar field accounts for this particular effect. This term can be related to the BD theory, as discussed later in this article.

\subsection{Variation of $x_4$ and saving the equivalence principle}

Previously, we have discussed that 
\begin{equation}
p_4 = \frac{m_g I}{\upsilon} \quad \implies \quad \frac{dx_4}{d\tau} = \left(\frac{m_g}{m_0}\right)\left(\frac{I}{\upsilon}\right) \,.    
\end{equation}

\noindent Now, $v$ is constant, but there is no reason for $I$ to remain constant. While it may have some background value, local inhomogeneities throughout the Universe can cause $I$ to vary. 

The Weak Equivalence Principle (WEP) has been rigorously tested by many researchers and states that the trajectory of a freely falling test body is independent of its internal structure or composition. In simpler terms, all objects fall at the same rate in a gravitational field, meaning the ratio between inertial and gravitational mass should be the same for all objects. WEP has been validated with remarkable precision, but it does not require that this ratio remains constant across different parts of the Universe.

Consider two objects, both falling from point $A$ to point $B$ under the same gravitational field. If the background mass distribution varies between these two points, then $I$, and hence the ratio $\frac{m_g}{m_0}$, may also vary. Although this ratio may differ across spacetime, it does so independently of the objects' composition, thus satisfying WEP. However, this does not imply that the mass ratio must be fixed everywhere in the Universe, thereby violating the strong equivalence principle and Einstein's equivalence principle.

We will now assume that the gravitational properties of matter are intrinsic. Therefore, the passive gravitational mass of an object, $m_g$, should not depend on the background matter distribution. However, the background distribution will naturally affect $I$, meaning that $p_4$ cannot remain constant and will vary at different positions in spacetime. If we allow $p_4$ to vary, then in Eq.~\ref{Eq:20}, the Hamiltonian can no longer be independent of $x^4$. This effect can be incorporated by making $g^{AB}$ dependent on $x^4$.

In fact, we can always transform to a coordinate system where $\frac{m_0}{m_g}$ remains fixed while allowing the metric to depend on $x^4$, thereby preserving the equivalence principle. To illustrate this, suppose that in different regions of the Universe, $I\left(\frac{m_g}{m_0}\right)$ varies. Let us assume we are in a primed coordinate system where the metric is $g'^{AB}$, a function of $x'_\mu$ for $\mu \in {0,3}$. Then $\frac{dx'_4}{d\tau}$ will vary along the particle’s path. However, we can always transform to an unprimed coordinate system where variation of $\frac{dx_4}{d\tau}$ exactly cancels the variation of $I$ and thereby keeping  $\frac{m_0}{m_g}$ ratios to be constant.

When changing to this coordinate system, the metric $g^{CD}$ will no longer be a function of $x_\mu$ for $\mu \in (0,3)$ alone but will also depend on the fourth variable $x_4$. Thus, $g^{CD}$ becomes a function of $x_A$ where $A \in (0,4)$. Initially, we introduced $x_4$ as a dummy variable, but we now see that we can use it to save the equivalence principles and it can have significant implications. We can now use this five-dimensional metric to calculate any equation of motion throughout the Universe without needing to account for changes in the gravitational-to-inertial mass ratio or background gravitational field.

\subsection{Developing the gravitational field equation}
\label{Section:FieldEquation}

As it is discussed above, the five-dimensional metric can be written in the 4+1 dimensional form as 
  
\begin{equation}
\widetilde{g}^{AB}=\left(\begin{array}{cc}
g^{\alpha\beta} & -\upsilon A^{\alpha}\\
-\upsilon A^{\beta} & \upsilon^2 A^{\beta}A_{\beta}+\frac{1}{\phi^{2}}
\end{array}\right)
\,.\label{eq:metric}
\end{equation}

\noindent Here, $g^{\alpha\beta}$ represents a 4-dimensional metric, $A^{\alpha}$ is a 4-dimensional vector, and $\phi$ is a scalar field. Previously, we have also argued that variations in the background gravitational field or violations of the strong equivalence principle can be addressed by varying $x^4$. However, when the above quantities depend on $x^4$, they will no longer behave as 4-dimensional vectors or tensors. Nevertheless, as we will see, this simple Kaluza-Klein type of breakdown is not strictly necessary for a gravity theory. Instead, this breakdown only helps us interpret the theory in scenarios where the Einstein's equivalence principle holds in 4-dimension (metric is independent on $x^4$) and helps in comparing the theory with other gravity theories. 

For developing the theory of gravity we can start from the Einstein-Hilbert action $\mathcal{S} =\int \sqrt{-\widetilde{g}} \left[\frac{1}{2\kappa}\widetilde{R} + \widetilde{\mathcal{L}}_M \right]d^5 x $ 
but in 5 dimension. Here $\kappa = 8\pi G$ is a constant and $\widetilde{\mathcal{L}}_M$ is Lagrangian from the matter component. Varying this action with respect the 5 dimensional metric will give us the Einstein's tensor for this 5 dimensional metric. In absence of any gravitating mass the field equation should be given by $\widetilde{G}_{AB}=0$, where $\widetilde{G}_{AB}$ is the 5 dimensional Einstein's tensor. In presence of gravitating mass we need the stress energy tensor, that we will discuss in a later section. Here we should note that we mention the gravitating mass instead of vacuum, because if there is complete vacuum in the Universe or there is no background matter, then according to the theory there will be no meaning of inertia. In other words we can think of it as if the gravitating mass at any point is something that is affecting the tress energy tensor. Rest of the Universe is the background.

Few straightforward calculations can show that $\widetilde{G}_{AB}=0$, in 5 dimension translate to the following equations in the four dimension

\begin{eqnarray}
\widetilde{G}_{\mu \nu}
&=& G_{\mu \nu}-\frac{1}{\phi}\left(\nabla_\mu \partial_\nu \phi-g_{\mu \nu} \square \phi\right) %\nonumber\\ && 
-\frac{1}{2} \upsilon^2 \phi^2\left(g^{\alpha \beta} F_{\mu \alpha} F_{\nu \beta}-\frac{1}{4} g_{\mu \nu} F_{\alpha \beta} F^{\alpha \beta}\right) \label{eq:Einstein-tensor-first}\nonumber\\
&& +\upsilon^2 A_\mu A_\nu \widetilde{G}_{44}+\upsilon A_\mu\left(\widetilde{G}_{\nu 4}-\upsilon A_\nu \widetilde{G}_{44}\right) %\nonumber\\ &&
+\upsilon A_\nu\left(\widetilde{G}_{\mu 4}-\upsilon A_\mu \widetilde{G}_{44}\right) +P_{\mu\nu} \label{eq:Einstein's tensor}\\
\widetilde{G}_{4 \nu} &=& \upsilon A_\nu \widetilde{G}_{44}+\frac{1}{2} \upsilon \phi^2 g^{\alpha \beta} \nabla_\beta F_{\nu \alpha}+\frac{3}{4} \upsilon F_{\nu \alpha} \partial^\alpha \phi^2 +Q_{\nu} \\
\widetilde{G}_{44} &=& \frac{3}{8} \upsilon^2 \phi^4 F_{\alpha \beta} F^{\alpha \beta}-\frac{1}{2} \phi^2 R + U \label{BDScalar}
\end{eqnarray}

\noindent Here $F_{\alpha\beta}=A_{\alpha;\beta}-A_{\beta;\alpha}$ is the field tensor. $G_{\mu\nu}$ is the four dimensional Einstein tensor. 
%$G_{\alpha\beta}$ is a four-dimensional Einstein's tensor.
$P_{\mu\nu}$, $Q_\nu$, $U$ are the derivatives with respect to $x^4$. A general expressions for $P_{\mu\nu}$, $Q_\nu$, $U$  can be extremely complicated. Therefore, general expressions for these are not given. However, I have shown the expressions for some particular metric in a later section.  

We can see that the 4-dimensional Einstein's tensor comes up with some terms on the right-hand side, even in the absence of any matter. These extra terms, in fact, behave in the same way as matter and curve the space-time~\cite{Overduin1998}. These terms can be interpreted as if some extra energy is coming from the background (i.e. from distant stars, galaxies, etc., due to their motion with respect to the chosen reference frame ). If the background of a particle changes, then the terms on the right-hand side will change. 
Therefore, any object sitting on that part of space-time will fill a force originating from the background, as shown before, and there is no need to consider any fictitious mathematical force.

As we have discussed previously in Sec.~\ref{sec322}, in the $K'$ reference frame, the distant objects are rotating. Therefore, according to Eq.~\ref{eq:Einstein's tensor}, the $G_{\alpha\beta}$ is nonzero. This gives rise to the inertial forces required to balance the equation. The need for these additional terms in the right-hand side of Einstein's field equations were realized much before by Hoyle and Narlikar. Therefore, they added an ad-hock scalar field with negative energy, which they term as the $C$-field (check Appendix~\ref{AppendixHN}). This became the basis for energy conservation in steady-state cosmology. However, as we can see that the terms are much more complicated than a simple scalar field.

\subsubsection*{Connection with the Brans Dicke theory}

In Sec.~\ref{sec:bd}, we discussed the BD theory and explained their reasoning for proposing a varying gravitational constant. The theory suggests that the Newtonian constant $G$ changes as the Universe expands. To incorporate this concept, in their theory they introduced a scalar field in the action and derives its equation of motion.

To replicate this scenario in our equations, we can assume that derivatives with respect to $x^4$ are zero and that the vector field  vanishes. These assumptions account for the effect of an expanding Universe case as described in BD theory. Interestingly, from Eq.~\ref{eq:Einstein's tensor}, we see that $G_{\mu\nu}$ acquires a term similar to the Brans-Dicke scalar field.

\begin{equation}
\widetilde{G}_{\mu \nu}
= G_{\mu \nu}-\frac{1}{\phi}\left(\nabla_\mu \partial_\nu \phi-g_{\mu \nu} \square \phi\right) \,.
\end{equation}

This is a special case of the Brans-Dicke equation, where the Dicke coupling constant is set to $w_D = 0$. One might be concerned that observational results from various experiments suggest $w_D$ should be very large to match empirical data. Consequently, $w_D = 0$ may seem inconsistent with observational evidence.

However, it is important to note that the introduction of $w_D$ in the original BD-theory was somewhat \textit{ad hoc}. This links the  equation of motion for the Brans-Dicke scalar field to the trace of the stress-energy tensor. In our case, the equation of motion for the $\phi$ field differs from Brans-Dicke framework, it does not require a large $w_D$. We will revisit this point in a later section (Sec.~\ref{Sec:BD}), where the stress-energy tensor for the theory is discussed in detail.

\subsection{A brief explanation of the theory }

Here, we briefly outline the theory that we have developed above. The Hamiltonian is given by Eq.~\ref{Eq:27}. Under the condition that the background matter has no influence on the equations, we can set $A^\mu = 0$ and $\phi = 1$, which gives, 

\begin{equation}
    \widetilde{g}^{AB}p_A p_B = g^{\alpha \beta}p_\alpha p_\beta +p_4^2 \,.
\end{equation}

Provided the inertial mass does not depend on the background, we have $p_4 = m_0$. This leads to the equation $-E^2 + p^2 + m_0^2 = 0$. In both special and general relativity, it is assumed that the inertial mass of a particle remains constant, allowing the four-dimensional spacetime to vary. However, in Machian gravity, inertia is assumed to arise from distant objects. Thus, in five dimensions, all particles are considered to move along a null manifold, and the five-dimensional spacetime is allowed to vary. The momentum in the fifth dimension represents the inertial mass of the particle, which remains constant in any local region.

In an inertial frame, the metric components corresponding to the fifth dimension are constant, recovering the results of special and general relativity. However, in non-inertial reference frames, these metric components are functions of space-time and hence their effects must be accounted for. We also allow the inertial mass to vary over very large spacetime scales. We use the variation in the fifth dimension to keep $m_0$ fixed and to recover the equivalence principle.

Another important point is that a particle is assumed to move within a five-dimensional null hyperspace. Therefore, it allows us to choose any affine parameter. In our calculations, we consistently use the four-dimensional line element as the affine parameter.

The beauty of the theory is that it tends to various known gravity theories under specific conditions in which they were developed. The proposed framework is built upon a five-dimensional metric involving three essential elements: a scalar field $\phi$, a vector field $A^\mu$, and an extra dimension $x^4$. (note that these two fields behave as scalar and vector field only if the metric is independent of $x^4$.)

%\begin{itemize}
    %\item 
    \subsubsection*{General Relativity ($A^{\mu} \xrightarrow{} \mathrm{const}$, $\varphi\xrightarrow{} 1$, $\widetilde{g}_{AB}$ independent of $x^4$):} In solar system scale, in a frame where distant cosmic bodies appear fixed, GR performs with excellent precision. In a reference frame where the distant objects are fixed, we should have $A^{\mu} \xrightarrow{} [1, 0, 0, 0]$. We also ignore the effects due to the expansion of the Universe  in GR. This gives $\varphi\xrightarrow{} 1$. Also GR is a 4-dimensional theory and therefore all metric elements are independent of $x^4$. Its straightforward to see that under such limit Eq.~\ref{eq:Einstein-tensor-first} gives $\widetilde{G}_{\mu\nu}=G_{\mu\nu}$. If for now we assume that under the above condition the 5D stress-energy tensor resembles stress-energy tensor of GR, then we can recover GR equations. %$\widetilde{T}_{\mu\nu} - \frac{1}{2}\widetilde{g}_{\mu\nu}\widetilde{T}=T_{\mu\nu}-\frac{1}{2}g_{\mu\nu}T$ then the theory becomes identical to GR.
    
   \subsubsection*{Non-inertial frame ($\varphi\xrightarrow{} 1$, $\widetilde{g}_{AB}$ independent of $x^4$):}
    
    Lets suppose the expansion of the Universe can be neglected but the reference frame experiences acceleration due to its motion with respect to the smoothed out Universe. We expect inertial forces to appear in such a reference frame. This aligns with Sciama's description of the coordinate system. As we are ignoring the effect due to the expansion of the Universe we should have $\varphi\xrightarrow{} 1$. Sciama also took a 4-dimensional spacetime and hence $\frac{\partial}{\partial x^4} \xrightarrow{} 0$. However, $A^\mu$ are not constant here. Under such circumstances our model gives Eq.~\ref{Eq:33}, where we saw that the right hand side gives the pseudo forces as explained by Sciama in his theory. Therefore, under such circumstances the model  acts as a dual to Sciama's model.

   \subsubsection*{Brans-Dicke case ($A^{\mu} \xrightarrow{} \mathrm{const}$, $\widetilde{g}_{AB}$ independent of $x^4$):}  Brans and Dicke argued that due to the expansion of the Universe as well as due to the  local inhomogeneities the gravitational constant should vary. They replaced the gravitational constant with a scalar field and introduced a kinetic term for the scalar field as discussed in Sec~\ref{sec:bd}. Our model replicates this scenario if we put the condition that $A^{\mu} \xrightarrow{} [1, 0, 0, 0]$, $dx^4$-dependence vanishes and $\phi$ is a function of spacetime. Our framework produces a similar scalar field solution. Although the field equation for the $\phi$ term  differ from the Brans-Dicke theory (field equation for $\phi$ is discussed in Sec.~\ref{Sec:BD}). Thus, although our model is not an exact dual of the BD theory, it shares important similarities and is consistent with the logic presented by Brans and Dicke in their paper.

    \subsubsection*{Galactic/Cosmology scales ($g_{\mu\nu}$ dependent on $x^4$):} 
    At galactic scales, we assume that the Einstein Equivalence Principle may be violated, which in our theory is manifested by allowing the metric to depend on $x^4$. In such cases, depending on the circumstances, the theory can reduce to various modified gravity models that have been proposed by different researchers to explain the dark components of the Universe. For example, to account for galactic rotation curves, we may assume a weak gravitational field. Under this assumption, our theory yields a Yukawa-like potential as described in the next section, which has been widely used in the literature. Similarly in cosmological scale we can get some additional geometric components which will behave as dark components as we will discuss in Sec.~\ref{sec:Cosmology}.

\section{Static, Spherically symmetric solution for weak gravitation field}

In a vacuum, the field equation for the theory is $\widetilde{G}_{AB} = 0$, which, after some rearrangements, can be written as $\widetilde{R}_{AB} = 0$, where $\widetilde{R}_{AB}$ is the Ricci tensor. Here, I would like to emphasize again that the vacuum does not include the background matter distribution as the theory is not applicable in an otherwise empty Universe. Similar to Einstein's equations, we also match $\widetilde{G}_{AB} = \widetilde{T}_{AB}$ at every point in space, where $\widetilde{T}_{AB}$ is the five-dimensional stress-energy tensor. At any point, apart from the matter density and pressure at that point, all other matter acts as background. Suppose there is a lone star, and all other stars and galaxies in the Universe are far away from it. We aim to determine the spacetime around this star. In this region, $\widetilde{T}_{AB} = 0$. Of course, the matter distribution is not zero far away. However, that can be ignored for calculating the stress energy tensor at that point. In Einstein's GR, we can always assume that distant matter has no effect on the local dynamics. However, in Machian Gravity theory, we can only state that distant objects will not directly affect the stress-energy tensor. Instead, they influence the coordinate system, determining which coordinates are inertial and which are non-inertial.

Let us assume that the perturbation in the metric due to the gravitational field is $\widetilde{\gamma}_{A B}$. For this weak field limit, only $\widetilde{R}_{00}=\widetilde{R}^C_{0 C 0}$ is the important term, where term on the right-hand side is the Riemann tensor. The rest of the terms of the Ricci tensor will be either ${\rm O}(1/c)$ or ${\rm O}(1/c^2)$ smaller and can be ignored in the weak field limit. We can break the Riemann tensor as  

\begin{equation}
\widetilde{R}_{0 A 0}^B=\partial_A \widetilde{\Gamma}_{00}^B-\partial_0 \widetilde{\Gamma}_{A 0}^B+\widetilde{\Gamma}_{A C}^B \widetilde{\Gamma}_{00}^C-\widetilde{\Gamma}_{0 C}^B \widetilde{\Gamma}_{A 0}^C \,.
\end{equation}

\noindent The second term here is a time derivative, which vanishes for static fields. The third and fourth terms are of the form $(\widetilde{\Gamma})^2$, and since $\widetilde{\Gamma}$ is first-order in the metric perturbation, the square terms contribute only at second order and can be neglected, giving 

\begin{eqnarray}
\widetilde{R}_{00} &=& \widetilde{R}_{0 A 0}^A = \partial_A\left(\frac{1}{2} \widetilde{g}^{A C}\left(\partial_0 \widetilde{g}_{C 0}+\partial_0 \widetilde{g}_{0 C}-\partial_C \widetilde{g}_{00}\right)\right)  \nonumber\\
&=&-\frac{1}{2}  \partial_A \left(\widetilde{g}^{A B}\partial_B \widetilde{g}_{00}\right) \,.
\label{Equation51}
\end{eqnarray}

\noindent Let us consider that the expansion of the Universe can be ignored, making $\phi=1$. Also, consider an inertial reference frame such as $A^\alpha$ is constant. We can take the space part of $A^\alpha$ to be zero. For the static solution, the time derivative also vanishes.

From Eq.~\ref{metric_lower}, we have $\widetilde{g}_{00} = g_{00} + v^2 \phi^2 A_0 A_0$. Under the above conditions, the second term in this expression should be constant. We also know that in a weak gravitational field, $g_{00} = -1 + \gamma_{00}$, where $\gamma_{00}$ represents a small perturbation in the four-dimensional metric. Additionally, for convenience in writing, let us take $x^A = \{t, x, y, z, \zeta\}$.

Under the above circumstances the Eq.~\ref{Equation51} becomes

\begin{equation}
\partial_{\zeta}^{2}\gamma_{00}+\partial_{x}^{2}\gamma_{00}+\partial_{y}^{2}\gamma_{00}+\partial_{z}^{2}\gamma_{00}=0\,.\label{eq:laplace equation}
\end{equation}

\noindent Under the assumption of spherical symmetry of the special part, it can be written as

\begin{equation}
\partial_{\zeta}^{2}(r \gamma_{00})+\partial_{r}^{2}(r \gamma_{00})=0\,.\label{eq:waveequation}
\end{equation}

\noindent Let us assume that we can use `separation of variables' for solving this equation. Considering $(r \gamma_{00})=\mathcal{R}(r)\chi(\zeta)$, we get 

\begin{equation}
\frac{1}{\mathcal{R}}\frac{\partial^{2}\mathcal{R}}{\partial r^{2}}=-\frac{1}{\chi}\frac{\partial^{2}\chi}{\partial\zeta^{2}}=\lambda^{2}\,,
\label{eq:waveequation1}
\end{equation}

\noindent where, $\lambda$ is a constant. It can be real or imaginary. In other words, $\lambda^2$ can be positive or negative. Provided $\lambda^2$ is positive, it gives

\begin{equation}
\mathcal{R}=P_{1}e^{\lambda r}+P_{2}e^{-\lambda r}\,, \qquad
\chi=Q_{1}\cos(\lambda\zeta)+Q_{2}\sin(\lambda\zeta)\,,
\end{equation}

\noindent where, $P_{1}$, $P_{2}$, $Q_{1}$ and $Q_{2}$ are constants. These above solutions give the particular integrals for these differential equations. The complementary functions are given by 

\begin{equation}
\mathcal{R}_{CF}=S+S_{1}r\,, \qquad
\chi_{CF}=S_{2} +S_{3}\zeta\,,
\end{equation}

\noindent Here, again $S$, $S_1$, $S_2$ and $S_3$ are constants. We can write the full solution as 

\begin{equation}
(r \gamma_{00})=\left(S+S_1 r + P_{1}e^{\lambda r} + P_{2}e^{-\lambda r}\right)\left(S_{2} +S_{3}\zeta + Q_{1}\cos(\lambda\zeta)+Q_{2}\sin(\lambda\zeta)\right)\,.
\label{Eq.50}
\end{equation}

\noindent This is an interesting solution. We introduced $p_4 = \left(\frac{m_g I}{\upsilon}\right)$ as a constant term and the metric  is initially considered independent of $x^4=\zeta$ term. However, since $I$ can vary across spacetime, maintaining a fixed ratio $m_g/m_0$ requires the metric to depend on $\zeta$. As long as the metric remains independent on $\zeta$, Eq.~\ref{eq:waveequation1} yields $\lambda =0$, which can eventually give us the standard Newtonian potential.

However, when the metric does depend on $\zeta$, a non-zero $\lambda$ arises. If $\lambda$ is real we get the solution that I have shown here. If it is imaginary, then the wavy part will be in the radial component and the exponential part will be in the $\zeta$ component.

We know that $\frac{dx_4}{d\tau} = \frac{m_g I}{m_0 \upsilon } \sim \mathcal{O}(1)$. Putting the values of the metric from Eq.~\ref{Eq:24} we can deduce that $\frac{d\zeta}{d\tau} = \frac{dx^4}{d\tau} \sim \mathcal{O}(1)$. Given the homogeneity and isotropy of the Universe, $I$ should not vary significantly, implying $\zeta \sim \tau \sim t$. In other words $\zeta$ is a term that increases with time.

We don't want the Newtonian potentials to grow with time. To prevent that from happening we must set $S_3 = 0$. Additionally, provided $\lambda$ is real we require $Q_1, Q_2 \ll S_2$, otherwise, these terms introduce unwanted temporal variation in the potential. If $\lambda$ is  imaginary number, the solution includes a growing term like $e^{\lambda \zeta}$, which must be eliminated by setting its coefficient to zero. Otherwise, the potential would increase over time, which is unphysical. The corresponding decaying term will naturally diminish. Therefore, in any case for a realistic solution, the second part of Eq.~\ref{Eq.50} should approach a constant $S_2$.

Regarding the radial dependence of the potential, lets assume $\lambda$ is real. We must avoid exponential growth of the potential with radius, by setting $P_1$ to $0$. If $\lambda$  was imaginary, the radial solution would include oscillatory behavior, which is not observed in nature. Therefore, we will not discuss the imaginary $\lambda$ case here. ( We can definitely avoid the  oscillating behavior by setting its coefficients, $P_1, P_2 \ll S$ eventually leading to a standard Newtonian like potential. However, that can be ignored for our discussion here. )   These gives

\begin{equation}
(r \gamma_{00})=S+S_1 r + P_{2}e^{-\lambda r}\,.\label{eq:potential-equation_0}
\end{equation}

\noindent Here we absorb $S_2$ within the other coefficients. $S_1$ term only adds a constant value to the potential and will not affect any of
the calculations. Therefore, we ignore this term in the rest of the paper. 

\begin{equation}
\gamma_{00}=\frac{S}{r} + P_{2}\frac{e^{-\lambda r}}{r}\,.\label{eq:potential-equation}
\end{equation}

\noindent The equation for a geodesic path is given by 

\begin{equation}
\frac{d^{2}x^{A}}{ds^{2}}-\widetilde{\Gamma}_{BC}^{A}\frac{dx^{B}}{ds}\frac{dx^{C}}{ds}=0\,.
\end{equation}

\noindent Under the above assumptions and under the weak field limit, its straight forward to show that  $\frac{d^{2}x^{A}}{dt^{2}}=\frac{1}{2} \partial_A \gamma_{00}$ (Check the values of the $\Gamma^A_{BC}$ in Appendix~\ref{AppendixC:Christoffel}). This is the same form as we see in standard GR.  Relating it with Newtonian gravity, we get $\gamma_{00}=2\varphi$, where $\varphi$ is the Newtonian potential of the gravitational field. 

Replacing these limiting values in Eq.~\ref{eq:potential-equation} and substituting $P_2=2KM$ and $S=2(1+K)M$ and replacing $\gamma_{00} = 2\varphi$ we can get the potential as 

\begin{equation}
\varphi=\frac{GM}{r}\left[1+K\left(1-e^{-\lambda r}\right)\right]\,.\label{eq:potential}
\end{equation}

\noindent Here, M is the mass at the center, and $G$ is Newton's gravitational constant. $\lambda$ and $K$ are the background-dependent quantities. They depend on the mass distribution around an object and not on the mass of the individual objects. 

To better understand the implications of these term, consider a scenario where a large object, such as a star is located at the center, and we want to calculate the gravitational field around it. Suppose all the other objects are far away from it. Now consider a smaller object, such as a planet, is orbiting it at a radius $r$. Assume that the gravitational mass of the planet is $m_g$. As all the other objects are far away, we can always assume that in the small enough region of spacetime $p_4 = m_g I/v$ should remain constant as $I$ must be constant in this region.  
Therefore, the metric is independent of $\zeta$ and hence  $\frac{\partial \gamma_{00}}{\partial \zeta} = 0$. This implies $\lambda = 0$, and consequently reduces the equation to the standard  Newtonian gravity. Naturally, in an expanding Universe, this term would be nonzero since $\phi$ cannot be assumed constant, as discussed earlier.

Now, consider the case of a very large extended object, e.g. a spiral galaxy. In this case, even if we assume that the mass distribution outside the galaxy is fixed in our coordinate system, the galaxy’s own mass contributes significantly to the total gravitational potential of the Universe ( Earlier in Sec.~\ref{LimitationsSciama} I discussed that the contribution from galaxy or galactic cluster can be of the order $10^{-7}-10^{-9}$. However, this can still be  significant for our purpose. ).  As a result, $\frac{\partial \gamma_{00}}{\partial \zeta} \ne 0$.

Galactic velocity profiles show that in galaxies, the $\lambda$
is of the order of few $\texttt{kpc}^{-1}$. When $r$ is small, $e^{-\lambda r}\sim1$. Therefore, $\Phi$ takes the form of Newtonian potential i.e. $\Phi=\frac{GM}{r}$.
This gives the Newtonian gravitational equation at the solar system
scale. In the asymptotic limit of $r\rightarrow\infty$, the
exponential term goes to $0$. Hence, for large values of $r$, it becomes $(1+K)$ times that of Newtonian potential and can provide additional gravitational force in large gravitationally balanced systems, such as galaxies or clusters. A similar form of potential has previously been used 
by other groups to explain the galactic velocity profile correctly \cite{Moffat2009,Moffat2005,Brownstein2005,Brownstein2005a,Moffat2005a}.

\subsection{The Tully–Fisher relation}

As the potential due to a static spherically symmetric gravitational 
field is given by Eq.~\ref{eq:potential}, we can calculate the acceleration due to the gravitational field as

\begin{equation}
\frac{\partial\Phi}{\partial r}=-\frac{GM}{r^{2}}\left[1+K\left(1-e^{-\lambda r}\left(1+\lambda r\right)\right)\right]\,.
\label{eq:Gravfield}
\end{equation}

\noindent Suppose a particle is orbiting a mass $M$ in a circular orbit of radius $r$, and its orbital velocity is $v$. The gravitational field should be equal to the centripetal acceleration of the particle, giving

\begin{equation}
v^2=\frac{GM}{r}\left[1+K\left(1-e^{-\lambda r}\left(1+\lambda r\right)\right)\right] \,.
\label{eq:velocity}
\end{equation}

\noindent This is an interesting equation. For small $r$, i.e. $\lambda r \ll 1$, this equation tends to the Keplarian velocity of a particle in a circular orbit. When $\lambda r \gg 1$, $\exp(-\lambda r) \rightarrow 0$, it gives 

\begin{equation}
    v^2 = \frac{GM}{r}(1+K)\,,
\end{equation}

\noindent which is again a Keplarian velocity with a multiplication constant. Large virial masses of galaxy clusters require additional dark matter components. This additional $(1+K)$ factor can help us explain the missing mass in galaxy clusters without addition any dark matter. 

\begin{figure}
    \centering
    \includegraphics[trim=0.3cm 1.0cm 1.3cm 1.5cm, clip=true,width=0.75\columnwidth]{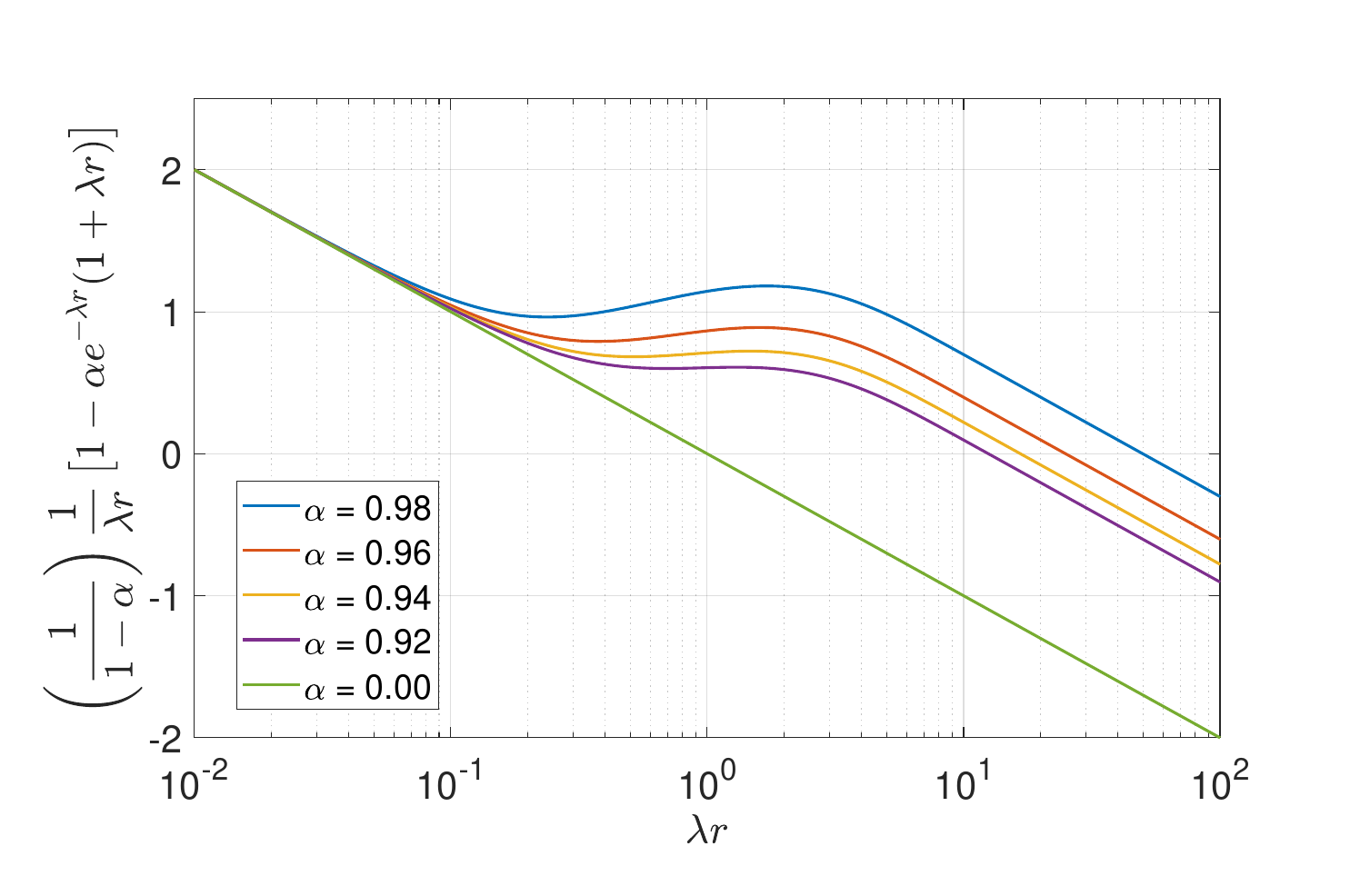}
    \caption{For different $\alpha$ there is a range of $\lambda r$ for which the curve simply flattens out.}
    \label{fig:alpha_values}
\end{figure}

Observations show that the velocity in the outer part of the spiral galaxies (rotationally bounded system) does not decrease with increasing radius, as suggested by Keplarian velocity. In fact, it is almost independent of radius $r$. Interestingly the above equation Eq.~\ref{eq:velocity} also has an attractive property. For a range of values of $r$ and $K$ the velocity becomes almost independent of $r$. This was first explained in \cite{sanders1984anti, sanders2002modified}. 

\begin{equation}
v^2=\frac{GM(1+K)\lambda}{\lambda r}\left[1- \alpha e^{-\lambda r}\left(1+\lambda r\right)\right] \,.
\label{eq:velocity1}
\end{equation}

\noindent where $\alpha = \frac{K}{1+K}$.  For $\alpha\in(0.92,0.95)$ and $\lambda r \in (0.4,2.5)$ the velocity becomes almost independent of $r$. This can be seen in Fig.~\ref{fig:alpha_values}. Therefore, this expression can explain the velocity of spiral galaxies too. From the range of $\alpha$ we can derive the range of $K$ to be $(11,19)$. 

Therefore, for this $r$ range, the velocity of the test particle behaves as $v^2 \sim GM(1+K)\lambda$. However, according to the Tully–Fisher relation, the mass of a spiral galaxy is linked to its asymptotic velocity as $M \sim v^\gamma$, where $\gamma \in (3.5,4)$. If we assume that $M\sim v^4$, then we can take 

\begin{equation}
    (1+K) \sim \frac{1}{\sqrt{M}} \qquad\implies\qquad K \sim \sqrt{\frac{M_c}{M}} - 1 \,.
    \label{eq:Kform}
\end{equation}

\noindent Here $M_c$ is some constant mass. For others  e.g. elliptical galaxy or clusters the value of $K$ can be different. Putting everything together, the expression for the final velocity becomes

\begin{equation}
v^2 \sim \frac{GM}{r}\left[1+\left(\sqrt{\frac{M_c}{M}} - 1\right)\left(1-e^{-\lambda r}\left(1+\lambda r\right)\right)\right] \,.
\label{eq:velocityFinal}
\end{equation}

Therefore, for a mass distribution similar to a spiral galaxy, this equation follows Newtonian velocity for a particle in orbit for $\lambda r \ll 1$. For $\lambda r \in (0.4, 2.5)$, velocity becomes constant and follows the Tully Fisher relation i.e., $v^4\sim M$. Finally, for $r\gg 2.5$, it behaves as $v^2\sim\frac{\sqrt{M}}{r}$. Of course, this is not an exact relation. However, the above equations give a rough form of the value of $v$. We don't have an exact form of $K$ yet. For different- types of mass distributions, the shape of $K$ may be different.

\section{Testing the theory against observations}

Observational results from distributed systems, such as galaxies and galaxy clusters, have shown that the baryonic matter calculated from the luminosity of these objects is not enough to explain their dynamical properties. We need additional matter, known as dark matter. 
Several relations are proposed by researchers %over the year 
to connect the dynamical properties of galaxies and galaxy clusters with their observed luminosity. The Tully-Fisher relation for spiral galaxies and the Faber-Jackson relation in elliptical galaxies are some of the earliest relations. Several other such relations have also been put forward~\cite{tully1977new, faber1976velocity, mcgaugh2004mass, tully1997ursa, sanders1994faber}. In this section, we try to briefly show whether the Machian gravity model can explain the observations for galaxy clusters and galaxies. The detailed analysis has been discussed in~\cite{das2023aaspects,das2023aspects5}.

\subsection{Galactic cluster mass}

Galaxy clusters are believed to consist of three principal components. As the name suggests, each cluster contains between 100 and 1000 galaxies, accounting for approximately $1\%$ of its total mass. The regions between these galaxies are not empty but are filled with substantial amounts of intracluster gas, which contributes about $9\%$ of the cluster's mass. The dominant component, comprising roughly $90\%$ of the total mass, is attributed to dark matter based on estimates using Newtonian mechanics. However, in modified gravity theories, this dark matter component is not required.

The density distribution of hot gas within a galaxy cluster and thereby the baryonic mass, can be effectively modeled using the King $\beta$-model. Analytically, the baryonic mass profile of a cluster can be calculated  as ~\cite{King1966,Brownstein2005a,Cavaliere1976}

\begin{equation}
 M_{b}(r)=4\pi \rho_0\int_0^r   \left[ 1+\left(\frac{r'}{r_c}\right)^2\right]^{-3\beta/2} r'^2dr'   \,,
 \label{eq:massformula}
\end{equation}

\noindent where $\rho_0$ is the central density and $r_c$ is known as the core radius of a cluster. $\beta$ is some fitting constant and can be found by fitting the the mean radial profile of X-ray surface brightness of the cluster. It is seen that $\beta$ is of the order of $\frac{2}{3}$.

On the other hand, the calculation of the dynamical mass of a galaxy cluster typically relies on the assumption of hydrostatic equilibrium. For isothermal clusters, which constitute the majority of observed systems, the dynamical mass can be calculated using the hydrostatic equilibrium condition and the assumption of a constant temperature profile as

\begin{figure*}[t!]
\centering
\includegraphics[trim=0.0cm 0.0cm 0.0cm 0.0cm, clip=true, width=0.48\textwidth]{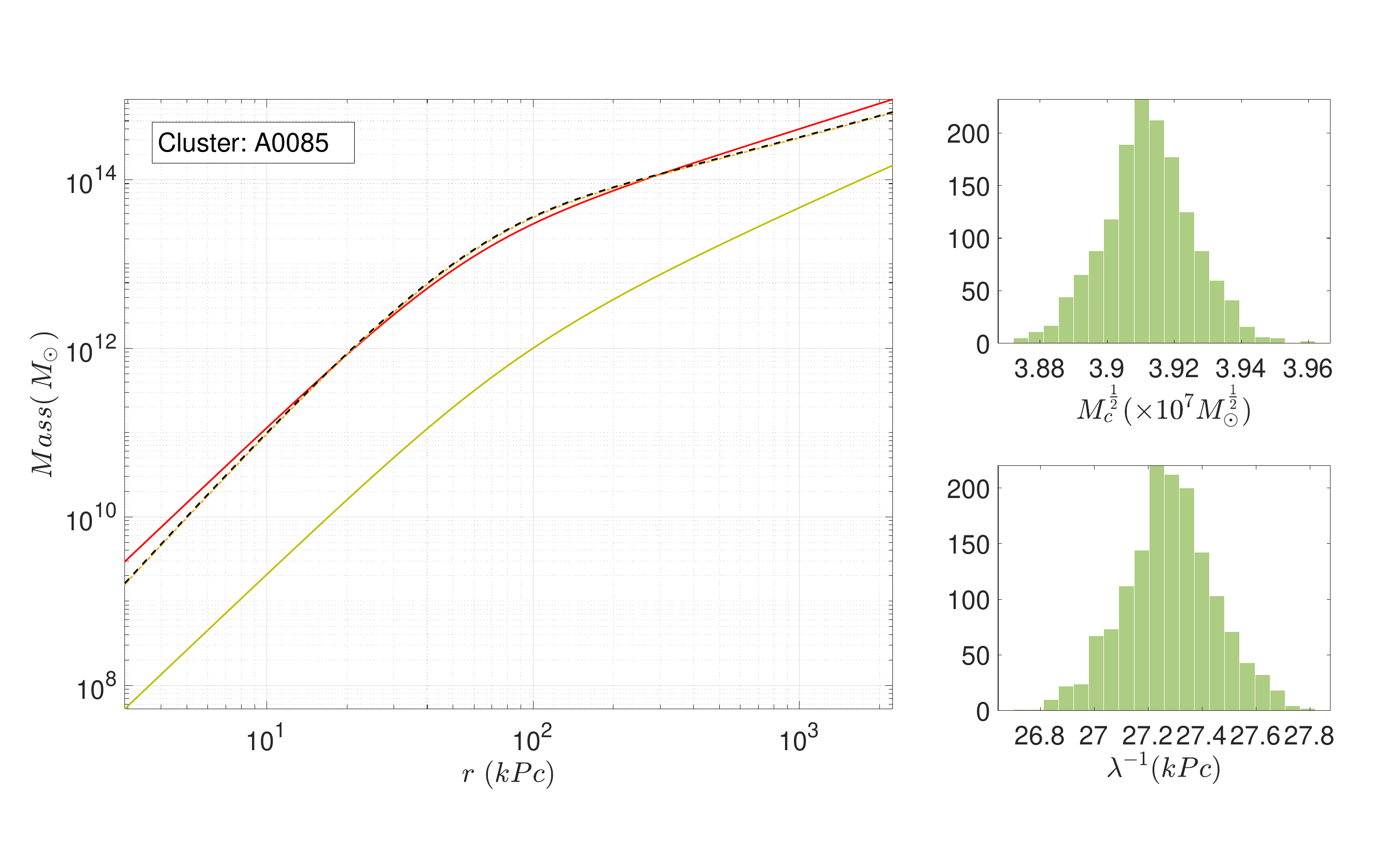}
\includegraphics[trim=0.0cm 0.0cm 0.0cm 0.0cm, clip=true, width=0.48\textwidth]{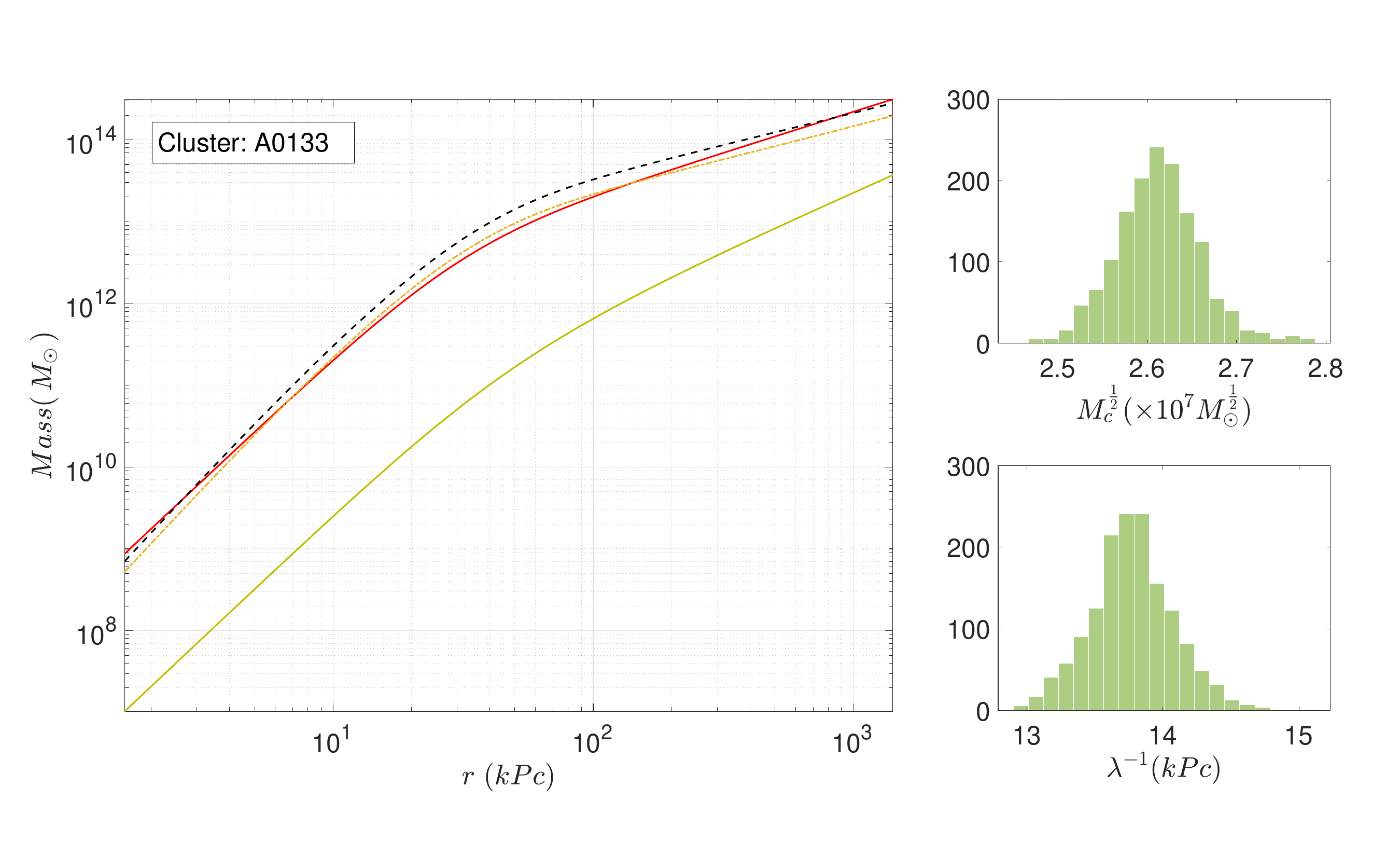}
\includegraphics[trim=0.0cm 0.0cm 0.0cm 0.0cm, clip=true, width=0.48\textwidth]{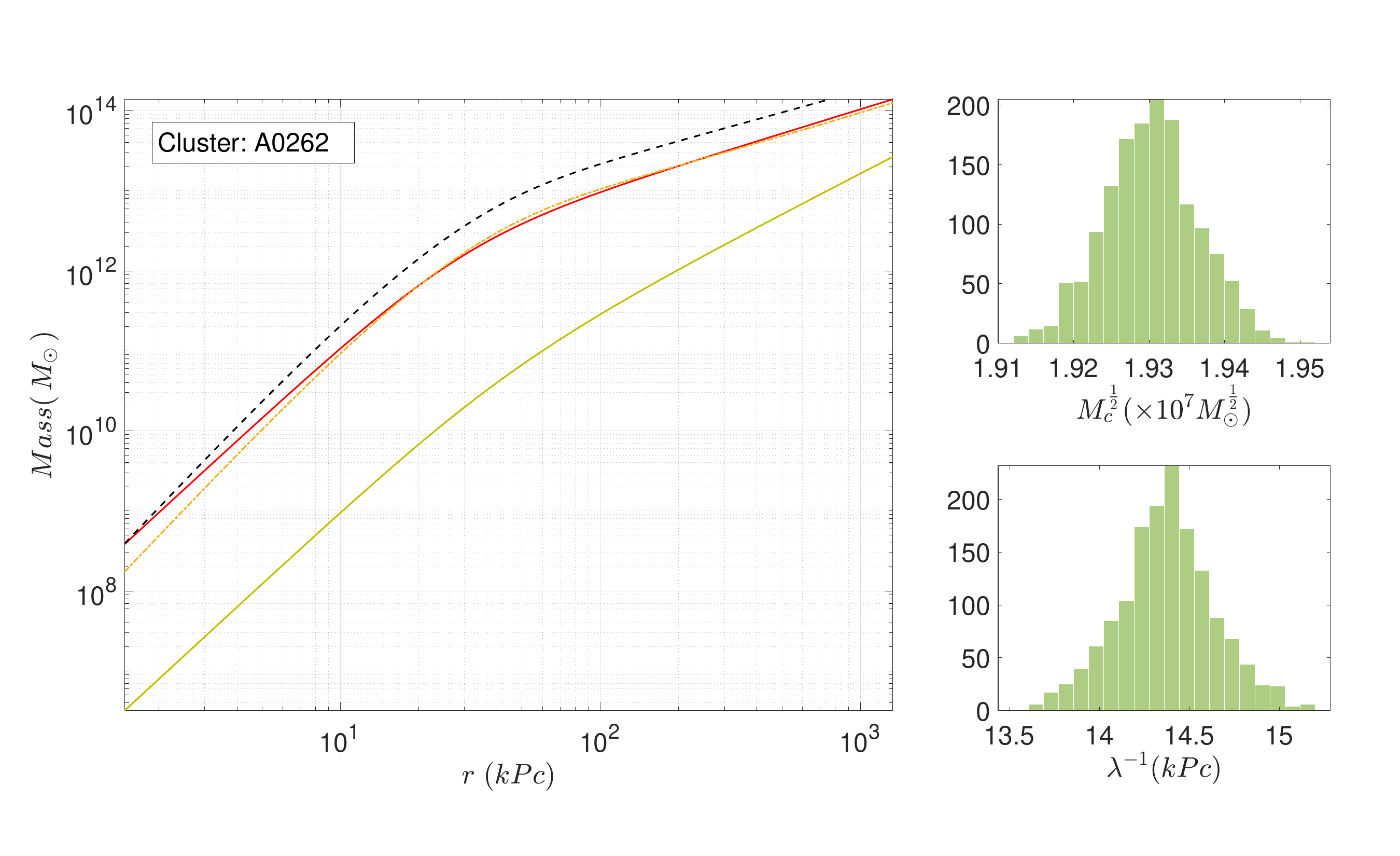}
\includegraphics[trim=0.0cm 0.0cm 0.0cm 0.0cm, clip=true, width=0.48\textwidth]{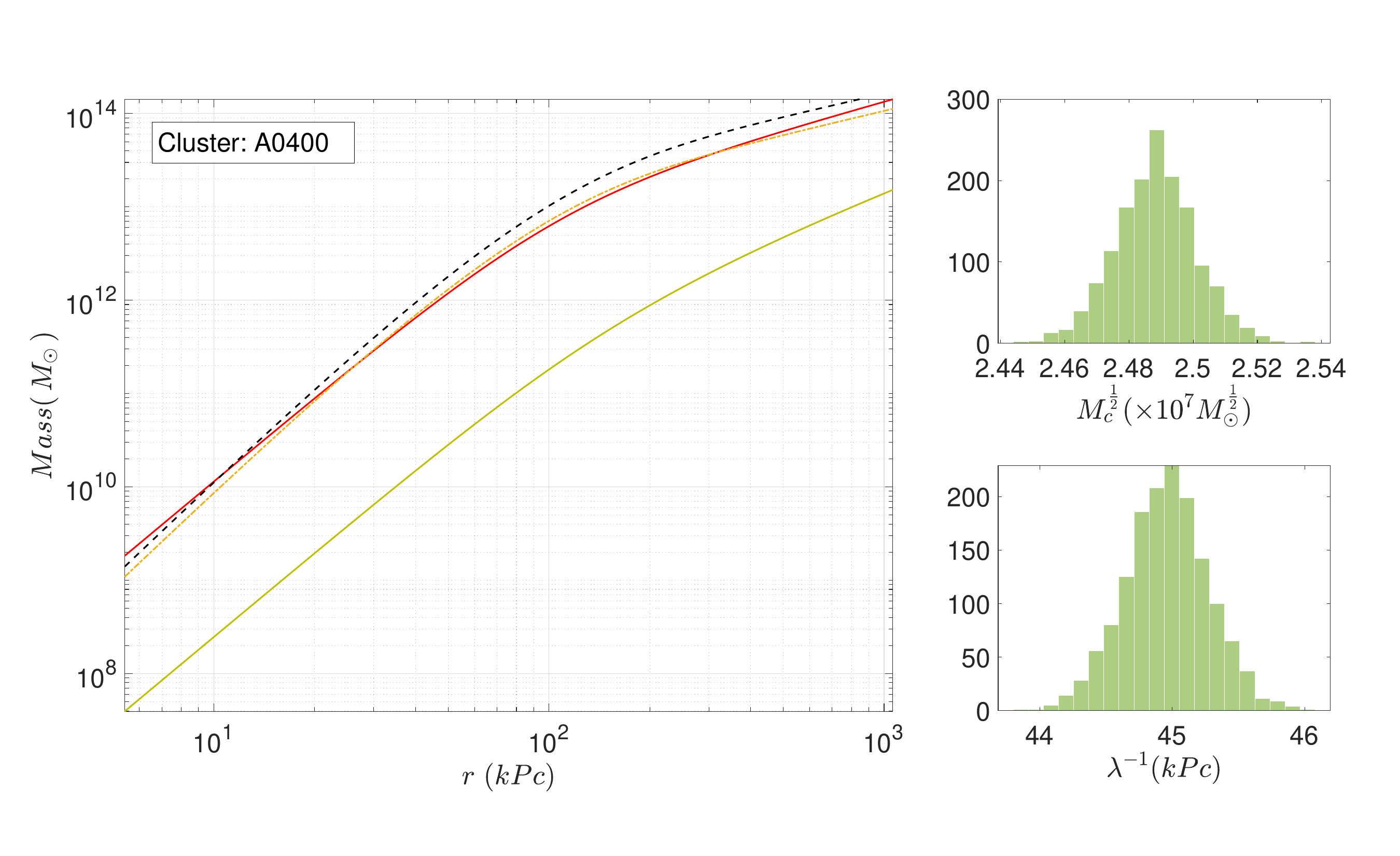}
\caption{\label{clusterdata}The plot displays the matter distribution in 4 galaxy clusters. The lime-colored solid plot represents the baryonic mass distribution calculated using King's-$\beta$ function (Eq.~\ref{eq:massformula}). The solid red curve shows the Newtonian dynamic mass calculated using Eq.~\ref{eq:M_N1}. The yellow dot-dashed line illustrates the quantity on the left-hand side of Eq.~ref{eq:}, where $\sqrt{M_c}$ and $\lambda^{-1}$ are chosen based on the best-fit values obtained from MCMC analysis. Additionally, the right-side plots depict the distributions of $\sqrt{M_c}$ and $\lambda^{-1}$ from the MCMC analysis. The black dashed line shows the same quantity but with best fit $\sqrt{M_c}$ and $\lambda^{-1}$ values calculated by analyzing 106 clusters.}
\end{figure*}

\begin{equation} 
M_d(r) = \frac{3\beta kT}{\mu m_pG}\left(\frac{r^3}{r^2+r_c^2}\right)\,.
\label{eq:M_N1}
\end{equation}

\noindent Here, $m_p$ is the mass of proton, $\mu = 0.609$ is mean atomic weight
 and $k$ is the Boltzmann constant. 

In Fig.~\ref{clusterdata}, we have shown the mass for 4 clusters as a function of radius $r$. The solid yellow curve shows the baryonic mass calculated using Eq.~\ref{eq:massformula}. The solid red curve shows that Newtonian dynamic mass, calculated using Eq.~\ref{eq:M_N1}. We can see that we need a significant amount of additional matter or dark matter to explain the dynamical properties of galaxy clusters. 

However, under Machian gravity theory the gravitational field is given by Eq.~\ref{eq:Gravfield}. Therefore, effectively the term on the right hand side of the equation would behave as the dynamic mass of the cluster, or 

%proposed in this paper, the Eq.~\ref{eq:M_N1} takes the form 

\begin{equation} 
M_d = M_b(r)\left[1+K\left( 1-\exp(-\lambda r)\left(1+\lambda r\right)\right)\right]
\label{eq:M_N}
\end{equation}

\begin{table*}[t]
\centering
\begin{tabular}{|l|c|c|c|r|r|c|c|}
\hline
Cluster & $T$ & $\rho_{0}$ & $\beta$ & $r_{c}$ & $r_{\rm out}$ & $M_{c}^{\frac{1}{2}}$ & $\lambda^{-1}$ \\[0.1cm]
\, & [keV] & [$10^{-25}\,{\rm g/cm}^{3}$] & \, & [kpc] & [kpc] & [$10^{7} M_{\odot}$] & [$r_c$]\\[0.1cm]
\hline \rule{0pt}{0.4cm}
$\texttt{A0085}$&$6.90^{0.40}_{-0.40}$&$0.34$&$0.532^{0.004}_{-0.004}$&$58.5^{3.3}_{-3.9}$&$2241.0^{139.0}_{-162.0}$&$3.91^{0.01}_{-0.01}$&$0.47^{0.003}_{-0.003}$\\[0.1cm]
\hline \rule{0pt}{0.4cm}
$\texttt{A0133}$&$3.80^{2.00}_{-0.90}$&$0.42$&$0.530^{0.004}_{-0.004}$&$31.7^{1.9}_{-2.3}$&$1417.0^{96.0}_{-109.0}$&$2.62^{0.04}_{-0.04}$&$0.43^{0.009}_{-0.009}$\\[0.1cm]
\hline \rule{0pt}{0.4cm}
$\texttt{A0262}$&$2.15^{0.06}_{-0.06}$&$0.16$&$0.443^{0.018}_{-0.017}$&$29.6^{8.5}_{-7.2}$&$1334.0^{432.0}_{-386.0}$&$1.93^{0.01}_{-0.01}$&$0.49^{0.009}_{-0.009}$\\[0.1cm]
\hline \rule{0pt}{0.4cm}
$\texttt{A0400}$&$2.31^{0.14}_{-0.14}$&$0.04$&$0.534^{0.014}_{-0.013}$&$108.5^{7.8}_{-8.8}$&$1062.0^{97.0}_{-108.0}$&$2.49^{0.01}_{-0.01}$&$0.41^{0.003}_{-0.003}$\\[0.1cm]
\hline
\end{tabular}
\caption{\label{tab:clusterProperties} The table listed the properties of the four X-ray Clusters. We have listed the temperature ($T$), central density ($\rho_0$), $\beta$, and the core radius $r_c$. $r_{\text{out}}$ is the radius where the gas density of the cluster falls below 250 times the mean cosmological baryonic density. $M_c^{\frac{1}{2}}$ and $\lambda^{-1}$ are the results from the MCMC run.}
\end{table*}

\noindent Here we have considered $K$ to be of the form given in Eq.~\ref{eq:Kform}, i.e. $K=\left(\sqrt{\frac{M_c}{M_b(r)}} -1 \right)$. However, there is no strong reason for $K$ to behave like this. A detailed study is shown in~\cite{das2023aspects5} and is beyond the scope of this article. We have calculated the best-fit values of the parameter $\sqrt{M_c}$ and $\lambda^{-1}$ for each of the clusters  through MCMC analysis. In Fig.~\ref{clusterdata} we have shown the results from our analysis. %The the  plots the lime colored solid line represents the baryonic mass distribution in the cluster that we have calculated using Eq.~\ref{eq:massformula}. 
The yellow dot-dashed line in the plots show the quantity $M_b(r)\left[1+K\left( 1-\exp(-\lambda r)\left(1+\lambda r\right)\right)\right]$. It can be seen that it matches with the solid red line, which shows the Newtonian dynamic mass profile of the cluster calculated using Eq.~\ref{eq:M_N1}. Therefore, we don't require any additional dark matter components. We have also shown the distribution of these two parameters from our MCMC analysis for each of the galaxy clusters in the figure. The values of different parameters are shown in the Table~\ref{tab:clusterProperties}. The background mass distribution in every cluster is different. Therefore, there is no reason for the best fit values of the parameters to be similar. However, we have also calculated the best-fit values of parameters $\sqrt{M_c}$ and $\lambda^{-1}$ by running a MCMC on all the 106 clusters given in~\cite{reiprich2003cosmological}. The black dashed curve in each of the plot shows the fitting using these bestfit values of $\sqrt{M_c}$ and $\lambda^{-1}$ from the full data set. We can see that even with these common values of the parameters the fitting to the clusters are remarkably good. 

\subsection{Spiral Galactic rotation curves}

Spiral galactic rotation curves can provide another essential test for the theory. The circular velocities of stars and gas further from the nucleus of the spiral galaxy generally do not decline following widely expected Keplerian fall-off. Observations confirmed that galaxy rotation curves are primarily flat, with some galaxies showing modestly declining and some accelerating circular velocities further away from a nucleus.

\begin{figure*}
    \centering
\includegraphics[trim=9.3cm 4.9cm 9.3cm 4.9cm, clip=true, width=0.325\textwidth]{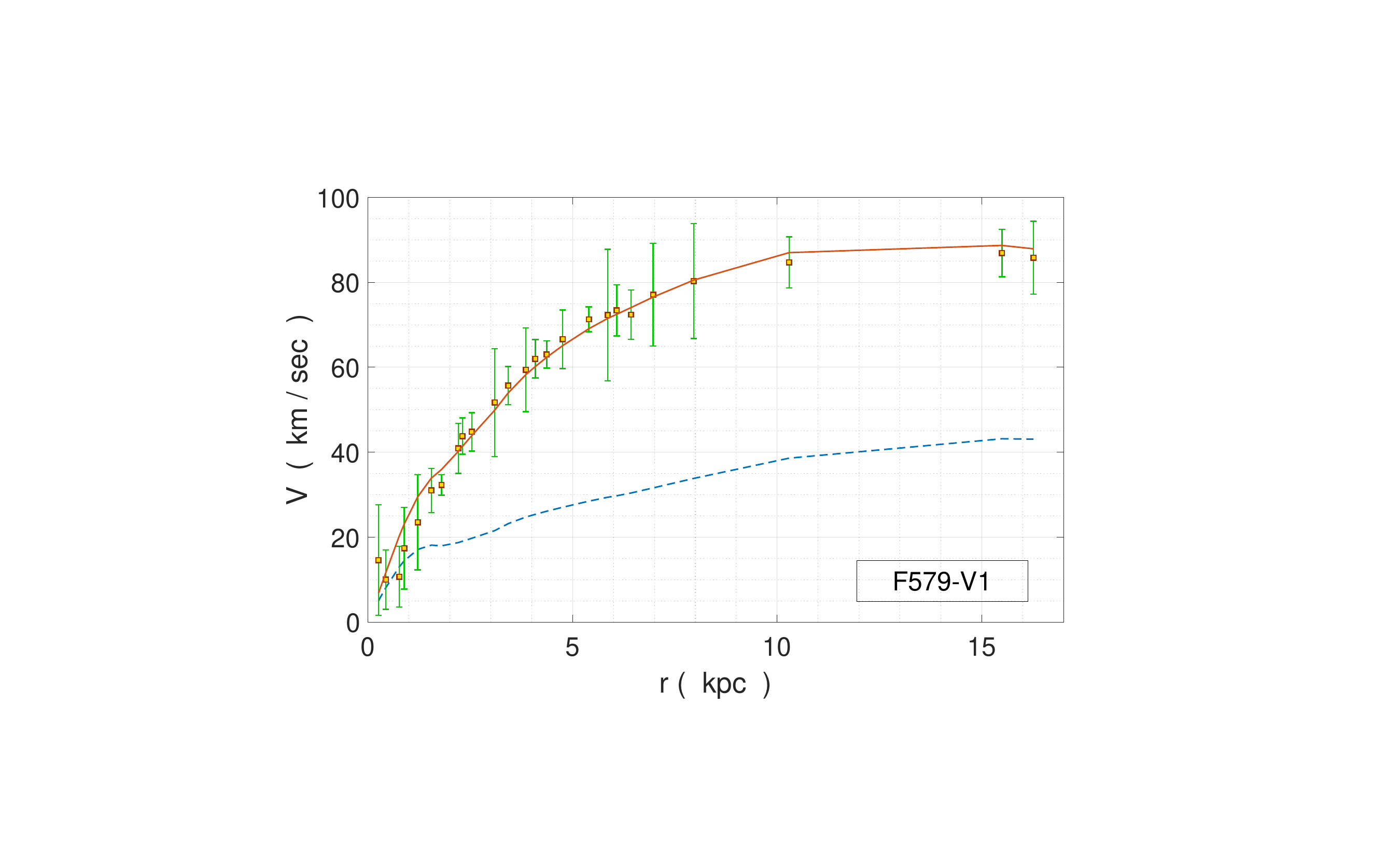}
\includegraphics[trim=9.3cm 4.9cm 9.3cm 4.9cm, clip=true, width=0.325\textwidth]{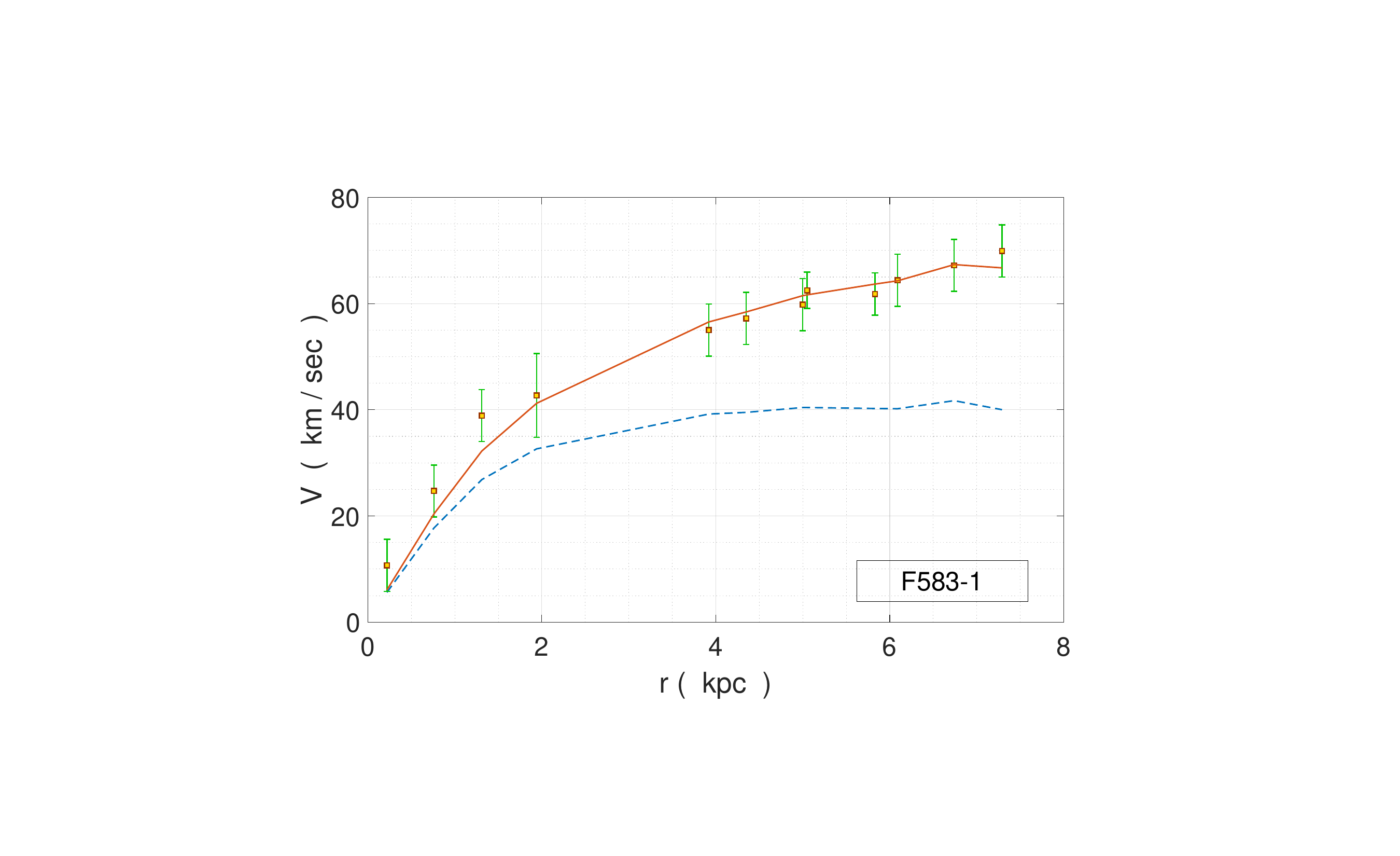}
\includegraphics[trim=9.3cm 4.9cm 9.3cm 4.9cm, clip=true, width=0.325\textwidth]{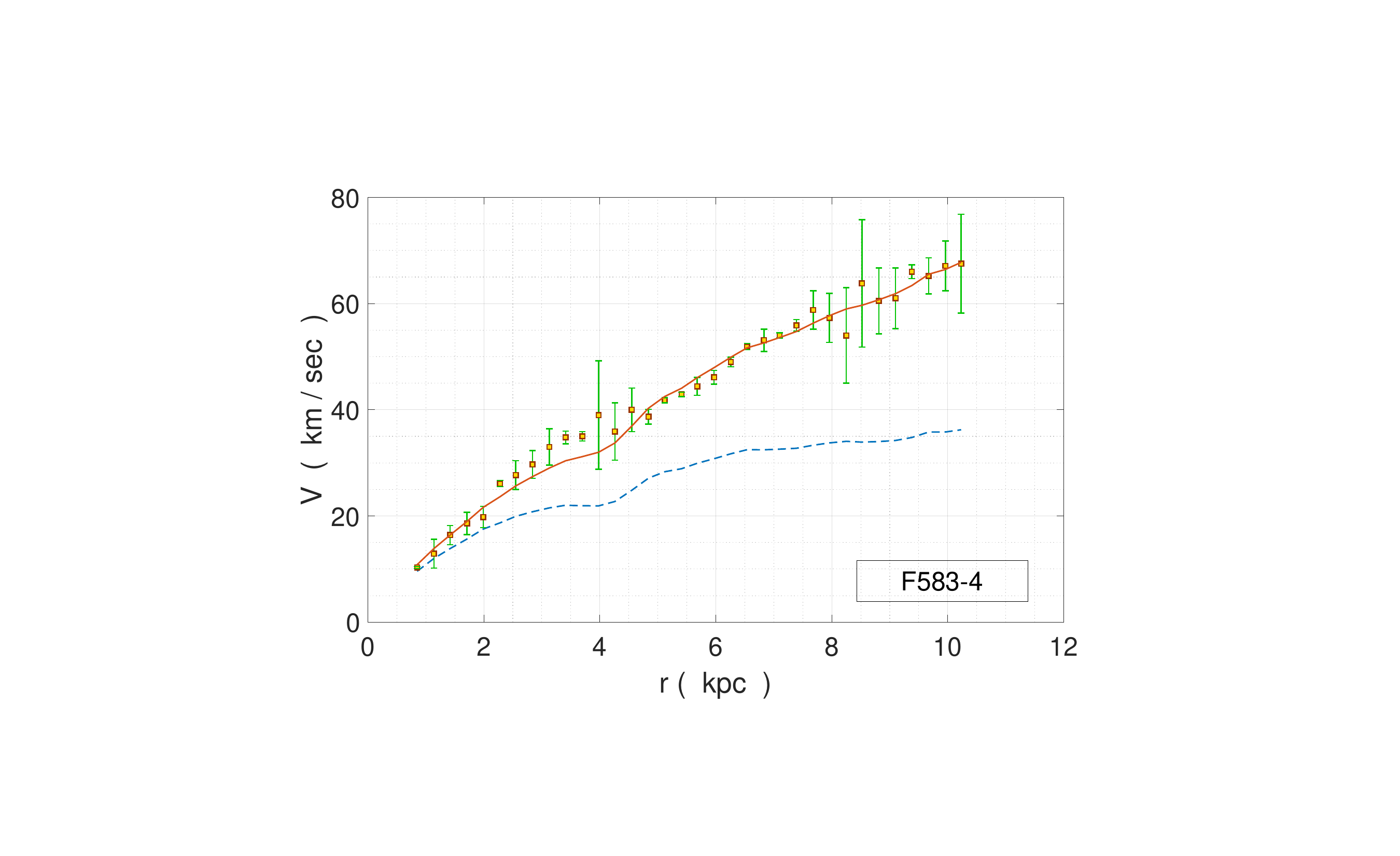}
\includegraphics[trim=9.3cm 4.9cm 9.3cm 4.9cm, clip=true, width=0.325\textwidth]{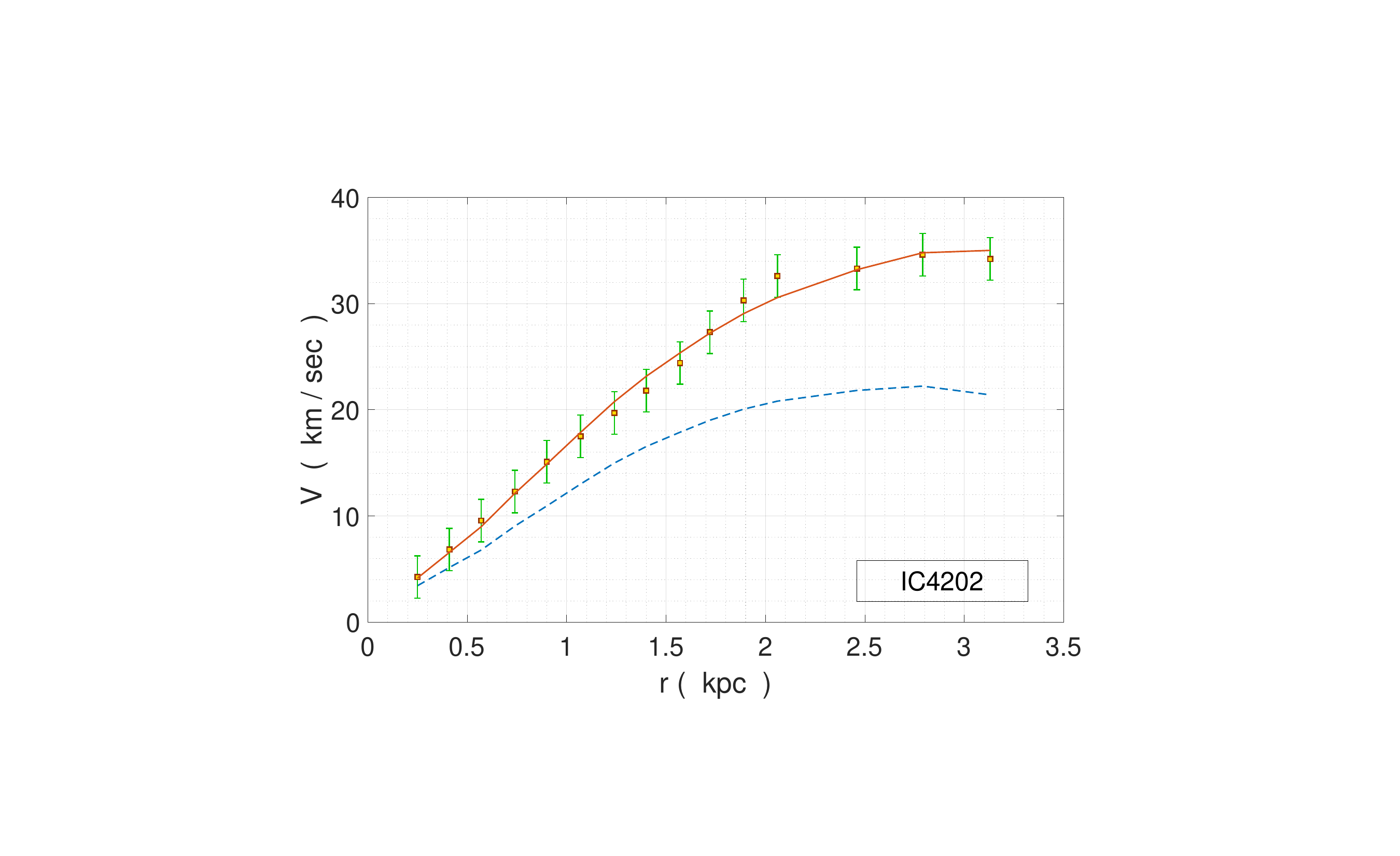}
\includegraphics[trim=9.3cm 4.9cm 9.3cm 4.9cm, clip=true, width=0.325\textwidth]{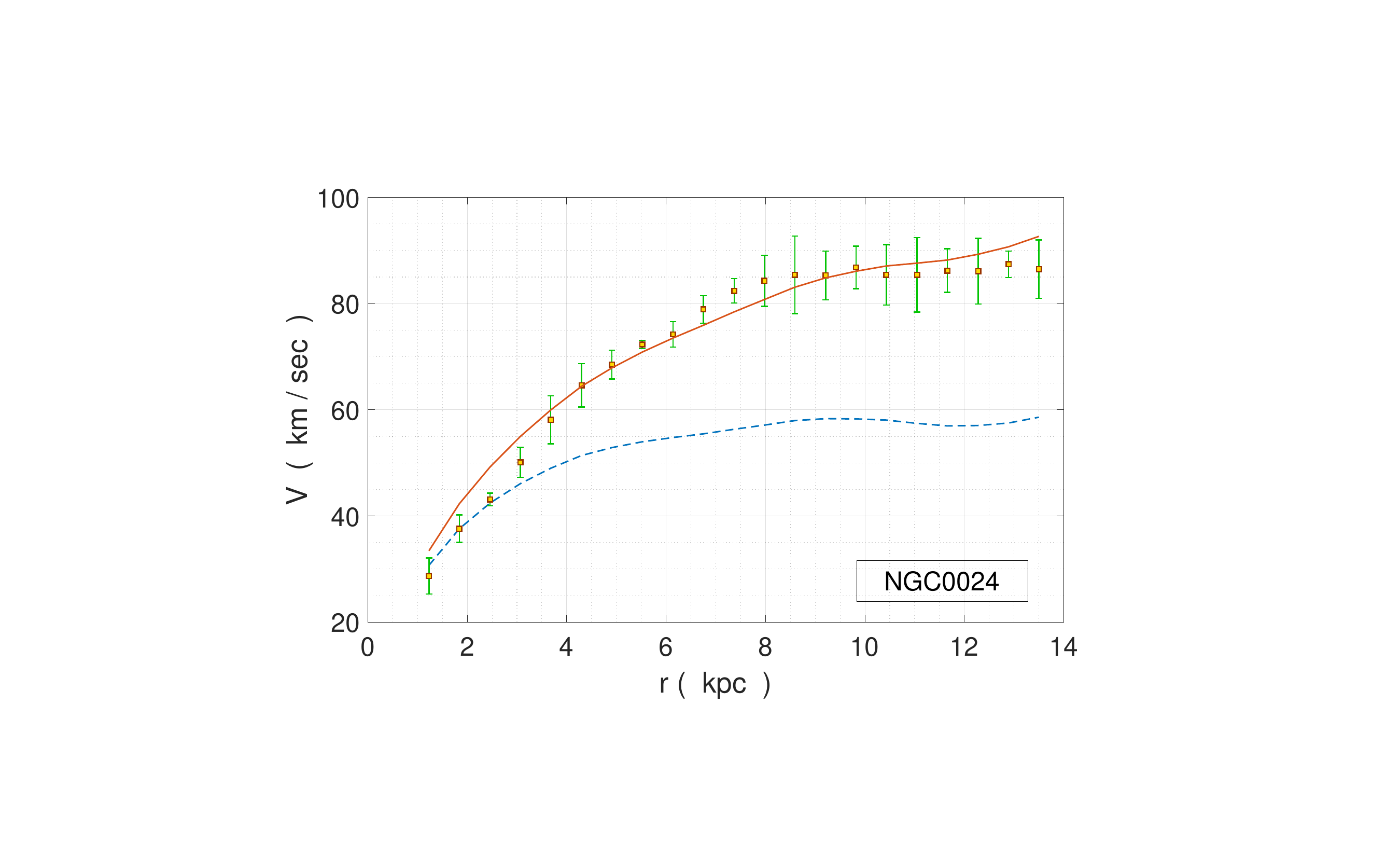}
\includegraphics[trim=9.3cm 4.9cm 9.3cm 4.9cm, clip=true, width=0.325\textwidth]{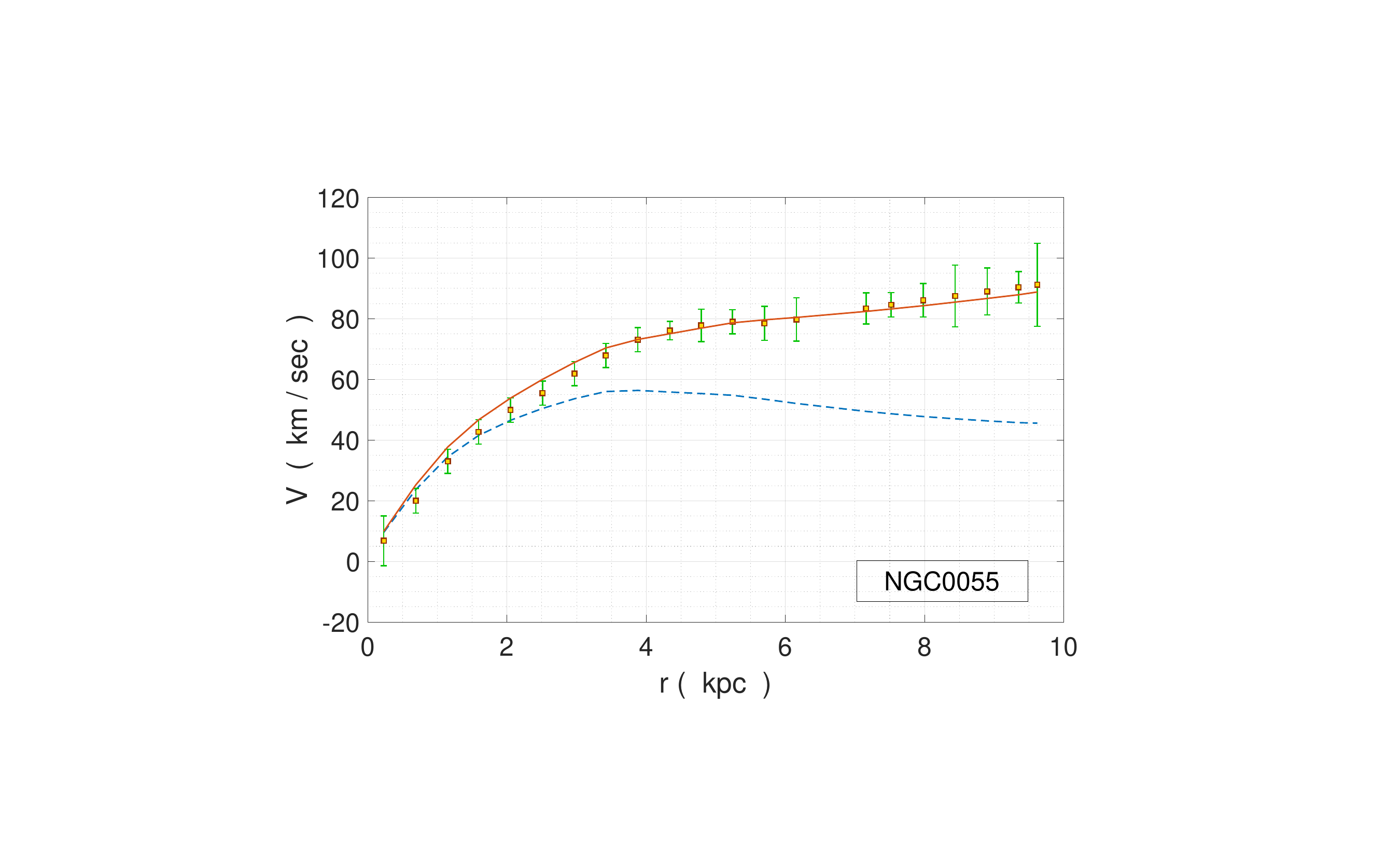}
\includegraphics[trim=9.3cm 4.9cm 9.3cm 4.9cm, clip=true, width=0.325\textwidth]{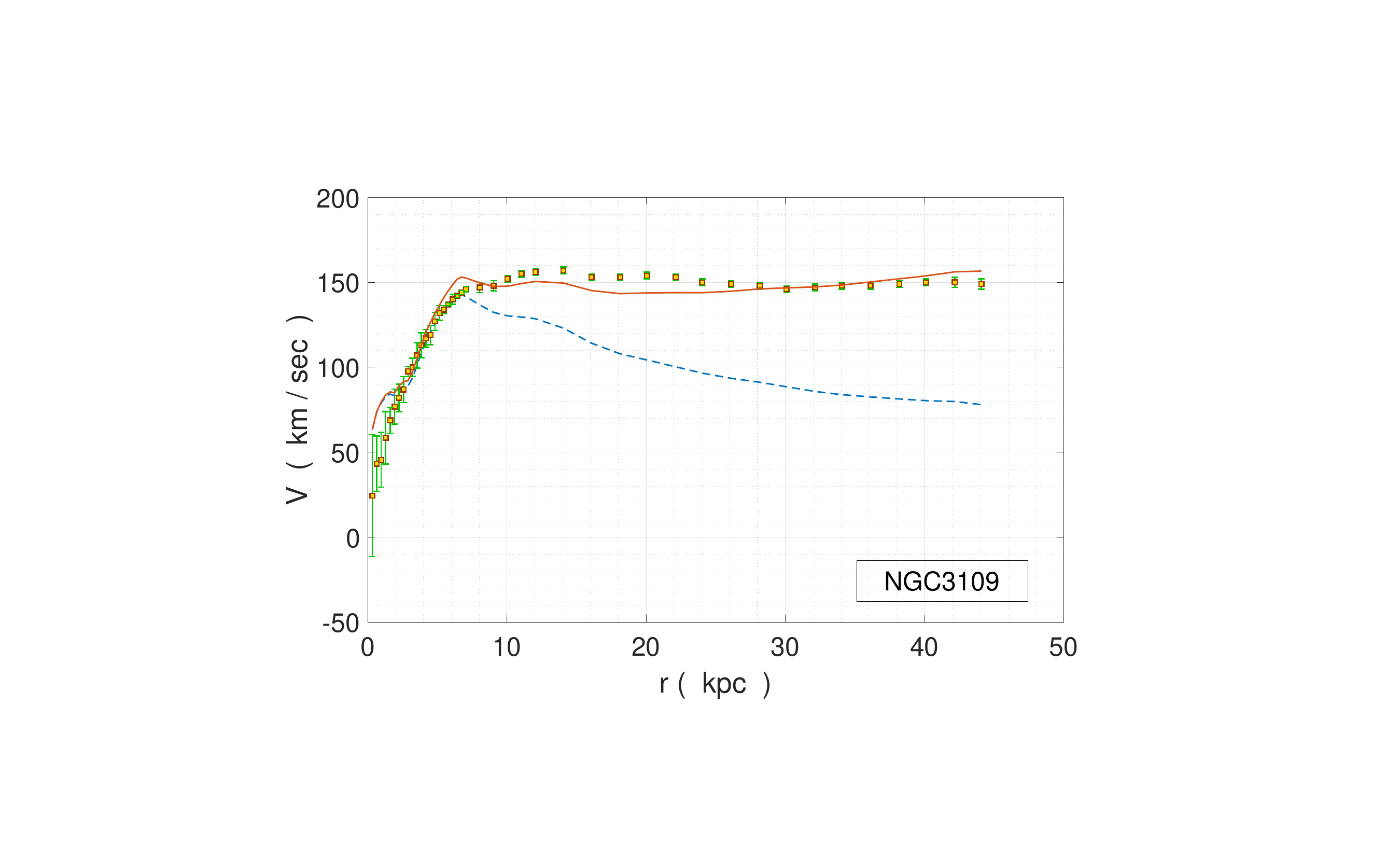}
\includegraphics[trim=9.3cm 4.9cm 9.3cm 4.9cm, clip=true, width=0.325\textwidth]{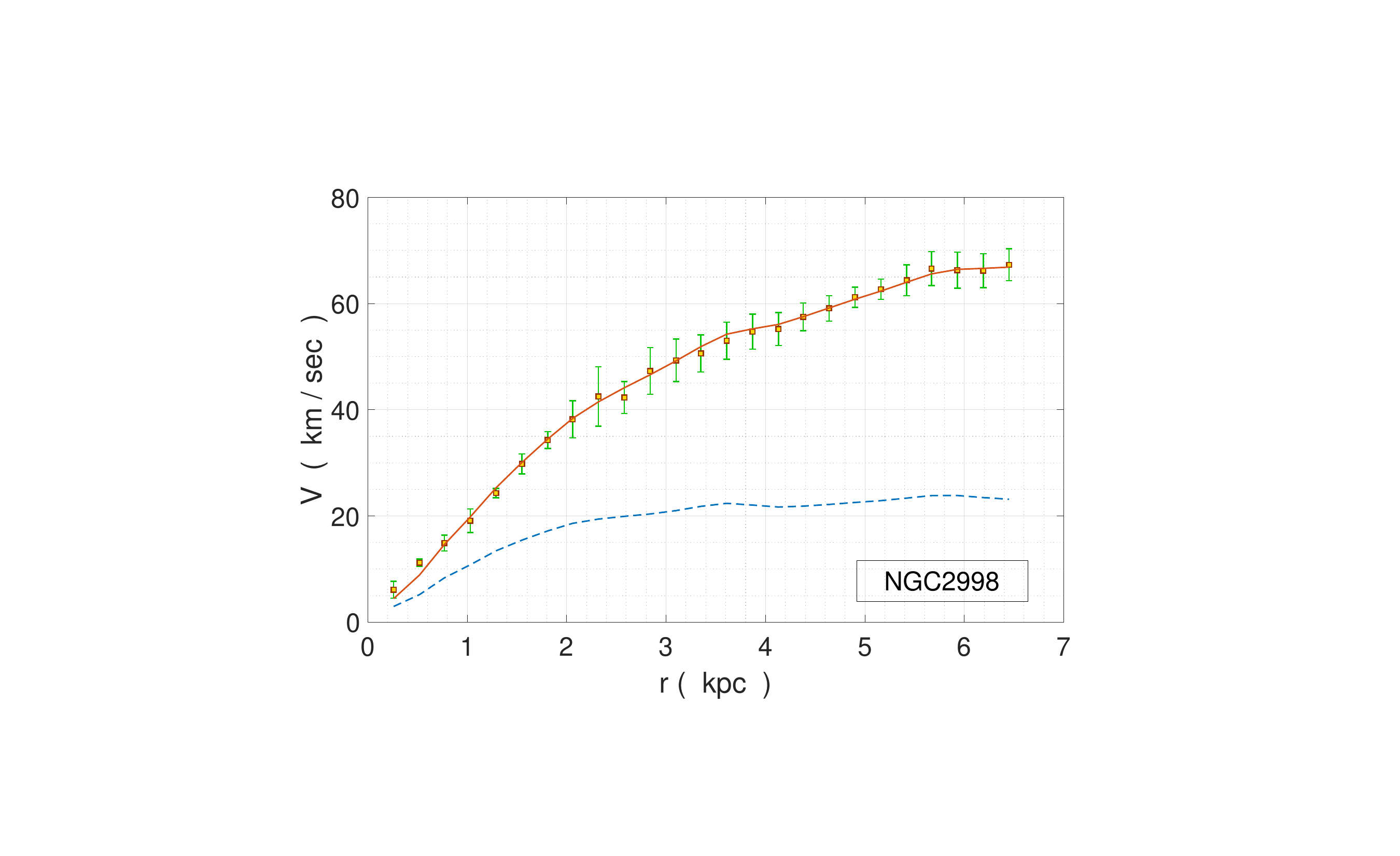}
\includegraphics[trim=9.3cm 4.9cm 9.3cm 4.9cm, clip=true, width=0.325\textwidth]{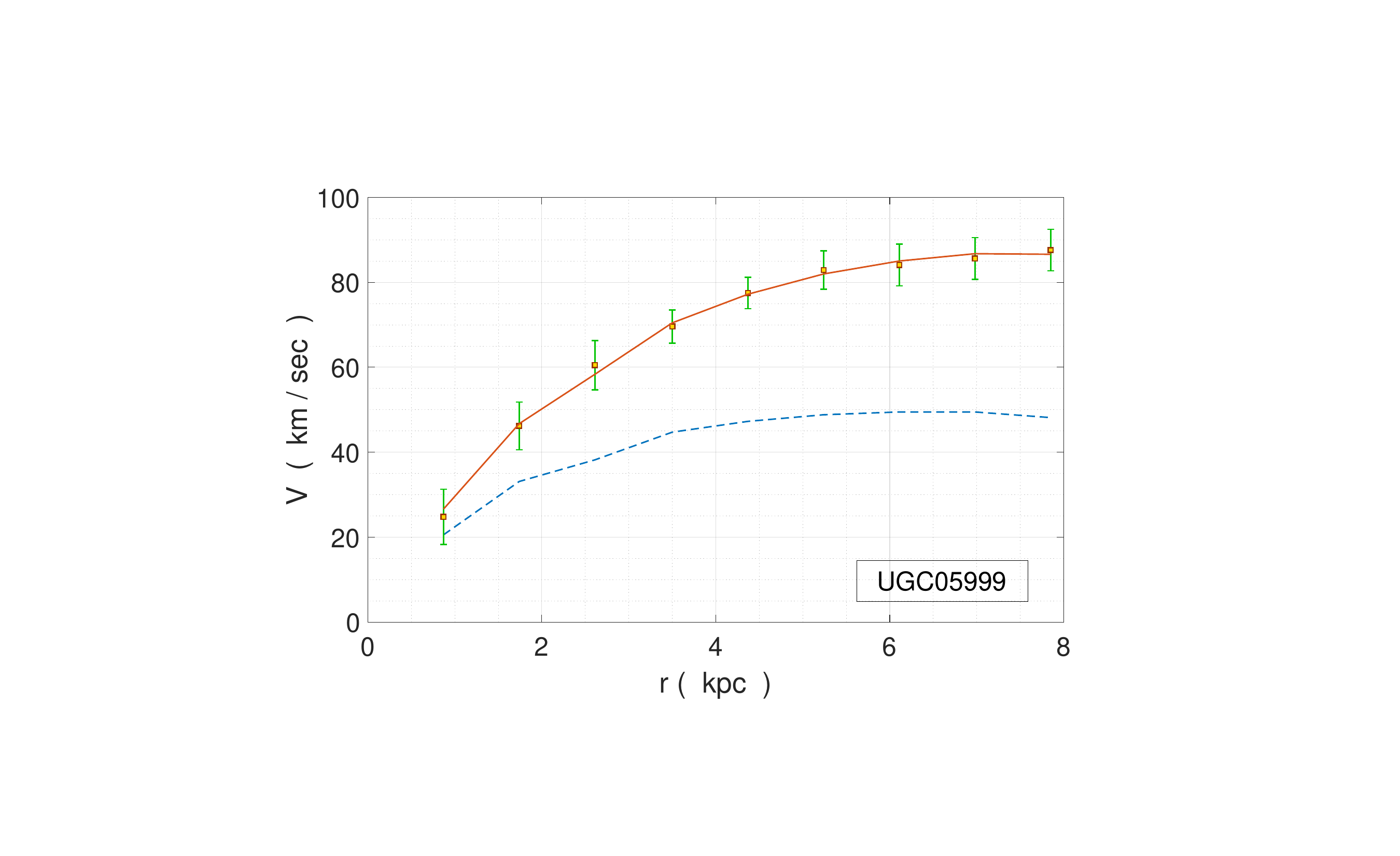}
    \caption{The graph displays the rotation curves of 9 galaxies taken from the SPARC dataset. The blue dotted curve represents the velocity profile calculated using Newtonian gravity based on the baryonic matter (measured from the surface brightness). However, it is evident that the baryonic mass alone is insufficient to explain the observed velocity profile. The solid red curve illustrates the velocity profile calculated using Machian Gravity. This model accurately accounts for each feature of the velocity profile, providing a much better fit than Newtonian gravity.}
    \label{fig:galacticvelocity}
\end{figure*}

The most widely accepted theory to explain this spiral galaxy phenomenon is considering the existence of non-baryonic dark matter. However, it is observed that for any feature in the luminosity profile, there is a corresponding feature in the rotation curve and vice versa, also known as Renzo's rule. Therefore, explaining the above phenomenon is impossible if the dark matter is an independent quantity unrelated to the baryonic matter in the galaxy. Several modified gravity models have been suggested. The most well-known is Mingrim's MOND which provides a phenomenological model for the galactic velocity curve. 

This section checks how the Machian gravity model can explain the galactic velocity profile. In Eq.~\ref{eq:velocityFinal}, we have shown how the velocity profile in a gravitationally bound galaxy is related to the mass. Therefore, we use this equation to check whether it can explain the galactic velocity profiles. 

We use the SPARC (Spitzer Photometry and Accurate Rotation Curves) data for our analysis~\cite{SPARC}. The dataset contains the accurate rotation curve for 175 spiral galaxies. It provides observed rotational velocity and the error bars as a function of radius from the galaxy's center. It also provides the surface brightness of the galaxy disk and bulge, which can be converted to mass by multiplying with the mass-to-light ratio. Once we know the enclosed mass within that radius, we can easily calculate the theoretical velocity using Eq.~\ref{eq:velocityFinal}. 

In Fig.~\ref{fig:galacticvelocity}, we have shown the observed and theoretical velocity profiles for nine galaxies. The full detail of 175 galaxies with detailed analysis is shown in~\cite{das2023aaspects}. The dotted blue curve shows the galactic velocity profile calculated from Newtonian mechanics using baryonic matter. Plots show that the Newtonian velocity using baryonic matter is much less than the observed velocities of the galaxies. However, in all the plots, we can see that the Newtonian velocity for all the galaxies is somehow related to the observed velocity. For example, the observed velocities grow with radius when the Newtonian velocity grows with radius. Similarly, the observed velocity also falls off when the Newtonian velocity starts falling. In the galaxy F583-4, we can see some features at 4kPc. The same feature is observed in the Newtonian velocity too. Therefore, while we can explain the galactic velocity profiles using dark matter independent of baryon, it's impossible to explain all these phenomena. Our analysis has two parameters $\sqrt{M_c}$ and $\lambda$. We tried to fit these parameters for different galaxies through MCMC analysis. The red curve shows the best-fit curve using  Eq.~\ref{eq:velocityFinal}. We can see that all the features can be explained exceptionally well using Machian gravity. Here I also like to inform the readers that the values of these parameters are different from those used for fitting the cluster mass profiles. This is expected because the matter distribution for galaxy cluster is completely different from those for the galaxies

Interestingly our analysis shows that there is an extremely strong correlation between the parameter $\sqrt{M_c}$ and $\lambda$, giving $\sqrt{M_c}\propto \lambda^{-1}$. Our analysis also shows that  $GM_c \lambda^2\sim a_0$, where $a_0$ is constant acceleration for each galaxy. Therefore, we only need to set a single parameter $a_0$ for each of these galaxies. On top of that, $a_0$ for all the galaxies are of the order of the acceleration of the Universe. The exact reason for this corelation is not known and its just an empirical result from our analysis. The detailed analysis of these data is beyond the scope of the present paper and is discussed in detail in~\cite{das2023aaspects}.

\section{Machian Gravity in presence of the source terms }

In the previous section, we discussed the gravitational field equations  from Machian gravity in vacuum under the weak-field approximation. However, to extend the theory to more general settings, it is necessary to derive the full stress-energy tensor for the theory. Starting from the Einstein–Hilbert action, as discussed before, the complete field equations can be obtained as:

\begin{equation}
    \widetilde{G}_{AB} = \kappa \widetilde{T}_{AB} \,.
\end{equation}

\noindent Here $\widetilde{T}_{AB}$ is the five dimensional stress energy tensor for the theory. We know $\widetilde{G}_{AB} = \widetilde{R}_{AB} - \frac{1}{2}\widetilde{g}_{AB}\widetilde{R}$. However, this  equation is in 5 dimension and in 5-dimension $\widetilde{g}_{AB}\widetilde{g}^{AB} = 5$. This gives

\begin{equation}
\widetilde{G} = g^{AB}\left[\widetilde{R}_{AB} - \frac{1}{2}\widetilde{g}_{AB}\widetilde{R}\right] = \left(1 -\frac{5}{2}\right)\widetilde{R} = -\frac{3}{2}\widetilde{R} \;.
\end{equation}

\noindent Therefore, we can relate the Ricci scalar with the stress energy tensor as %This leads us to 

\begin{equation}
\widetilde{R}_{AB} = \kappa \left(\widetilde{T}_{AB} - \frac{1}{3} \widetilde{g}_{AB}\widetilde{T} \right)  \;.  
\end{equation}

\noindent Note that there is a factor of $\frac{1}{3}$ multiplied with the trace part instead of the usual factor of $\frac{1}{2}$ as we see in the 4-dimensional equations. 

To derive the stress-energy tensor, we begin by doing a coordinate transformation to a frame in which $p_4$ is constant. In such a coordinate system, the metric tensor and thereby the Ricci tensor and the Einstein tensor becomes independent of the $x^4$ coordinate. This transformation helps us defining the stress-energy tensor as we will see later. Let us denote this as the primed coordinate system, with the transformation given by $dx'^A = \Lambda^A_K , dx^K$. In this frame, the field equations take the form:

%For deriving the stress energy tensor, first we can do some coordinate transform and go to coordinate system where $p_4$ is a constant, which implies that the Ricci tensor is independent of the $x^4$ coordinate. Let us assume that this is the primed coordinate system. ${dx'}^A = \Lambda^A_K dx^K$. So the above equation will take the form 

\begin{equation}
    {\widetilde{G}'}_{KL} = {\widetilde{G}}_{AB}\Lambda^A_K\Lambda^B_L = \kappa {\widetilde{T}}_{AB}\Lambda^A_K\Lambda^B_L = \kappa {\widetilde{T}}'_{KL} \,.
\end{equation}

\noindent Next let us represent the  5-velocity in this 
frame in  terms of the 4-velocities. By definition $\widetilde{U}^{\prime \mu} = U^\mu$ and $\widetilde{U}^{\prime 4} = 1$. This  gives us 

\begin{eqnarray}
    \widetilde{U}^{\prime}_4 &=& \upsilon \phi^2 A^\prime_\beta U^{\prime\beta} + \phi^2 \,, \\
    \widetilde{U}^\prime_\mu  &=& (g^\prime_{\mu\nu} + 
    \upsilon^2 \phi^2 A^\prime_\mu A^\prime_\nu )U^{\prime\nu} + \upsilon \phi^2 A^\prime_\mu  
    = U^\prime_\mu + \upsilon \widetilde{U}^\prime_4 A^\prime_\mu \label{Eq69} \,.
\end{eqnarray}

The 5-D Einstein's tensor is given by Eq.\ref{eq:Einstein's tensor} - Eq.~\ref{BDScalar}. Since our metric is independent on $x^4$,  the terms $P_{\mu\nu}$, $Q_\mu$ and $R$ vanish. We now define some relevant stress energy tensor terms as 

\begin{eqnarray}
T^{\prime (T)}_{\mu\nu} &=& \widetilde{T}^\prime_{\mu\nu}-v^2 A^\prime_\mu A^\prime_\nu \widetilde{T}^\prime_{44}- \upsilon A^\prime_\mu\left(\widetilde{T}^\prime_{\nu 4}-v A^\prime_\nu \widetilde{T}^\prime_{44}\right) - \upsilon A^\prime_\nu\left(\widetilde{T}^\prime_{\mu 4}-v A^\prime_\mu \widetilde{T}^\prime_{44}\right) \,,   \label{Eq70}  \\
J^{\prime (V)}_\nu &=& \widetilde{T}^\prime_{4 \nu} - v A^\prime_\nu \widetilde{T}^\prime_{44} \,, \label{Eq71}\\
S &=& \widetilde{T}^\prime_{44} - \frac{1}{2}\phi^2{T^\mu_\mu}^{\prime (T)} \;. \label{Eq72}
\end{eqnarray}

We can associate the first tensor component with the standard stress energy tensor for a perfect fluid in 4 dimension because this is related to $G'_{\mu \nu}$. Therefore, we have

\begin{equation}
T^{\prime (T)}_{\mu\nu} = (\rho + p) U^\prime_\mu U^\prime_\nu + p g^\prime_{\mu\nu} \,.
\label{4Dstress}
\end{equation}

The second term can be interpreted as a source term for the vector field. For example, consider a perfect fluid that is static. Therefore, it contributes only to the zeroth component of the Newtonian vector potential of the background. However, if the fluid component has some velocity, it will affect all the components of the vector potential from the background. 

We can compare with the  mechanism in Kaluza–Klein theory, where the charge density serves as a source for the Electromagnetic vector field. From Kaluza–Klein theory, we know that to relate energy density to charge density, one multiplies the mass density by the charge per unit mass, represented by fourth component of the velocity. In our case, we can apply similar logic  to obtain the effective gravitational mass density i.e. multiply $(\rho + p)$ by $U^\prime_4$ (recall that $U^\prime_4 = \left(\frac{m_g}{m_0}\frac{I}{\upsilon}\right)$). Therefore, the source term for the vector field can be written as:

\begin{equation}
J^{\prime(V)}_{\nu} = (\rho + p) \widetilde{U}^\prime_4 U^\prime_\nu \,.
\label{4Dsource}
\end{equation}

Assigning the final term is most difficult, as it does not correspond directly to any known term in the Kaluza–Klein formulation. However, we can approach this by considering the physical condition we aim to satisfy. Suppose the vector field is constant. Then, according to Eq.~\ref{BDScalar}, we should have $\widetilde{G}^\prime_{44} = -\frac{1}{2} \phi^2 R$. From our definition of $S$, this implies that $S \sim 0$.

From Eq.~\ref{4Dstress} and Eq.~\ref{4Dsource} we can intuitively assume that $\widetilde{T}_{AB}$ can be written in the form 

\begin{equation}
 \widetilde{T}^\prime_{AB} = (\rho + p ) \widetilde{U}^\prime_A \widetilde{U}^\prime_B + p \widetilde{g}^\prime_{AB} + (A\rho + B p) \widetilde{X}^\prime_{AB} \,,   
\end{equation}

\noindent where $\widetilde{X}^{\prime}_{AB} = \widetilde{g}^\prime_{A4}\widetilde{g}^\prime_{4B} / \widetilde{g}^\prime_{44}$ and $A$, $B$ are some constants which we will determine below.  The justification for this form of the stress-energy tensor is as follows. If we replace the 4-velocities in Eq.~\ref{4Dsource} and Eq.~\ref{4Dstress} from Eq.~\ref{Eq69}, the equations for $\widetilde{T}_{AB}$ naturally takes the form as suggested above. The pressure term will have only the 4 dimensional $g'_{\mu\nu}$ as we have assigned the only that part of pressure in Eq.~\ref{4Dsource}. Therefore, in 5-dimension we can intuitively take it to be $p \widetilde{g}'_{AB}$. The last term linked with $\widetilde{X}^{\prime}_{AB}$ is added to account for the $\widetilde{G}^{\prime}_{44}$ term.  Similar technique is  also used  in~\cite{coquereaux1990theory} for defining the stress-energy tensor for Jordan-Thiry type scalar field. It can be easily checked that this particular form of stress-energy tensor will not change anything in $T^{\prime (T)}_{\mu\nu}$ and $J^{\prime(V)}_{\nu}$ definition.

In our theory the 5-dimensional manifold is a null manifold, i.e.  $\widetilde{g}^{\prime AB}\widetilde{U}^{\prime}_A\widetilde{U}^{\prime}_B = 0$. Therefore, we can calculate $\widetilde{T} = A\rho + (B+5) p$. Also in the 4 dimension we have  $T^{(T)} = {T^\mu_\mu}^{\prime (T)} = -\rho + 3p$. Therefore, We can write

\begin{equation}
S = \widetilde{T}^{\prime}_{44} + \frac{1}{2}\phi^2 {T^\mu_\mu}^{\prime (T)} = (\rho + p)\widetilde{U}^{\prime}_4\widetilde{U}^{\prime}_4 + \phi^2 p + (A\rho + B p) \phi^2 -\frac{1}{2}\phi^2  (-\rho + 3p) \,. 
\end{equation}

\noindent When $\widetilde{U}^{\prime}_4 \sim 1$ and $\phi \sim 1$ and $A_\mu$ constant we should have $S \sim 0$. This results in $A=-\frac{3}{2}$ and $B = - \frac{1}{2}$. It can be easily verified that for the above condition it recovers GR.

Of course, this definition has been obtained in a reference frame where the metric tensor is independent of $\zeta$. However, we can always transform back to the coordinate system in which $m_0/m_g$ is constant. In that frame, the form of the stress-energy tensor remains unchanged, as the transformation involves only rotating metric tensors and the velocities and does not contain any derivatives. Therefore, we get the 5-dimensional stress energy tensor to be

\begin{equation}
 \widetilde{T}_{AB} = (\rho + p ) \widetilde{U}_A \widetilde{U}_B + p \widetilde{g}_{AB} - \frac{1}{2}(3\rho +  p) \widetilde{X}_{AB} \,.   
\end{equation}

\subsection{Exploring the Brans-Dicke Theory}
\label{Sec:BD}
Earlier in Sec.~\ref{Section:FieldEquation}, we discussed that the field equation for the theory yields a Brans-Dicke-like scalar field with the Dicke coupling constant $\omega_D = 0$. However, observations indicate that the Dicke coupling constant should be very large % approach infinity 
to align with experimental data. 

The main challenge here is that the laplacian of scalar field must be very small to explain observations. %Since the equation of motion for the scalar field is given by Eq.~\ref{Eq21_BD_Scalar}, achieving a small $\phi$ requires $\omega_D \rightarrow 0$.
However, in the BD theory the field equation for the scalar field is given by 

\begin{equation}
\square \phi=\frac{8 \pi}{3+2 \omega_D} T  \qquad\qquad\dots\qquad \text{(in BD Theory)}
\end{equation}

\noindent Therefore, unless $\omega_D$ is very large, the value of $\square \phi$ should be large and can not explain the observations.

However, in the Machian gravity theory, that is not the case. Our theory reduces to the Brans-Dicke type theory when the metric is independent of $x^4$ and the vector field $A^\alpha$ is constant. Under such conditions, the equation of motion for the scalar field is given by $R_{44} = -\phi \square \phi$. Again, from the stress energy tensor, we can calculate $\widetilde{R}_{44} = \kappa \left[\widetilde{T}_{44} - \frac{1}{3}\widetilde{g}_{44}\widetilde{T}\right] $. Putting all the terms together we get

\begin{equation}
    \phi \square \phi = -\frac{1}{2} (\rho + p) \left(\widetilde{U}_4^2 - \phi^2 \right) \,.
\end{equation}

\noindent Now in the right hand side $\widetilde{U}_4^2 - \phi^2 \sim 0$. Also, in an expanding Universe we can consider $\phi \sim 1$, because somehow $\phi$ represents the variation in the gravitational constant which is significantly small. Therefore, the Laplacian term is very small and hence it should be able to explain the  observational results. The model don't need any additional $\omega_D$ parameter to explain the observations. 

It is important to note that although Machian gravity introduces a scalar field similar to that in BD theory, it is not dual to BD theory in any limit, as the field equation for $\phi$ is fundamentally different.

\section{Cosmological solution from a generalized metric}
\label{sec:Cosmology}
%{\color{red} Probably need some justification why the vector potential is taken as zero. }
For understanding the cosmological implications of the theory we start with a general spherically symmetric line element, given by 
%We can write the metric for a homogeneous and isotropic space as

\begin{equation}
ds^{2}=-e^{\omega}dt^{2}+e^{\kappa}dr^{2}+R^{2}\left(d\theta^{2}+\sin^{2}\theta d\phi^{2}\right)+ e^{\mu}d\zeta^{2}\,.\label{eq:line-element-2}
\end{equation}

\noindent The exponentials are taken to ensure that these quantities cannot be negative. The coefficient for the two-sphere term is taken as $R^2$ such that individual $R$ can assume both positive or negative signature while $R^2$ should always be positive. While this is a general spherically symmetric metric, for a flat isotropic and homogeneous case, as our Universe is, we can always replace $R = e^{\kappa/2} r$. Additionally, the fact that the metric is diagonal ensures that it satisfies the Weyl postulate, which states that the worldlines of fluid particles in the Universe should be orthogonal to spacelike hypersurfaces. This condition is essential for any line element to be physically meaningful in a cosmological context.  %Also for Universe we have taken 

For this line element, we can calculate $\widetilde{G}_{AB}$ and express them in terms of 4-dimensional Einstein tensor $G_{\alpha\beta}$ along with additional terms. For a complete calculation, check the Appendix~\ref{Appcosmology}. The components of the 5-dimensional Einstein's tensor for the above line element are given by

\begin{widetext}
\begin{eqnarray}
\widetilde{G}_{0}^{0} & = & G_{0}^{0}+e^{-\omega}\Bigg(\frac{\dot{\mu}\dot{\kappa}}{4}+\frac{\dot{\mu}\dot{R}}{R}\Bigg)-e^{-\kappa}\Bigg(\frac{R'\mu'}{R}-\frac{\kappa'\mu'}{4}+\frac{\mu''}{2}+\frac{\mu'^{2}}{2}\Bigg)\nonumber 
 - e^{-\mu}\left(\frac{\kappa^{**}}{2}+\frac{\kappa^{*2}}{4}\right. \\
& &\left. -\frac{\kappa^{*}\mu^{*}}{4}+\frac{R^{*}}{R}\left(\kappa^{*}-\mu^{*}\right)+\frac{R^{*2}}{R^{2}}+\frac{2R^{**}}{R}\right)\label{eq:G-0-0}\,, \nonumber\\
\widetilde{G}_{0}^{1} &=& G_{0}^{1}+e^{-\kappa}\left(\frac{\dot{\mu}'}{2}+\frac{\dot{\mu}\mu'}{4}-\frac{\omega'\dot{\mu}}{4}-\frac{\dot{\kappa}\mu'}{4}\right)\label{eq:G-1-0}\,, \nonumber \\
\widetilde{G}_{1}^{1} & = & G_{1}^{1}+e^{-\omega}\left(\frac{\ddot{\mu}}{2}+\frac{\dot{\mu}^{2}}{4}-\frac{\dot{\omega}\dot{\mu}}{4}+\frac{\dot{R\dot{\mu}}}{R}\right)-e^{-\kappa}\left(\frac{\mu'\omega'}{4}+\frac{\mu'R'}{R}\right)\nonumber 
- e^{-\mu}\left(\frac{\omega^{**}}{2}+\frac{\omega^{*2}}{4}\right. \\
& &\left.+\frac{R^{*2}}{R^{2}}+\frac{2R^{**}}{R}+\frac{R^{*}}{2R}\left(\omega^{*}-\mu^{*}\right)-\frac{\mu^{*}\omega^{*}}{4}\right)\,,\label{eq:G-1-1} \nonumber \\
\widetilde{G}_{2}^{2} & = & G_{2}^{2}+e^{-\omega}\left(\frac{\dot{R}\dot{\mu}}{2R}-\frac{\dot{\omega}\dot{\mu}}{4}+\frac{\dot{\mu}\dot{\kappa}}{4}+\frac{\ddot{\mu}}{2}+\frac{\dot{\mu}^{2}}{4}\right)-e^{-\kappa}\Bigg(\frac{R'\mu'}{2R}+\frac{\mu''}{2}\nonumber 
 +\frac{\mu'^{2}}{4}+\frac{\omega'\mu'}{4} \\
 &  &-\frac{\mu'\kappa'}{4}\Bigg)
- e^{-\mu}\Bigg(\frac{R^{**}}{R}+\frac{R^{*}\omega^{*}}{2R}+\frac{R^{*}\kappa^{*}}{2R}-\frac{R^{*}\mu^{*}}{2R}\nonumber 
 +\frac{\omega^{**}}{2}+\frac{\omega^{*2}}{4}+\frac{\kappa^{**}}{2}+\frac{\kappa^{*2}}{4} \nonumber \\
 &  &+\frac{\kappa^{*}\omega^{*}}{4}-\frac{\kappa^{*}\mu^{*}}{4} 
-\frac{\mu^{*}\omega^{*}}{4}\Bigg)\label{eq:G-2-2}\,, \nonumber \\
\widetilde{G}_{3}^{3} &=& \widetilde{G}_{2}^{2}\label{eq:G-3-3}\,.
\label{equation:G5d4d}
\end{eqnarray}
\end{widetext}

\noindent Here $\dot{(\ldots)}$, $(\ldots)'$  and  $(\ldots)^{*}$ represent the derivative with respect to the $t$, $r$ and $\zeta$ respectively.

As the metric don't have the off-diagonal components, its easy to see that $\widetilde{T}^\alpha_\beta = T^\alpha_\beta$. Therefore, from the field equation we can write

%Assuming that $\widetilde{T}_{A 4} = \widetilde{T}_{4 A} \approx 0$ and $\widetilde{T}_{\alpha \beta} \approx T_{\alpha \beta}$, we can write 

\begin{equation}
\widetilde{G}^\alpha_{ \beta} = T^\alpha_{\beta} \,,\qquad\qquad G^\alpha_{\beta} = T^\alpha_{\beta} + Q^\alpha_{\beta} \,.
\end{equation}

\noindent Here $Q^\alpha_\beta$ are the expressions in the right-hand side of Eq.~\ref{eq:G-0-0}. $Q^\alpha_\beta$ are some geometric terms. However, these terms behave as if there is some additional matter component and contribute to the 4-dimensional Einstein tensor. $Q^{\alpha}_{\beta}$ can be treated as the stress-energy tensor from these geometric components. These are purely geometric terms.
To simplify these components, we associate a density and the pressure to these geometric components as $\rho_{g}$ and $p_{g}$. For time-dependent spherical symmetry, the usual stress-energy tensor in 4-dimension is given as 

\begin{equation}
Q^\alpha_\beta =(\rho_g+p_g)u^{\alpha}u_{\beta}+p_g g^\alpha_\beta \,,
\end{equation}

\noindent $u^{\alpha}$ is the four-velocity of the fluid. In our case $u^0\ne 0$, $u^1\ne 0$ and $u^2 = u^3 = 0$. %Also we have $u^{\alpha} u_{\alpha} =1$. 
Putting these values in the above expression one can obtain

\begin{equation}
\rho_{g}=Q_{0}^{0}+Q_{1}^{1}-Q_{2}^{2} \,, \qquad\qquad
p_{g}=-Q_{2}^{2}\label{eq:pg}\,.
\end{equation}

\noindent The subscript $g$ represents that these terms are purely geometric. 
Using the expressions from  Eq.~\ref{eq:G-0-0} we can get a simplified expression for $\rho_g$. However, the expression for $p_g$ will still remain complex. To simplify this, we need to use the expression $\widetilde{R}_{4}^{4}\approx 0$. As $\widetilde{R}_{4}^{4}\approx 0$ we can add it with the expression of $p_g$, and simplify the expression for $p_g$. (For detailed calculation, please refer Appendix~\ref{Appcosmology}).

After a few algebraic manipulations, we can obtain

\begin{eqnarray}
\rho_{g} & = & \frac{3}{2}\left(\frac{e^{-\kappa}\mu'R'}{R}-\frac{e^{-\omega}\dot{\mu}\dot{R}}{R}\right)-\frac{3}{2} e^{-\mu}\left(\frac{R^{*}\mu^{*}}{R}-\frac{2R^{**}}{R}\right) \nonumber\\
&& + e^{-\mu}\frac{R^{*2}}{R^{2}}- e^{-\mu}\left(\frac{\omega^{*}\kappa^{*}}{4}\right)+ e^{-\mu}\frac{R^{*}}{2R}\left(\kappa^{*}+\omega^{*}\right)\label{eq:density_1}\,, \\
p_{g} & = & \frac{1}{2}\left(\frac{e^{-\kappa}\mu'R'}{R}-\frac{e^{-\omega}\dot{\mu}\dot{R}}{R}\right)-\frac{1}{2} e^{-\mu}\left(\frac{R^{*}\mu^{*}}{R}-\frac{2R^{**}}{R}\right) \nonumber\\
& &- e^{-\mu}\left(\frac{\omega^{*}\kappa^{*}}{4}\right) - e^{-\mu}\frac{R^{*}}{2R}\left(\kappa^{*}+\omega^{*}\right)\label{eq:pressure_1}\,.
\end{eqnarray}

\noindent These expressions show that there are four different types of components of these geometric pressure and density components, i.e.

\begin{eqnarray}
\rho_{gr}&=&3p_{gr}=\frac{3}{2}\left(\frac{e^{-\kappa}\mu'R'}{R}-\frac{e^{-\omega}\dot{\mu}\dot{R}}{R}\right) %\nonumber\\ & &
-\frac{3}{2} e^{-\mu}\left(\frac{R^{*}\mu^{*}}{R}-\frac{2R^{**}}{R}\right)\label{eq:radiation}\,, \\
\rho_{gd}&=& e^{-\mu}\frac{R^{*2}}{R^{2}}\label{eq:dust}\,, \\
\rho_{gs}&=&p_{gs}=- e^{-\mu}\left(\frac{\omega^{*}\kappa^{*}}{4}\right)\label{eq:stiff-matter}\,, \\
\rho_{g\Lambda}&=&-p_{g\Lambda}= e^{-\mu}\frac{R^{*}}{2R}\left(\kappa^{*}+\omega^{*}\right)\label{eq:dark-energy}\,. 
\end{eqnarray}

\noindent

The pressure and density of $p_{gr}$ or $\rho_{gr}$ by Eq.~\ref{eq:radiation} follows the relation $p=\frac{\rho}{3}$. Therefore, they behave exactly as photons or massless neutrinos in the Universe and can be treated as dark radiation in standard cosmology. 
The $2^{nd}$ component, given by Eq.~\ref{eq:dust}, behaves as a non-relativistic matter with $0$ pressure. This fulfills all the criteria for cold dark matter. The third component, i.e., Eq.~\ref{eq:stiff-matter}, is another exciting component where pressure and density are equal. This is the stiffest equation of state that a fluid can have because, after this, the speed of sound inside a fluid will exceed the speed of light, violating the consistency relation. This kind of fluid was once proposed by Zeldovich and named as stiff matter. The last component, given by Eq.~\ref{eq:dark-energy}, has the $\rho=-p$. Thus it behaves as dark energy of the standard cosmological model.

Our equation shows that, in Machian gravity, the dark components emerge automatically from geometry. Thus, the theory can provide a cosmological model exactly similar to the standard cosmological model without demanding any external dark matter or dark energy. This solution was initially proposed by Wesson in his induced matter hypothesis and has been explored by various researchers~\cite{de1988cosmological,wesson2005equivalence,moraes2015cosmological, halpern2000behavior}. It is interesting to see that the density given by Eq.~\ref{eq:radiation} \-- Eq.~\ref{eq:dark-energy} are mostly dependent on the $\zeta$ derivative of the variables, except that there are some terms in the radiation part which are dependent on the spatial and the temporal derivative. %On a constant $\zeta$ hyperspace, if we redefine some of the terms, it will behave as a FLRW metric. 

\section{Discussion and Conclusion}
While GR explains gravity exceptionally well, several flaws have been observed in the past by researchers. Multiple theories have been suggested to address some of these issues. However, most of these theories are designed to explain the observational data. 

In this paper, a new theory of gravitation based on Mach's principle is introduced by unifying multiple well-known gravity theories into a single framework. This metric theory is derived from the action principle, ensuring compliance with all conservation laws. It is designed to reduce to various existing theories under specific conditions. For example, in scenarios where the BD theory applies, this framework produces a scalar field similar to that of BD theory. In a non-inertial reference frame, where Sciama proposed a vector potential-based theory, the Machian gravity theory provides an analogous formulation. It accounts for the effects of reference frame acceleration and generates inertial forces resulting from the motion of distant objects in the background Universe. The theory can also account for the additional terms in Einstein's tensor that were required to explain Mach's principle, as proposed by Hoyle and Narlikar.

On the galactic scale, in the weak gravitational limit, the theory aligns with various phenomenological models explored by researchers to explain galactic velocity profiles. Therefore, it successfully accounts for galactic rotation curves and the mass distribution of galaxy clusters without the need for additional dark matter candidates. Similarly, on the cosmological scale, there are theories like Space-Time-Matter (STM) that explain cosmological models without invoking any dark sector. The Machian gravity theory exhibits behavior similar to STM theories in cosmology, providing an explanation for the expansion history of the Universe without requiring any additional dark matter or dark energy components.

While the Machian Gravity theory shows promise in explaining various observational results, it is still in its preliminary stage, necessitating further in-depth analysis with additional observational data. In particular, the bullet cluster data is recognized as the most compelling evidence for the existence of dark matter, and testing the theory against such data is crucial.

Furthermore, the efficacy of the theory needs to be tested against other significant cosmological phenomena, such as CMB, Intensity mapping and Big Bang Nucleosynthesis (BBN), as well as other observational data. These comprehensive tests will ultimately provide a definitive assessment of the theory's validity and its potential to advance our understanding of gravity and the fundamental workings of the Universe.

\appendix

\section{An approximation of Newtonian potential of the Universe }
\label{gravitypotentialuniverse}

The Newtonian potential at a point, arising from all the matter in the Universe, is a concept that arises when discussing Machian gravity theory. As explained in the text, the Newtonian potential should be given by

\begin{equation}
    \Phi = -G \int_V \frac{\rho}{r}dV \,.
\end{equation}

\noindent To get the potential from the entire Universe we need to integrate over the volume inside the Hubble radius (as everything outside will not have an effect).

For a flat Universe the radial distance can be calculate as $\eta = \int \frac{dt}{a}$ (we assume $c=1$) and $a$ is the scale factor. Thus the volume between radius $\eta$ and $\eta + d\eta$ can be written as 
\begin{equation}
    dV = 4\pi a^3 \eta^2 d\eta \,.
\end{equation}

The Newtonian potential is, of course, valid only when the gravitational field is not time-varying, the field is weak, and the speeds involved are much smaller than $c$. Under these conditions, the general theory of relativity reduces to Newtonian gravity.  In the present case, while the gravitational field is weak, it varies with time. Also the speeds involved near the edge of the Hubble radius are very high. As a result, the simple Newtonian potential cannot yield accurate results. Nevertheless, we can still check whether it provides a reasonable approximation.

An important point to consider is the choice of radius in the Newtonian gravitational potential equation. For a static spacetime, the standard radial distance would be appropriate. However, in an expanding Universe, if we consider the perspective of gravitons, the gravitons emitted at a certain scale factor $a$ will be redshifted by the time they reach the present epoch. Therefore, it is more reasonable to use the luminosity distance, which accounts for the change in the wavelength of the gravitons. The luminosity distance is given by $r = \eta / a$.

Putting all these together we can get 

\begin{eqnarray}
\Phi = -G \int_V \frac{\rho}{r}dV
= -4\pi G \rho_{cr} \int_V a^4 \left( \frac{\Omega_m}{a^3}+\frac{\Omega_r}{a^4} + \Omega_\Lambda \right) \eta d\eta \nonumber \\
= -4\pi G \rho_{cr} \int_1^0 a^4 \left( \frac{\Omega_m}{a^3}+\frac{\Omega_r}{a^4} + \Omega_\Lambda \right) \left(\int_1^a \frac{d\eta}{da'}da'\right) \frac{d\eta}{da}da \,.  \nonumber
\end{eqnarray}

\noindent We also have 
\begin{equation}
\frac{1}{a^2}\frac{da}{d\eta} = H_0\sqrt{\frac{\Omega_m}{a^3}+\frac{\Omega_r}{a^4} + \Omega_\Lambda} ,.
\end{equation}

\noindent Therefore, replacing it in the potential equation we get 
\begin{eqnarray}
\Phi = -\frac{4\pi G \rho_{cr}}{H_0^2} \int_1^0 a^2 \sqrt{\left( \frac{\Omega_m}{a^3}+\frac{\Omega_r}{a^2} + \Omega_\Lambda \right)} \times %\nonumber\\
\left(\int_1^a \frac{1}{a'^2 \sqrt{\frac{\Omega_m}{a'^3}+\frac{\Omega_r}{a'^4} + \Omega_\Lambda} }da'\right) da \,.
\end{eqnarray}

For the values $\Omega_m = 0.27$, $\Omega_\Lambda = 0.73$, and $\Omega_r = 0.0000824$, the value of $\Phi$ approaches $0.416$. Of course, Newtonian mechanics is not applicable in such scenarios. However, we observe that the result is quite close to $1$. Its very interesting to see that this value is exactly $1$ for a completely matter dominated Universe. In the Machian gravity theory presented in this paper, the dark components come only from the geometry of the Universe. Therefore, this value needs detailed analysis. 

It is also worth noting that if we were to use the real distance instead of the luminosity distance, we would obtain $\Phi \rightarrow 2.9$, which is still relatively close to $1$. Thus, we can see that the gravitational potential tends to remain approximately of order $1$. 

Of course, neither general relativity nor Newtonian mechanics adheres to Mach's principle, so there is no inherent reason why $\Phi$ should be close to $1$. Even in Sciama's theory, while $\Phi$ is required to equal $1$, ideally it can be slightly modified to fit any value as long as it's of the order of $1$. In our Machian Gravity theory there is no requirement for the Newtonian potential to be a unity or even constant. We can always adjust the $v$ parameter and take care of any potential.

\section{An illustrative understating of the origin of pseudo forces}
\label{generalizedinertia}
In Section~\ref{SciamaTheory}, we discuss the Sciama's approach to solve the Mach's principle. Here, I explain it in a more organized way in generalized coordinate system and with some example. This will help the readers to understand the  theory better.  

\subsection{Pseudo forces in a non-relativistic frame}
Let's say there are two reference frames, $K$ and $K'$. $K$ is fixed to the background, and it conventionally serves as the inertial reference frame. Let’s assume that the origins of these coordinate systems coincide. Now, consider that the coordinate frame $K'$ has some rotational velocity and some linear acceleration $\vec{a}$. Suppose that $K'$ is rotating with an angular velocity $\Omega$ relative to $K$ and that the motion is non-relativistic. If we denote the position vectors of the particle in $K$ and $K'$ as $\vec{r}$ and $\vec{r}'$, respectively, we can then write

\begin{equation}
\frac{d^2 \vec{r}}{dt^2}=\frac{d^2 \vec{r}'}{dt^2}+ \vec{\mathbf{a}} +2 \vec{\boldsymbol{\Omega}} \times \frac{d \vec{r'}}{dt}+\frac{d \vec{\boldsymbol{\Omega}}}{d t} \times \vec{r'}+\vec{\boldsymbol{\Omega}} \times\left(\vec{\boldsymbol{\Omega}} \times \vec{r'}\right) \;.
\label{Realforce}
\end{equation}

Here the first term on the right-hand side represents the acceleration of the particle in the $K'$ reference frame. The second term gives the acceleration of the $K'$ frame with respect to the $K$ frame. The third term is the Coriolis force, the fourth term is the Euler force, and the fifth term is the centripetal force. All of these accelerations are observed from the perspective of an observer in the $K$ reference frame, which is either fixed or moving with a constant linear velocity relative to the background. Therefore, an observer in the $K$ frame can calculate that an observer in the $K'$ frame would perceive these accelerations.

Now consider from the perspective of an observer sitting in the $K'$ frame. In this case, $\vec{\mathbf{a}} = \vec{\mathbf{\Omega}} = 0$, which means that all terms from the second onward in the previous equation vanish. Thus, according to this observer, the particle's acceleration would simply be $\frac{d^2 \vec{r}'}{dt^2}$. Therefore, the observer may think that the force that is acting on the particle is $m_0\frac{d^2 \vec{r}'}{dt^2}$. However, we know that this is not true as the observer in $K'$ frame will experience pseudo forces. We need to add the pseudo forces for calculating the actual acceleration experienced by the particle. According to Sciama's theory this force is coming due to the motion of the distant objects in the background. 

We can use the Sciama's theory to calculate these pseudo forces. In this $K'$ frame, the velocity of the background mass distribution near the particle is 

\begin{equation}
\vec{V}_B = \vec{V} - \vec{\boldsymbol{\Omega}} \times \vec{r}'   
\end{equation}

\noindent where $\vec{V} = \frac{d(\vec{r}-\vec{r'})}{dt}$ and $a = \frac{d\vec{V}}{dt}$ are the velocity and acceleration of $K'$ frame with respect to the $K$ frame. Therefore, we can calculate the gravito-magnetic fields as

\begin{eqnarray}
\vec{E} &=& - \vec{\nabla} \Phi - \frac{\partial \vec{\mathcal{A}}}{\partial t}\,, \qquad \vec{B} = \vec{\nabla} \times \vec{\mathcal{A}}\,. \\
%\end{eqnarray}
%\begin{eqnarray}
\vec{F} &=& \left(\vec{E}+\vec{V}' \times \vec{B}\right) %\nonumber\\
%&=&
= \left(- \vec{\nabla} \Phi - \frac{\partial \vec{\mathcal{A}}}{\partial t}+\vec{V}' \times \left( \vec{\nabla} \times \vec{\mathcal{A}} \right)\right)\,.
\end{eqnarray}

\noindent Now, $\Phi$ and $\vec{\mathcal{A}}$ are given by 

\begin{eqnarray}
\vec{\mathcal{A}} = m_g I \vec{V}_B\,, \qquad \Phi = - m_g I \sqrt{1+\vec{V}_B . \vec{V}_B}\,.
\end{eqnarray}

\noindent Therefore, we can calculate 

\begin{eqnarray}
\vec{\nabla}\Phi / (m_g I) &=& -\frac{(\vec{V}_B . \vec{\nabla} )\vec{V}_B}{\sqrt{1+\vec{V}_B . \vec{V}_B}} \approx -(\vec{V}_B . \vec{\nabla} )\vec{V}_B  %\nonumber \\
=  - \vec{\boldsymbol{\Omega}} \times (\vec{\boldsymbol{\Omega}} \times \vec{r}' ) \nonumber \\
\vec{V}' \times (\vec{\nabla} \times \vec{\mathcal{A}}) / (m_g I) &=& - 2 (\vec{\boldsymbol{\Omega}} \times \vec{V}') \nonumber \\
\frac{d \vec{\mathcal{A}}}{d t} / (m_g I) &=& \frac{d\vec{V}}{dt} + \frac{d \vec{\boldsymbol{\Omega}}}{dt} \times \vec{r}' \;.
\end{eqnarray}

\noindent So, putting everything back in the force equation we can get

\begin{equation}
a_{pseudo} = \left(\frac{m_g I}{m_0}\right)\left[ - \vec{a} - \frac{d \vec{\boldsymbol{\Omega}}}{dt} \times \vec{r}' - 2 \vec{\boldsymbol{\Omega}} \times \vec{V}' - \vec{\boldsymbol{\Omega}} \times (\vec{\boldsymbol{\Omega}} \times \vec{r}' ) \right] \,.
\label{Pseudoforce}
\end{equation}

\noindent Therefore, in the $K'$ reference frame, the gravitational field on the particle due to the rest of the Universe produces a force equal to $F_{\text{pseudo}} = m_0 a_{\text{pseudo}}$. 
We can see that this force is exactly equal and opposite to the force in Eq.~\ref{Realforce}. 
Although these two expressions originate from entirely different derivations \---   Eq.~\ref{Realforce} from a vector algebra giving the accelerations with respect to $K'$ frame, while, Eq.~\ref{Pseudoforce} gives the gravitational force from on the particle due to the motion of the distant objects in the Universe \--— their final forms are identical, which is interesting.
%Although these two expressions arise from completely different derivations—Eq.~\ref{Realforce} from a vector product and the second set from vector calculus, the final results are exactly the same, which is quite surprising. 

To illustrate, consider the example of a merry-go-round. Imagine two people sitting on opposite sides of its diameter. If one throws a ball toward the other, a stationary observer in the $K$ frame will see the ball moving in a straight line. However, as the merry-go-round rotates with angular velocity $\boldsymbol{\Omega}$, the person on the merry-go-round will perceive a force $2\boldsymbol{\Omega}\times V'$, corresponding to the Coriolis force in Eq.~\ref{Realforce}.

Now, in the rotating reference frame of those on the merry-go-round, let’s imagine it is isolated in space, with no nearby objects visible. To these individuals, the merry-go-round has zero angular velocity relative to their own reference frame. Consequently, they would expect zero Coriolis force, yet they observe the ball’s trajectory curving rather than moving in a straight line. Thus, the ball appears to experience a force in their frame. In their view, it is the entire Universe that rotates, producing a gravitomagnetic force acting on the ball and causing it to change direction. This force is given by Eq.~\ref{Pseudoforce}.

\subsection{An illustration of pseudo forces in a relativistic case}
\label{rotating_frame}

Now let us consider the relativistic case. First, assume that $K$ is an inertial reference frame. In this frame, the background created by distant stars and galaxies is fixed. Therefore, a particle in this frame does not experience any force. Let the coordinates in this frame be $(t, x, y, z)$. Next, consider another reference frame, $K'$, which rotates with an angular velocity $\omega$ in the $x$-$y$ plane. The origins of both coordinate systems coincide. Let the coordinates in the $K'$ frame be $(t', x', y', z')$.

The coordinate transformation between these two reference frames is given by

\begin{eqnarray}
t &=& t'\,, \nonumber \\
x &=& x' \cos(\omega t') - y' \sin(\omega t') \,, \nonumber\\
y &=& x' \sin(\omega t') + y' \cos(\omega t') \,, \nonumber \\
z &=& z'\,.
\end{eqnarray}

\noindent The proper acceleration of a particle in any coordinate system is given by the geodesic equation

\begin{equation}
a^\mu = \frac{d^2 x'^\mu}{d\tau^2} - \Gamma^\mu_{\nu\lambda}\frac{dx'^\nu}{d\tau}\frac{dx'^\lambda}{d\tau} \,.
\label{amu}
\end{equation}

Here, the first term $\frac{d^2 x'^\mu}{d\tau^2}$ is the acceleration of the particle with respect to someone in the $K'$ reference frame. The second term represents the acceleration caused by some inertial forces. So when we look at $K$, we will see that we need to add these acceleration components with the acceleration observed by the observer in the reference frame $K'$.

In the $K$ reference frame the metric is a Minkowski metric. Therefore, the line element is given by $d\tau^2 = -dt^2 + dx^2 + dy^2 + dz^2$. As the $K'$ reference frame is rotating at a speed $\omega$ with respect to $K$, the line element in the $K'$ reference frame is given by $d\tau^2 = - dt'^2 + dx'^2 + dy'^2 + dz'^2 + \omega^2 (x'^2 + y'^2) dt'^2 + 2\omega (x' \, dy' - y' \, dx') dt'$.

Here, for simplicity I have assumed that $\omega$ is not a function of time. The Christoffel symbols for this metric are given by

\begin{eqnarray}
& \Gamma^x_{tt} = -\omega^2 x' \,,\qquad\qquad
\Gamma^x_{ty} = \Gamma^x_{yt} = -\omega \,,\nonumber\\
& \Gamma^y_{tt} = -\omega^2 y' \,, \qquad\qquad
\Gamma^y_{tx} = \Gamma^y_{xt} = \omega  \,,\nonumber\\
& \text{all others} = 0 \,.
\end{eqnarray}

\noindent We can use these to calculate the components of Eq.~\ref{amu} as 

\begin{eqnarray}
\frac{d^2 t}{d\tau^2} &=&\frac{d^2 t'}{d\tau^2} \,,  \nonumber \\
\frac{d^2 x}{d\tau^2} &=&\frac{d^2 x'}{d\tau^2} - \omega^2 x' \left( \frac{dt'}{d\tau} \right)^2 - 2 \omega \frac{dy'}{d\tau} \frac{dt'}{d\tau} \,,\nonumber \\
\frac{d^2 y}{d\tau^2} &=&\frac{d^2 y'}{d\tau^2} - \omega^2 y' \left( \frac{dt'}{d\tau} \right)^2 + 2 \omega \frac{dx'}{d\tau} \frac{dt'}{d\tau} \,,\nonumber \\
\frac{d^2 z}{d\tau^2} &=&\frac{d^2 z'}{d\tau^2}  \,.
\end{eqnarray}

Here, the terms proportional to $\omega^2$ correspond to the centripetal acceleration, while those proportional to $2\omega$ represent the Coriolis force. Additionally, if we had included the $\dot{\omega}$ terms, we would have obtained an Euler-type force. Therefore, these are the standard equations that we use in the special or the general theory of relativity. Of course as there are no external forces as assumed in the beginning, in $K$ frame the accelerations are $0$. However, the observer who is sitting in the $K$ reference frame can calculate that the observer at $K'$ reference frame will experience this kind of centripetal and Coriolis acceleration.

Now from the prospective of an observer in $K'$ reference frame $\omega=0$. So anyway, the observer in the $K'$ reference frame will assume that its reference frame is some Minkowski type of reference frame. Therefore, all the Crystoffel symbols will be $0$.  However, she is seeing some force being applied on the particle.

Therefore, according to Sciama's theory, it should have happened that the observer in the $K'$ reference frame should see that the background behind it is moving with a velocity, and due to that there will be some gravito-magnetic tensor (similar to the electro-margentic tensor) which is responsible for the pseudo force. From Eq.~\ref{Eq:33}, we can see that we can create the $F_{\mu\nu}$ for this particular situation as

\begin{equation}
F_{\mu\nu} =
\left(\frac{dt'}{d\tau}\right)\begin{bmatrix}
0 & -\omega^2 x' & -\omega^2 y' & 0 \\
\omega^2 x' & 0 & -2\omega & 0 \\
\omega^2 y' & 2\omega & 0 & 0 \\
0 & 0 & 0 & 0
\end{bmatrix}.
\end{equation}

\noindent Using this $F_{\mu\nu}$ and using the fact that $r^2 = x'^2 + y'^2$ is constant we can obtain the same acceleration using equation  $\frac{d^2 x'^\mu}{d\tau^2}=\left(\frac{m_g I}{m_0}\right) \eta^{\mu\nu} F_{\nu\alpha} \frac{dx'^\alpha}{d\tau}$. 
Therefore, We can see that using Kaluza-Klein type of metric we can obtain Eq.~\ref{Eq:33}, that can be used to explain Sciama's hypothesis. If a particle rotates around a center point or a particle is fixed and the entire Universe rotates around a center point both can give same equation of motion.

\section{A brief discussion about Kaluza-Klein mechanism}
\label{appendix_Crystoffel}

In this work, we have extensively used the Kaluza-Klein mechanism to project the five-dimensional geometry into a four-dimensional metric, along with a scalar and vector field. In this section, we provide a step-by-step derivation of the Kaluza-Klein equations. These detailed mathematical expressions are used in various calculations throughout the paper.

The 5-dimensional metric and its inverse can be expressed in terms of a 4-dimensional metric (If the five-dimensional metric depends on the fifth dimension, it will not yield the four-dimensional metric and vector field. However, this kind of separation can be achieved in any case.) as 

\begin{eqnarray}
\widetilde{g}_{A B} &=& \left(\begin{array}{cc}
g_{\alpha \beta}+\upsilon^2 \phi^2 A_\alpha A_\beta & \upsilon \phi^2 A_\alpha \\
\upsilon \phi^2 A_\beta &  \phi^2
\end{array}\right) \quad \text { or } \nonumber\\  \widetilde{g}^{A B}&=&\left(\begin{array}{cc}
g^{\alpha \beta} & -\upsilon A^\alpha \\
-\upsilon A^\beta & \upsilon^2 A^\beta A_\beta+\frac{1}{\phi^2}
\end{array}\right)\,.
\label{appendixA:gmunu}
\end{eqnarray}

\subsection{Cristoffel Symbols}
We can calculate the Christoffel symbols as 
\begin{equation}
 \widetilde{\Gamma}_{A B}^C=\frac{1}{2} \widetilde{g}^{C D}\left(\partial_B \widetilde{g}_{D A}+\partial_A \widetilde{g}_{D B}-\partial_D \widetilde{g}_{A B}\right)   
\end{equation}
and write those in terms of the 4 dimensional Christoffel symbols as

\begin{widetext}
\begin{eqnarray}
\widetilde{\Gamma}_{44}^4
&=& -\frac{1}{2} \partial_\alpha \widetilde{g}_{44} \widetilde{g}^{4 \alpha} + \left[ \partial_4 \widetilde{g}_{4 \alpha} \widetilde{g}^{4 \alpha}+\frac{1}{2} \partial_4 \widetilde{g}_{44} \widetilde{g}^{44}\right] \nonumber \\
&=&\frac{1}{2} \upsilon A^\nu \partial_\nu \phi^{2}   +
\left[ -\upsilon^2 \phi^{2}A^\alpha {\partial_4 A_\alpha} +  \frac{1}{2}\partial_4(\ln\phi^{2}) - \frac{1}{2} \upsilon^2 A^\alpha A_\alpha\partial_4\phi^{2}\right], \\
\widetilde{\Gamma}_{4 \nu}^4 
&=& \frac{1}{2} \widetilde{g}^{4 \alpha}\left(\partial_\nu \widetilde{g}_{\alpha 4}-\partial_\alpha \widetilde{g}_{4 \nu}\right)+\frac{1}{2} \widetilde{g}^{44} \partial_\nu \widetilde{g}_{44}  + \left[\frac{1}{2} \partial_4 \widetilde{g}_{\nu \alpha} \widetilde{g}^{4 \alpha}\right] \nonumber \\
%\end{eqnarray}
%\newpage
%\begin{eqnarray}
&=&\frac{1}{2}\upsilon^2 \phi^2 A^\alpha F_{\alpha \nu}  +\frac{1}{2 \phi^{2}} \partial_\nu \phi^{2} +\frac{1}{2}\upsilon^2 A_\nu A^\alpha \partial_\alpha \phi^2  +  \nonumber \\
&&  \left[-\frac{1}{2} \upsilon A^\alpha \partial_4 g_{\nu \alpha} - \frac{1}{2}\upsilon^3 A_\nu A_\alpha A^\alpha \partial_4 \phi^2  -\frac{1}{2} \upsilon^3 A_\alpha A^\alpha \partial_4 A_\nu \phi^2-\frac{1}{2} \upsilon^3 A_\nu A^\alpha \partial_4 A_\alpha \phi^2  \right]\,, \\
\widetilde{\Gamma}_{44}^\nu &=& \widetilde{g}^{\nu \alpha} \partial_4 \widetilde{g}_{4 \alpha} +\left[\frac{1}{2} \widetilde{g}^{4 \nu} \partial_4 \widetilde{g}_{44} -\frac{1}{2} \widetilde{g}^{\nu \alpha} \partial_\alpha \widetilde{g}_{44} \right]
\noindent \nonumber 
= -\frac{1}{2} g^{\nu \alpha} \partial_\alpha \phi^2 + \left[ \frac{1}{2} \upsilon \partial_4 \phi^2 A^\nu + \upsilon \phi^2 g^{\nu \alpha} \partial_4 A_\alpha   \right] \\
\widetilde{\Gamma}_{\alpha \nu}^4 &=& \frac{1}{2} \tilde{g}^{4 \beta}\left(\partial_\alpha \tilde{g}_{\beta \nu}+\partial_\nu \widetilde{g}_{\beta \alpha}-\partial_\beta \widetilde{g}_{\alpha \nu}\right) +\frac{1}{2} \widetilde{g}^{44}\left(\partial_\alpha \widetilde{g}_{4 v}+\partial_\nu \widetilde{g}_{4 \alpha}\right) -\left[\frac{1}{2}\partial_4 \widetilde{g}_{\alpha\nu} 
\widetilde{g}^{44}\right] \nonumber \\
&=& -\upsilon A_\beta \Gamma_{\alpha \nu}^\beta+\frac{1}{2} \upsilon^3 A^\beta A_\nu \phi^{2} F_{\beta \alpha} +\frac{1}{2} \upsilon^3 A_\alpha A^\beta \phi^{2} F_{\beta \nu} +\frac{1}{2} \upsilon^3 A_\alpha A^\beta A_\nu \partial_\beta \phi^{2} +\frac{1}{2}\upsilon \left(\partial_\alpha A_\nu+\partial_\nu A_\alpha\right) \nonumber\\
&\,& +\frac{1}{2 \phi^{2}}\upsilon\left( A_\nu \partial_\alpha \phi^{2}+A_\alpha \partial_\nu \phi^{2}\right) - \left[\frac{1}{2} ( \phi^{-2}+ \upsilon^2 A^\beta A_\beta) \partial_4(  g_{\alpha \nu} +  \upsilon^2 \phi^2 A_\alpha A_\nu ) \right] \\
%\end{eqnarray}
%\begin{eqnarray}
\widetilde{\Gamma}_{4 \alpha}^\nu &=& \frac{1}{2} \widetilde{g}^{\nu \mu}\left(\partial_\alpha \tilde{g}_{\mu 4}-\partial_\mu \tilde{g}_{4 \alpha}\right)+\frac{1}{2} \widetilde{g}^{v 4} \partial_\alpha \tilde{g}_{44} +\left[ \frac{1}{2} \widetilde{g}^{\nu \beta}\partial_4 \widetilde{g}_{\alpha \beta} \right]\nonumber \\
&=& \frac{1}{2} \upsilon g^{\nu \mu}\left(\phi^{2} F_{\alpha \mu}-A_\alpha \partial_\mu \phi^{2}\right) + \left[ \frac{1}{2} g^{\nu\beta}\partial_4(  g_{\alpha \beta} +  \upsilon^2\phi^2 A_\alpha A_\beta ) \right] \text {, } \\
\widetilde{\Gamma}_{\mu \nu}^\beta &=& \frac{1}{2} \tilde{g}^{\beta \alpha}\left(\partial_\mu \tilde{g}_{\alpha \nu}+\partial_\nu \tilde{g}_{\alpha \mu}-\partial_\alpha \tilde{g}_{\mu \nu}\right) +\left[\frac{1}{2} \tilde{g}^{\beta 4}\left(\partial_\mu \tilde{g}_{4 v}+\partial_\nu \tilde{g}_{4 \mu}\right) -\frac{1}{2} \partial_4 \tilde{g}_{\mu \nu} \tilde{g}^{4 \beta}\right] \nonumber\\
&=& \Gamma^\beta_{\mu \nu} + \frac{1}{2} \upsilon^2 g^{\beta \alpha} \left(\phi^2 A_\mu  F_{\nu \alpha}+  \phi^2 A_\nu  F_{\mu \alpha} -\partial_\alpha \phi^2 A_\mu A_\nu \right) \nonumber\\
&&  + \left[ \frac{1}{2} \upsilon A^\beta (\partial_4 g_{\mu \nu}  + \upsilon^2 \partial_4 \phi^2 A_\mu A_\nu ) +  \frac{1}{2} \upsilon^3 \phi^2 A^\beta \partial_4(A_\mu A_\nu) \right] \,.
\end{eqnarray}
\end{widetext}

\subsection{Kaluza-Klein mechanism}
\label{KaluzaKleinEquationC2}

In Sec.~\ref{sec322}, we demonstrated that inertial forces, such as the Centrifugal Force, Coriolis Force, Euler Force, and others, can arise due to the motion of distant objects in the Universe. These forces are derived using the Kaluza-Klein theory, assuming that all variables are independent of $x^4$, i.e. the $p_4$ is constant. To calculate the equation of motion for a particle, we begin with the Lagrangian describing its motion.

\begin{equation}
\mathcal{L} = \frac{1}{2} g_{\mu\nu} \dot{x}^\mu\dot{x}^\nu +\frac{1}{2} \phi^2\left( \dot{x}^4 + \upsilon A_\mu \dot{x}^\mu \right)^2 \,.
\end{equation}

\noindent Therefore, the equation of motion for $p_4$ will be 

\begin{eqnarray}
\frac{p_4}{m_0} = \frac{\partial\mathcal{L}}{\partial\dot{x}^4} = \phi^2 \left( \dot{x}^4 + \upsilon A_\mu \dot{x}^\mu \right) \,.
\label{eqB9}
\end{eqnarray}

%\begin{eqnarray}
%&& \frac{d}{d\tau} \left(A_\lambda  \frac{d x^\lambda}{\mathrm{~d} \tau}  +  \frac{d x^4}{\mathrm{~d} \tau}\right) = -\widetilde{\Gamma}^4_{AB}\frac{dx^A}{d\tau}\frac{dx^B}{d\tau} \nonumber\\&& 
%- A_\alpha \widetilde{\Gamma}^\alpha_{AB} \frac{dx^A}{d\tau}\frac{dx^B}{d\tau} + \partial_A A_B \frac{dx^A}{d\tau}\frac{dx^B}{d\tau} = 0\,.
%\end{eqnarray}

%\noindent Here we just replace the geodesic equations. Also $A_4 = 1$, and hence its derivative is $0$. This makes $\left(A_\lambda  \frac{d x^\lambda}{\mathrm{~d} \tau}  +  \frac{d x^4}{\mathrm{~d} \tau}\right) = K$, where $K$ is a constant of integration. Without the loss of generality, we can take $K=1/\phi^2$ by redefining the coordinate system. Note that in this case, we assume the $\phi$ is a constant. If it's not a constant, then this redefinition will not work. 

\noindent We can use the above condition to get the equation of motion to be

\begin{eqnarray}
&& \frac{\mathrm{d}^2 x^\mu}{\mathrm{~d} \tau^2}+\widetilde{\Gamma}_{B C}^\mu \frac{\mathrm{d} x^B}{\mathrm{~d} \tau} \frac{d x^C}{\mathrm{~d} \tau} \nonumber\\
&=& \frac{\mathrm{d}^2 x^\mu}{\mathrm{~d} \tau^2}+\widetilde{\Gamma}_{\nu \lambda}^\mu \frac{\mathrm{d} x^\nu}{\mathrm{~d} \tau} \frac{d x^\lambda}{\mathrm{~d} \tau} \nonumber  
+2\widetilde{\Gamma}_{\nu 4}^\mu \frac{\mathrm{d} x^\nu}{\mathrm{~d} \tau} \frac{d x^4}{\mathrm{~d} \tau}+\widetilde{\Gamma}_{4 4}^\mu \frac{\mathrm{d} x^4}{\mathrm{~d} \tau} \frac{d x^4}{\mathrm{~d} \tau} \nonumber\\
&=& \frac{\mathrm{d}^2 x^\mu}{\mathrm{~d} \tau^2} + 
\Gamma_{\nu \lambda}^\mu   \frac{\mathrm{d} x^\nu}{\mathrm{~d} \tau} \frac{d x^\lambda}{\mathrm{~d} \tau}
+  \left(\frac{\upsilon p_4}{m_0}\right)g^{\mu \alpha}F_{\nu \alpha} \frac{\mathrm{d} x^\nu}{\mathrm{~d} \tau} \nonumber \\
&& - \left(\frac{p_4}{m_0}\right)^2 g^{\mu\beta} \frac{\partial_\beta \phi}{\phi^3} \,.
\end{eqnarray}

Here for the last step we put the values of the Christoffel's symbols and then use Eq.~\ref{eqB9}. 
We can see that, we get an extra vector and scalar fields in the geodesic equation coming from the 5-dimensional coordinate system. As we discussed in the text, we can use these fields to generate all the pseudo-inertial forces. 

\subsection{Calculating the Einstein's Tensor}

For now we will consider that the metric does not depend on $x^4$.  Therefore, the derivatives with respect to $x^4$ are $0$. A general expression will make expressions extremely complicated. Therefore, a general expression is only given for some specific metric in  Appendix~\ref{Appcosmology}. 

We can calculate the Ricci tensor and Ricci scalar using the expressions for Cristoffel's symbols. The Ricci tensors are given by~\cite{Williams2015} 

%\url{https://arxiv.org/pdf/1204.3455}
\begin{eqnarray}
\widetilde{R}_{\mu \nu} &=& \widetilde{R}_{\mu 4 \nu}^4+\widetilde{R}_{\mu \alpha \nu}^\alpha \nonumber\\
&=& R_{\mu \nu}-\frac{1}{2} \upsilon^2 \phi^2 g^{\alpha \beta} F_{\mu \alpha} F_{\nu \beta}-\frac{1}{\phi} \nabla_\mu \nabla_\nu \phi \nonumber\\
&& + \upsilon^2 A_\mu A_\nu \widetilde{R}_{44}+\upsilon A_\mu\left(\widetilde{R}_{\nu 4}-\upsilon A_\nu \widetilde{R}_{44}\right) %\nonumber\\ && 
+\upsilon A_\nu\left(\widetilde{R}_{\mu 4}-\upsilon A_\mu \label{B11} \widetilde{R}_{44}\right) \\
\widetilde{R}_{4\alpha} &=& \frac{1}{2} \upsilon \phi^2 g^{\beta \mu} \nabla_\mu F_{\alpha \beta}+\frac{3}{4} \upsilon F_{\alpha \beta} \partial^\beta \phi^2+\upsilon A_\alpha \widetilde{R}_{44} \label{B12}\\
\widetilde{R}_{44} &=& -\phi \square \phi+\frac{1}{4} \upsilon^2 \phi^4 F^{\alpha \beta} F_{\alpha \beta} \,.
\label{B13}
\end{eqnarray}

\noindent The Ricci scalar can be derived as 

\begin{eqnarray}
\widetilde{R}  =g^{\mu \nu} \widetilde{R}_{\mu \nu}-2 \upsilon A^\mu \widetilde{R}_{\mu 4}+\left(\upsilon^2 A^2+\frac{1}{\phi^2}\right) \widetilde{R}_{44} %\nonumber \\
 =R-\frac{1}{4} \upsilon^2 \phi^2 F^{\alpha \beta} F_{\alpha \beta}-\frac{2}{\phi} \square \phi \,.
 \label{R114}
\end{eqnarray}

\noindent We can calculate the Einstein's tensor from the Ricci tensors as $\widetilde{G}_{A B} \equiv \widetilde{R}_{A B}-\widetilde{g}_{A B} \widetilde{R} / 2$. The values for the Einstein tensor are given by 

\begin{eqnarray}
\widetilde{G}_{\mu \nu} &=& \widetilde{R}_{\mu \nu}-\frac{1}{2} \widetilde{g}_{\mu \nu} \widetilde{R} \nonumber\\
&=& G_{\mu \nu}-\frac{1}{\phi}\left(\nabla_\mu \partial_\nu \phi-g_{\mu \nu} \square \phi\right) %\nonumber\\&& 
-\frac{1}{2} \upsilon^2 \phi^2\left(g^{\alpha \beta} F_{\mu \alpha} F_{\nu \beta}-\frac{1}{4} g_{\mu \nu} F_{\alpha \beta} F^{\alpha \beta}\right) \nonumber\\
&& +\upsilon^2 A_\mu A_\nu \widetilde{G}_{44}+\upsilon A_\mu\left(\widetilde{G}_{\nu 4}-\upsilon A_\nu \widetilde{G}_{44}\right) %\nonumber\\ && 
+\upsilon A_\nu\left(\widetilde{G}_{\mu 4}-\upsilon A_\mu \widetilde{G}_{44}\right) 
\label{G11}
\,,\\
\widetilde{G}_{4 \nu} &=& \widetilde{R}_{4\nu}-\frac{1}{2} \upsilon \phi^2 A_\nu \widetilde{R} %\nonumber\\ && 
=\upsilon A_\nu \widetilde{G}_{44}+\frac{1}{2} \upsilon \phi^2 g^{\alpha \beta} \nabla_\beta F_{\nu \alpha}+\frac{3}{4} \upsilon F_{\nu \alpha} \partial^\alpha \phi^2
\label{G12} \,,\qquad\\
\widetilde{G}_{44} &=& \widetilde{R}_{44}-\frac{1}{2} \phi^2 \tilde{R} %\nonumber\\ &=& 
= \frac{3}{8} \upsilon^2 \phi^4 F_{\alpha \beta} F^{\alpha \beta}-\frac{1}{2} \phi^2 R 
\label{G13}\,.
\end{eqnarray}

\section{Understanding the Hoyle-Narlikar's argument with C field}
\label{AppendixHN}
In this section, I discuss the argument put forward by Hoyle and Narlikar in~\cite{hoyle1963mach} to explain Mach's principle and how the Machian gravity model satisfies the view.

Newton, in his work, discussed his experiments of a rotating water-filled bucket suspended from a  twisted thread. The crucial point was that whenever rotation occurred relative to some particular reference frame, the surface of the water became depressed \--- a clear absolute effect, not merely a relative one. 

It was also clear that the reference frame, relative to which inertial forces were observed,  coincided within experimental error with the frame in which distant objects in the Universe were non-rotating. More accurate later experiments have confirmed this coincidence. Since the coincidence can scarcely be accidental, it is necessary to attempt an explanation of it.

Mach suggested that the correlation between the water curvature in Newton's bucket and the rotation of distant matter in the Universe can be addressed if we consider that the distant matters of the Universe are affecting the inertial properties of matter. 

The standard model of cosmology is based on general relativity along with two postulates, namely,  
\begin{enumerate}
    \item The Weyl postulate, which says that the world lines of matter form a geodesic congruence are normal to the spacelike hypersurface, which leads us to the line element of the form $ds^2 = dt^2 -g_{i j}dx^i dx^j$. and 
    \item 
The cosmological principle that says that at $t=constant$ hypersurface, the Universe is isotropic and homogeneous. 
\end{enumerate}

\noindent This leads us to the Robertson-Walker line element. 

Since $\theta$ and $\phi$ are not changed by transformation (isotropy), we can fix the  $\theta$ and $\phi$ coordinates by looking at a distant galaxy. Then we can use Einstein's equation 

\begin{equation}
G^{\mu\nu} + \Lambda g^{\mu\nu} = T^{\mu\nu}\,.
\label{EinsteinGREq}
\end{equation}

\noindent Under the previous considions we can show that the stress-energy tensor takes the form 

\begin{equation}
    T^{\mu\nu} = (\rho+p)\frac{dx^\mu}{ds}\frac{dx^\nu}{ds} + p g^{\mu\nu}
\label{FRWTmunu}
\end{equation}

\noindent where, $\rho$ and $p$ are the density and the pressure of the matter content. 

However, this is in direct contradiction to Mach's principle. Mach's principle requires us to read the Eq.~\ref{EinsteinGREq} from the right, i.e., given the $T^{\mu\nu}$, is it possible to get some unique line element from the equations. If the answer is affirmative, only then we can say that the theory can explain the observations related to the rotating frame. However, Gödel (1949)~\cite{godel1949example} showed that for the normal form of $T^{\mu \nu}$, the answer is not affirmative. Gödel obtained an explicit solution in which the line element is of the form

\begin{equation}
 \quad \mathrm{d} s^2=\mathrm{d} t^2+2 \mathrm{e}^{x^1} \mathrm{~d} t \mathrm{~d} x^2-\left(\mathrm{d} x^1\right)^2+\frac{1}{2} \mathrm{e}^{2 x^1}\left(d x^2\right)^2-\left(\mathrm{d} x^3\right)^2   
\end{equation}

\noindent and where $T^{i k}$ is given by Eq.~\ref{FRWTmunu} with $u=p=0, \rho=1 / \kappa, \Lambda=-\frac{1}{2} \kappa \rho$. This solution is fundamentally different from the Robertson-Walker line element, i.e., it cannot be obtained from FRW metric by a coordinate transformation. The importance of Gödel's solution is that it exhibits a vorticity of matter. 

So, in general relativity, the maximum that we can do is to take a spacelike surface and, on it, define the coordinate systems. This simply removes the arbitrariness of the coordinate systems. On top of this, define the matter and the kinematical situations and the quantities 

\begin{equation}
    g_{\mu\nu}, \frac{\partial g_{\mu\nu}}{\partial x^\mu}, \frac{\partial^2 g_{\mu\nu}}{\partial x^\mu\partial x^\nu}
\end{equation}

\noindent consistently with Eq.~\ref{EinsteinGREq}. Then Eq.~\ref{EinsteinGREq}. allows one, in principle, to calculate both the dynamical situation and the form of the metric tensor at points of the initial surface. While the specifications can be made such that we can get the Robertson-Walker line element, we need to put conditions on the $g^{\mu\nu}$. In other words, Newton's absolute space has been replaced in the GR theory by initial boundary conditions on the metric tensor.

To avoid setting these initial conditions, Hoyle and Narlikar~\cite{hoyle1963mach} introduced an additional scalar field with negative density. By doing so, we don't need to fix the initial conditions independently. In other words $G^{\mu\nu}$ and $T^{\mu\nu}$ can be anything. The rest can be absorbed into the $C$-field.

They use action principle to deduce the field equations as 

\begin{eqnarray}
R^{\mu \nu}-\frac{1}{2} g^{\mu \nu} R &=& -\kappa\left[T^{\mu \nu}-f\left\{C^\mu C^\nu-\frac{1}{2} g^{\mu \nu} C^\kappa C_\kappa\right\}\right] \nonumber\\
C_{; \mu}^\mu &=& (1 / f) j_{;}^\mu, \quad T_{; \nu}^{\mu \nu}=f C^\mu C_{; \nu}^\nu,   
\end{eqnarray}

\noindent where $C$ is a scalar field, $f$ is a coupling constant that determines the expansion rate of the Universe. $j^\mu = \rho \frac{dx^\mu}{ds}$ is the mass current.  While their logic seems correct, the choice of the scalar $C$-field is completely arbitrary. They also failed to explain how this additional field will recover all Mach's ideas, e.g., the different pseudo forces. 

In Machian Gravity theory, instead of an ad hoc single field, there is a scalar field and a vector field and also several additional terms with derivatives with respect to the $\zeta$ terms. We have discussed before how these terms give rise to all the pseudo forces. These terms also complement the logic put forward by Hoyle and Narlikar, as we don't require any special boundary condition at the beginning.  

\section{Calculations for Cosmology}
\label{Appcosmology}

In Sec.~\ref{sec:Cosmology} we describe the field equations involved in the cosmology calculation. This section describes the detailed calculations involved in projecting the five-dimensional field equation to four-dimensional space~\cite{Overduin1998,Ponce1993,Wesson1992}.

Any generalized five-dimensional metric can be written in terms of a four-dimensional metric, a scalar field, and a vector field as Eq.~\ref{appendixA:gmunu}. To simplify the calculations, we can choose a coordinate system so that the off-diagonal terms corresponding to the fifth dimension become 0. We can write the metric as

\begin{equation}
\tilde{g}_{A B}=\left(\begin{array}{ccccc} 
& & & & 0 \\
& g_{\alpha \beta} & & & 0\\
& & & & 0\\
& & & & 0\\
0 & 0 & 0 & 0 & g_{44}
\end{array}\right)\,.
\end{equation}

\noindent Here $g_{\alpha\beta}$ is the 4-dimensional metric; hence, it doesn't have any $g_{44}$ component. We use the term $g_{44}$ for notation simplification, which is same as $g_{44} = \widetilde{g}_{44}$. The nonzero components of Christoffel's symbols are given by
\newline \;
\newline \;
\newline \;
\begin{widetext}
\begin{align}
&\widetilde{\Gamma}_{44}^4=\frac{g^{44}\partial_4 g_{44}}{2}\,, & &
\widetilde{\Gamma}_{4\nu}^4=\frac{g^{44}\partial_\nu g_{44}}{2}\,, & &
\widetilde{\Gamma}_{44}^\nu=-\frac{1}{2} g^{\nu \mu}  \partial_\mu g_{44} \,, 
\nonumber \\
&\widetilde{\Gamma}_{\alpha \beta}^4=-\frac{1}{2} g^{44}\partial_4 g_{\alpha \beta}\,, & &
\widetilde{\Gamma}_{\beta 4}^\alpha=\frac{1}{2} g^{\alpha \gamma} \partial_4 g_{\beta \gamma}\,, & &
\widetilde{\Gamma}_{\mu \nu}^\alpha= \Gamma_{\mu \nu}^\alpha\,.
\label{AppendixC:Christoffel}
\end{align}

%\end{widetext}
\noindent The five-dimensional Ricci tensor i.e., $\widetilde{R}_{A B}$ can be written in terms of Christoffel's symbol as 

\begin{equation}
\widetilde{R}_{A B}=\left(\widetilde{\Gamma}_{A B}^C\right)_{, C}-\left(\widetilde{\Gamma}_{A C}^C\right)_{, B}+\widetilde{\Gamma}_{A B}^C \widetilde{\Gamma}_{C D}^D-\widetilde{\Gamma}_{A D}^C \widetilde{\Gamma}_{C B}^D\,.
\end{equation}

\noindent Separating the indices $(A, B, \ldots)$, into $(\alpha,\beta,\ldots)$ and $(4)$,  we can get the Ricci tensor as

%\begin{widetext}
\begin{eqnarray}
\widetilde{R}_{\alpha \beta} &=&\left(\widetilde{\Gamma}_{\alpha \beta}^\gamma\right)_{, \gamma}-\left(\widetilde{\Gamma}_{\alpha \gamma}^\gamma\right)_{, \beta}+\widetilde{\Gamma}_{\alpha \beta}^\gamma \widetilde{\Gamma}_{\gamma \delta}^\delta  -\widetilde{\Gamma}_{\alpha \delta}^\gamma \widetilde{\Gamma}_{\gamma \beta}^\delta \nonumber \\
&\,&+\left(\widetilde{\Gamma}_{\alpha \beta}^4\right)_{, 4}-\left(\widetilde{\Gamma}_{\alpha 4}^4\right)_{, \beta}+\widetilde{\Gamma}_{\alpha \beta}^4 \widetilde{\Gamma}_{4 \delta}^\delta+\widetilde{\Gamma}_{\alpha \beta}^\gamma \widetilde{\Gamma}_{\gamma 4}^4-\widetilde{\Gamma}_{\alpha \delta}^4 \widetilde{\Gamma}_{4 \beta}^\delta-\widetilde{\Gamma}_{\alpha 4}^\gamma \widetilde{\Gamma}_{\gamma \beta}^4 \nonumber \\
&=& R_{\alpha \beta}+\left(\widetilde{\Gamma}_{\alpha \beta}^4\right)_{, 4}-\left(\widetilde{\Gamma}_{\alpha 4}^4\right)_{, \beta}+\widetilde{\Gamma}_{\alpha \beta}^4 \widetilde{\Gamma}_{4 \delta}^\delta+\widetilde{\Gamma}_{\alpha \beta}^\gamma \widetilde{\Gamma}_{\gamma 4}^4-\widetilde{\Gamma}_{\alpha \delta}^4 \widetilde{\Gamma}_{4 \beta}^\delta-\widetilde{\Gamma}_{\alpha 4}^\gamma \widetilde{\Gamma}_{\gamma \beta}^4 \,.
\end{eqnarray}

\noindent Note that the 4 dimensional $\Gamma$ matrices are same as their 5 dimensional counterpart for the spacetime component as shown in Eq.~\ref{AppendixC:Christoffel}. Replacing the values of Christoffel's symbols from Eq.~\ref{AppendixC:Christoffel}, we can obtain the Ricci tensor as

\begin{eqnarray}
\widetilde{R}_{\alpha \beta}&=&R_{\alpha \beta}-\frac{\partial_4 g^{44} \partial_4 g_{\alpha \beta}}{2}-\frac{\partial_4\partial_4 g^{44} g_{\alpha \beta}}{2}-\frac{g^{44}, \beta g_{44, \alpha}}{2}-\frac{g^{44} g_{44, \alpha \beta}}{2} +\frac{g^{44} g_{44, \lambda} \Gamma_{\alpha \beta}^\lambda}{2}  \nonumber \\ 
&\,&-\frac{\partial_4 g^{\mu v} g_{\mu \nu} \partial_4 g^{44} g_{\alpha \beta}}{4}
-\frac{\left(g^{44}\right)^2 \partial_4 g_{\alpha \beta} \partial_4 g_{44}}{4} +\frac{g^{\lambda \mu} g^{44} \partial_4 g_{\alpha \lambda} \partial_4 g_{\beta \mu}}{2}-\frac{\left(g^{44}\right)^2 g_{44, \alpha} g_{44, \beta}}{4} \,,\nonumber \\
\widetilde{R}_{44}&=&
-\frac{g^{\lambda \beta}{ }_{, \lambda} g_{44, \beta}}{2}-\frac{g^{\lambda \beta} g_{44, \beta \lambda}}{2}-\frac{\partial_4 g^{\lambda \beta}\partial_4 g_{\lambda \beta}}{2}-\frac{\partial_4\partial_4 g^{\lambda \beta} g_{\lambda \beta}}{2}  -\frac{g^{\lambda \beta} g_{44, \beta} g^{\mu \sigma} g_{\mu \sigma, \lambda}}{4} \nonumber \\
&\,& +\frac{g^{44} \partial_4 g_{44}  g^{\lambda \beta}\partial_4 g_{\lambda \beta}}{4} -\frac{g^{\mu \beta} \partial_4 g_{\lambda \beta} g^{\lambda \sigma} \partial_4 g_{\mu \sigma}}{4}+\frac{g^{44} g_{44, \lambda} g^{\lambda \beta} g_{44, \beta}}{4}\,, \nonumber \\
\widetilde{R}_{4 \alpha}&=&\sqrt{g_{44}} P_{\alpha ; \beta}^\beta \,.
\end{eqnarray}
%\end{widetext}

\noindent In the last equation, ';' represents the covariant derivative.  $P_\alpha^\beta$ is a $2^{n d}$ rank tensor and given by 

\begin{equation}
 P_\alpha^\beta=\frac{1}{2 \sqrt{g_{44}}}\left(g^{\beta \lambda} \partial_4 g_{\lambda \alpha}-\delta_\alpha^\beta g^{\mu \nu} \partial_4 g_{\mu \nu}\right).   
\end{equation}
%$$.

Earlier, we discussed the stress-energy tensor for the 5-dimensional equation. If we consider that the  variation of $\phi$ is significantly small then the component of the stress energy tensor corresponding to the 5-th dimension will be significantly small, giving $\widetilde{T}_{\alpha \beta} \approx T_{\alpha \beta}$. This gives

\begin{equation}
\widetilde{R}_{4 \alpha} = \widetilde{R}_{4 4} \approx 0 \,.
\end{equation}

\noindent As we can see that the 4-dimensional Ricci tensor contains additional terms. Therefore in the field equation when projected in 4-dimensiona, it gives 

\begin{equation}
R_{\alpha\beta}=T_{\alpha\beta} - g_{\alpha\beta}T + {\rm Additional\; geometric \;terms}.
\end{equation}

\noindent If we can show that these terms have the same property as that of dark matter and dark energy, then our theory can predict everything in the same way as that of the standard cosmology without demanding any form of dark matter and dark energy.

\subsection{Calculating the components for a diagonal metric}

The most general line element for explaining cosmology should satisfy two postulates, the Weyl postulate and the isotropy condition. Similar to the Robertson-Walker line element,  we can choose the line element as

\begin{equation}
d s^2=e^\omega d t^2-e^\kappa d r^2-R^2\left(d \theta^2+\sin ^2 \theta d \phi^2\right)+\epsilon e^\mu d \zeta^2
\end{equation}

\noindent The line element being a diagonal line element automatically satisfies the Weyl postulate. Although this line element is a spherically symmetric line element it does not by default satisfy the  isotropy condition. However, we can take $R = r e^{\kappa/2}$ for get a flat isotropic space-time. Similarly for a non-flat spacetime also we can chose $e^\kappa$ and $R$ accordingly. Therefore, if we derive the field equations for this general metric we can replace various elements to get the equations for any specific line element. The exponentials are taken to make sure that these quantities can't be negative. The extra parameter $\epsilon$ gives us a library of changing the signature of the background dimension, as well as we can put $\epsilon = 0$ to get the four-dimensional components of the Ricci tensor. The nonzero Christoffel's symbols from this metric can be calculated as

%\begin{widetext}
\begin{align}
& \Gamma_{00}^0=\frac{\dot{\omega}}{2}, & & 
\Gamma_{00}^1=\frac{\omega^{\prime}}{2} e^{\omega-\kappa}, & &
\Gamma_{00}^4=-\frac{\epsilon}{2} \omega^* e^{\omega-\mu}, & &
\Gamma_{01}^0=\frac{\omega^{\prime}}{2}, \nonumber \\
& \Gamma_{01}^1=\frac{\omega}{2}, & &
\Gamma_{02}^2=\frac{\dot{R}}{R}, & &
\Gamma_{03}^3=\frac{\dot{R}}{R}, & &
\Gamma_{04}^0=\frac{\omega^*}{2},  \nonumber \\
& \Gamma_{11}^0=\frac{\kappa}{2} e^{\kappa-\omega}, & &
\Gamma_{11}^1=\frac{\kappa^{\prime}}{2}, & &
\Gamma_{11}^4=\frac{\epsilon}{2} \kappa^* e^{\kappa-\mu}, & &
\Gamma_{12}^2=\Gamma_{13}^3=\frac{R^{\prime}}{R}, \nonumber \\
& \Gamma_{14}^1=\frac{\kappa^*}{2}, & &
\Gamma_{41}^4=\frac{\mu^{\prime}}{2}, & &
\Gamma_{22}^0=R \dot{R} e^{-\omega}, & &
\Gamma_{22}^1=-R R^{\prime} e^{-\kappa}, \nonumber \\
& \Gamma_{22}^4=\epsilon R R^* e^{-\mu}, & &
\Gamma_{23}^3=\cot \theta, & &
\Gamma_{24}^2=\Gamma_{34}^3=\frac{R^*}{R}, & &
\Gamma_{33}^0=R \dot{R} e^{-\omega} \sin ^2 \theta, \nonumber \\
& \Gamma_{33}^1=-R \dot{R} e^{-\kappa} \sin ^2 \theta, &  &
\Gamma_{33}^2=-\sin \theta \cos \theta, &  &
\Gamma_{04}^4=\frac{\dot{\mu}}{2} \,, & &
\Gamma_{33}^4=\epsilon R R^* e^{-\mu} \sin ^2 \theta \,, \nonumber \\
& \Gamma_{44}^0=-\frac{\epsilon}{2} \dot{\mu} e^{\mu-\omega}, & & 
\Gamma_{44}^1=\frac{\epsilon}{2} \mu^{\prime} e^{\mu-\kappa}, & &
\Gamma_{44}^4=\frac{\mu^*}{2} \,.
\end{align}

In these expressions, $\dot{x}$ represents the derivative with respect to the normal time, $x^{\prime}$ represents the derivative with respect to the radius vector, i.e. $r$, and finally, $x^*$ represents the derivatives with respect to the fifth-dimensional coordinate $\zeta$. As we know the Christoffel's symbols, the 5D Ricci tensor for this metric can be calculated that are given by

\begin{eqnarray}
\widetilde{R}_{00} &=& -\frac{\ddot{\kappa}}{2}-\frac{\ddot{\mu}}{2}-2 \frac{\ddot{R}}{R}+\frac{\dot{\omega} \dot{\kappa}}{4}+\frac{\dot{\omega} \dot{\mu}}{4}+\frac{\dot{\omega} \dot{R}}{R}-\frac{\dot{\kappa}^2}{4}-\frac{\dot{\mu}^2}{4} 
+e^{\omega-\kappa}\left(\frac{\omega^{\prime \prime}}{2}+\frac{\omega^{\prime 2}}{4}-\frac{\omega^{\prime} \kappa^{\prime}}{4}+\frac{\omega^{\prime} \mu^{\prime}}{4}+\frac{\omega^{\prime} R^{\prime}}{R}\right) \nonumber \\
&\,& +\epsilon e^{\omega-\mu}\left(-\frac{\omega^{* *}}{2}-\frac{\omega^{* 2}}{4}+\frac{\omega^* \mu^*}{4}-\frac{\omega^* \kappa^*}{4}-\frac{\omega^* R^*}{R}\right) \,, \\
\widetilde{R}_{01} &=& -\frac{\dot{\mu}^{\prime}}{2}-\frac{\dot{\mu} \mu^{\prime}}{4}+\frac{\omega^{\prime} \dot{\mu}}{4}+\frac{\dot{\kappa} \mu^{\prime}}{4}+\frac{\dot{\kappa} R^{\prime}}{R}+\frac{\omega^{\prime} \dot{R}}{R}-\frac{2 \dot{R}^{\prime}}{R} \,, \\
\widetilde{R}_{04} &=& -\frac{\dot{\kappa}^*}{2}-\frac{\dot{\kappa} \kappa^*}{4}+\frac{\dot{\kappa} \omega^*}{4}+\frac{\kappa^* \dot{\mu}}{4}+\frac{\dot{\mu} R^*}{R}+\frac{\omega^* \dot{R}}{R}-\frac{2 \dot{R}^*}{R} \,, %\\
\end{eqnarray}
\begin{eqnarray}
\widetilde{R}_{11} &=& -\frac{\omega^{\prime \prime}}{2}-\frac{\mu^{\prime \prime}}{2}-\frac{\omega^{\prime 2}}{4}-\frac{\mu^{\prime 2}}{4}+\frac{\kappa^{\prime} \omega^{\prime}}{4}+\frac{\kappa^{\prime} \mu^{\prime}}{4}+\frac{\kappa^{\prime} R^{\prime}}{R}-\frac{2 R^{\prime \prime}}{R}  +e^{\kappa-\omega}\left(\frac{\ddot{\kappa}}{2}+\frac{\dot{\kappa}^2}{4}-\frac{\dot{\kappa} \dot{\omega}}{4}+\frac{\dot{\kappa} \dot{\mu}}{4}+\frac{\dot{\kappa} \dot{R}}{R}\right) \nonumber \\
&\,& +\epsilon e^{\kappa-\mu}\left(\frac{\kappa^{* *}}{2}+\frac{\kappa^{* 2}}{4}+\frac{\kappa^* \omega^*}{4}-\frac{\kappa^* \mu^*}{4}+\frac{\kappa^* R^*}{R}\right) \,, \\
%\end{eqnarray}
%\begin{eqnarray}
\widetilde{R}_{14} &=& -\frac{\omega^{\prime *}}{2}-\frac{\omega^{\prime} \omega^*}{4}+\frac{\kappa^* \omega^{\prime}}{4}+\frac{\mu^{\prime} \omega^*}{4}+\frac{\kappa^* R^{\prime}}{R}+\frac{\mu^{\prime} R^*}{R}-\frac{2 R^{\prime *}}{R} \,, \\
\widetilde{R}_{22} &=& 1+R^2 e^{-\omega}\left(\frac{\dot{R}^2}{R^2}+\frac{\ddot{R}}{R}-\frac{\dot{R}}{2 R}(\dot{\omega}-\dot{\kappa}-\dot{\mu})\right) 
-R^2 e^{-\kappa}\left(\frac{R^{\prime 2}}{R^2}+\frac{R^{\prime \prime}}{R}+\frac{R^{\prime}}{2 R}\left(\omega^{\prime}-\kappa^{\prime}+\mu^{\prime}\right)\right) \nonumber \\
&\,& \epsilon R^2 e^{-\mu}\left(\frac{R^{* 2}}{R^2}+\frac{R^{* *}}{R}+\frac{R^*}{2 R}\left(\omega^*+\kappa^*-\mu^*\right)\right) \,, \\
\widetilde{R}_{33}&=&\widetilde{R}_{22} \sin ^2 \theta \,. \\
\widetilde{R}_{44} &=& \left(-\frac{\omega^{**}}{2}-\frac{\omega^{*2}}{4}-\frac{\kappa^{**}}{2}-\frac{\kappa^{*2}}{4}+\frac{\mu^{*}\omega^{*}}{4}+\frac{\mu^{*}\kappa^{*}}{4}+\frac{\mu^{*}R^{*}}{R}
-\frac{2R^{**}}{R}\right)- \epsilon e^{\mu-\omega}\left(\frac{\ddot{\mu}}{2}+\frac{\dot{\mu}^{2}}{4}\right. \nonumber \\ 
& & \left.-\frac{\dot{\mu}\dot{\omega}}{4}+\frac{\dot{\mu}\dot{\kappa}}{4}+\frac{\dot{\mu}\dot{R}}{R}\right)
 + \epsilon e^{\mu-\kappa}\left(\frac{\mu''}{2}+\frac{\mu'^{2}}{4}+\frac{\mu'\omega'}{4}-\frac{\mu'\kappa'}{4}+\frac{\mu'R'}{R}\right)\,.
\end{eqnarray}

\noindent Using the components of the 5D Ricci tensor the Ricci scalar can be calculated as

\begin{eqnarray}
\widetilde{R} &=& -\frac{2}{R^2}-e^{-\omega}\left(\ddot{\kappa}+\frac{\dot{\kappa}^2}{2}+\ddot{\mu}+\frac{\dot{\mu}^2}{2}-\frac{\dot{\omega} \dot{\kappa}}{2}-\frac{\dot{\omega} \dot{\mu}}{2}-\frac{2 \dot{R}}{R}(\dot{\omega}-\dot{\kappa}-\dot{\mu})+\frac{\dot{\mu} \dot{\kappa}}{2}+\frac{2 \dot{R}^2}{R^2}+\frac{4 \ddot{R}}{R}\right) \nonumber \\
&\,& +e^{-\kappa}\left(\omega^{\prime \prime}+\frac{\omega^{\prime 2}}{2}+\mu^{\prime \prime}+\frac{2 R^{\prime}}{R}\left(\omega^{\prime}-\kappa^{\prime}+\mu^{\prime}\right)+\frac{\mu^{\prime 2}}{2}-\frac{\omega^{\prime} \kappa^{\prime}}{2}+\frac{\omega^{\prime} \mu^{\prime}}{2}-\frac{\mu^{\prime} \kappa^{\prime}}{2}+\frac{2 R^{\prime 2}}{R^2}+\frac{4 R^{\prime \prime}}{R}\right) \nonumber \\
&\,& -\epsilon e^{-\mu}\left(\omega^{* *}+\frac{\omega^{* 2}}{2}+\kappa^{* *}+\frac{\kappa^{* 2}}{2}+\frac{\kappa^* \omega^*}{2}-\frac{\kappa^* \mu^*}{2}+\frac{2 R^*}{R}\left(\omega^*+\kappa^*-\mu^*\right)\right. \nonumber \\
&\,& \left.-\frac{\mu^* \omega^*}{2}+\frac{2 R^{* 2}}{R^2}+\frac{4 R^{* *}}{R}\right) .
\end{eqnarray}

The above expressions give the Ricci tensor and Ricci scalar in a five-dimensional Universe. To get the 4-dimensional components of the Ricci tensor and Ricci scalar we can drop all the derivatives with respect to the $\zeta$ to $0$. We can also set $\epsilon = 0$. This gives

\begin{eqnarray}
R_{00} &=& -\frac{\ddot{\kappa}}{2}-2 \frac{\ddot{R}}{R}+\frac{\dot{\omega} \dot{\kappa}}{4}+\frac{\dot{\omega} \dot{R}}{R}-\frac{\dot{\kappa}^2}{4}+e^{\omega-\kappa}\left(\frac{\omega^{\prime \prime}}{2}+\frac{\omega^{\prime 2}}{4}-\frac{\omega^{\prime} \kappa^{\prime}}{4}+\frac{\omega^{\prime} R^{\prime}}{R}\right) \,,\\
R_{01} &=& \frac{\dot{\kappa} R^{\prime}}{R}+\frac{\omega^{\prime} \dot{R}}{R}-\frac{2 \dot{R}^{\prime}}{R} \,,\\
%\end{eqnarray}
%\begin{eqnarray}
R_{11} &=& -\frac{\omega^{\prime \prime}}{2}-\frac{\omega^{\prime 2}}{4}+\frac{\kappa^{\prime} \omega^{\prime}}{4}+\frac{\kappa^{\prime} R^{\prime}}{R}-\frac{2 R^{\prime \prime}}{R}+e^{\kappa-\omega}\left(\frac{\ddot{\kappa}}{2}+\frac{\dot{\kappa}^2}{4}-\frac{\dot{\kappa} \dot{\omega}}{4}+\frac{\dot{\kappa} \dot{R}}{R}\right) \,,\\
R_{22} &=& 1+R^2 e^{-\omega}\left(\frac{\dot{R}^2}{R^2}+\frac{\ddot{R}}{R}-\frac{\dot{R}}{2 R}(\dot{\omega}-\dot{\kappa})\right)-R^2 e^{-\kappa}\left(\frac{R^{\prime 2}}{R^2}+\frac{R^{\prime \prime}}{R}+\frac{R^{\prime}}{2 R}\left(\omega^{\prime}-\kappa^{\prime}\right)\right) \,,\\
R_{33} &=& R_{22} \sin ^2 \theta \,.
\end{eqnarray}

\noindent From these four-dimensional Ricci tensor the Ricci scalar can be calculated as

\begin{eqnarray}
R &=& -\frac{2}{R^2}-e^{-\omega}\left(\ddot{\kappa}+\frac{\dot{\kappa}^2}{2}-\frac{\dot{\omega} \dot{\kappa}}{2}-\frac{2 \dot{R}}{R}(\dot{\omega}-\dot{\kappa})+\frac{2 \dot{R}^2}{R^2}+\frac{4 \ddot{R}}{R}\right) \nonumber \\
&\,& +e^{-\kappa}\left(\omega^{\prime \prime}+\frac{\omega^{\prime 2}}{2}-\frac{\omega^{\prime} \kappa^{\prime}}{2}+\frac{2 R^{\prime}}{R}\left(\omega^{\prime}-\kappa^{\prime}\right)+\frac{2 R^{\prime 2}}{R^2}+\frac{4 R^{\prime \prime}}{R}\right) .
\end{eqnarray}
%\end{widetext}
\noindent We can use these above expressions and a few algebraic manipulations to obtain the expressions for $G^\mu_\nu$ used in Eq.~\ref{equation:G5d4d}.

\subsection{Calculating the stress-energy tensor}
According to the field equations $\widetilde{G}_{A B}=\kappa \widetilde{T}_{A B}$. %Earlier we have assumed that the  fifth dimensional component of the stress energy tensor can be very small. Therefore, here the 5th component of the stress-energy tensor i.e. $\widetilde{\tau}_{44}$ and $\widetilde{\tau}_{4\mu}$ are assumed to be $0$ for the calculations. 
Earlier in Sec.~\ref{Sec:BD} we have shown that $\widetilde{R}_{44} \sim 0$ as long as $g_{44} \sim 1$. As $\widetilde{g}_{AB}$ is a diagonal matrix, raising or lowering some components in a tensor is same as multiplying with the respective diagonal component of the metric, i.e. $\widetilde{R}^4_4 = \widetilde{R}_{44}\widetilde{g}^{44}$.  

From the field equation, we can write $\widetilde{G}^{\mu}_\nu = \kappa \widetilde{T}^{\mu}_\nu$, which gives $G^{\mu}_\nu = \kappa T^{\mu}_\nu + Q^{\mu}_\nu$, where $ Q^{\mu}_\nu = G^{\mu}_\nu - \widetilde{G}^{\mu}_\nu$, are the additional geometric terms while projecting 5 dimensional Einstein's tensor to the 4-dimensional format. (Note that as we are using a diagonal metric, in the 5 dimensional stress-energy tensor formula we can replace the vector field with $0$ which will lead us to the above equation.) As we discussed earlier, if we assume that these geometric quantities, $Q^\mu_\nu$, are getting generated from some geometric fluid,  
we can define the density and the pressure of these geometric fluids as $\rho_g$ and $p_g$ (check Eq.~\ref{eq:pg}). These can be calculated from $Q^\mu_\nu$ as

%\begin{widetext}
\begin{equation}
\rho_g=Q_{0}^0+Q_{1}^1-Q_{2}^2,
\qquad\qquad \text{and} \qquad\qquad 
p_g=-Q_{2}^2\,,
\end{equation}
%\end{widetext}
\noindent While the above expression provides a simplified equation for $\rho_g$, the components of $p_g$ are not simplified. However, we know $\widetilde{R}_{44} \sim 0$. If we add the expression for $\widetilde{R}^4_4$ with the $Q^2_2$, it will not change $p_g$. However, this can simplify the equations, giving the expressions for the density and pressure as
%\begin{widetext}
\begin{eqnarray}
\rho_g &=& \frac{3}{2}\left(\frac{e^{-\kappa} \mu^{\prime} R^{\prime}}{R}-\frac{e^{-\omega} \dot{\mu} \dot{R}}{R}\right)+\frac{3}{2} \epsilon e^{-\mu}\left(\frac{R^* \mu^*}{R}-\frac{2 R^{* *}}{R}\right)-\epsilon e^{-\mu} \frac{R^{* 2}}{R^2}+\epsilon e^{-\mu}\left(\frac{\omega^* \kappa^*}{4}\right) \nonumber \\
&\,& -\epsilon e^{-\mu} \frac{R^*}{2 R}\left(\kappa^*+\omega^*\right), \nonumber \\
p_g &=& \frac{1}{2}\left(\frac{e^{-\kappa} \mu^{\prime} R^{\prime}}{R}-\frac{e^{-\omega} \dot{\mu} \dot{R}}{R}\right)+\frac{1}{2} \epsilon e^{-\mu}\left(\frac{R^* \mu^*}{R}-\frac{2 R^{* *}}{R}\right)+\epsilon e^{-\mu}\left(\frac{\omega^* \kappa^*}{4}\right)+\epsilon e^{-\mu} \frac{R^*}{2 R}\left(\kappa^*+\omega^*\right) \nonumber
\end{eqnarray}

\noindent  These equations show that pressure and density consist of 4 clearly defined components, a radiation-like component, a matter-like component, a stiff matter-like component, and a dark energy-like component. 
\end{widetext}

\section*{Data Availability}
The data for the galactic velocity profile shown in Fig.~\ref{fig:galacticvelocity}  is openly available in SPARC database at \url{http://astroweb.cwru.edu/SPARC/}. The dataset used for cluster mass profile calculation shown in table~\ref{tab:clusterProperties} and Fig.~\ref{clusterdata} can be found in reference~\cite{reiprich2003cosmological}.

%\section*{Funding Declaration}
%The author received no financial support for the research, authorship, and/or publication of this article.

%\subsubsection{Centrifugal force}

%\setcounter{enumiv}{ 1 }
%\bibliographystyle{mnras}
%\bibliographystyle{natbib}
\bibliographystyle{JHEP}
\bibliography{reference}

\providecommand{\href}[2]{#2}\begingroup\raggedright\begin{thebibliography}{100}

\bibitem{MARTIN_1998}
S.P.~Martin, \emph{{A Supersymmetry primer}}, \href{https://doi.org/10.1142/9789812839657_0001}{\emph{Adv. Ser. Direct. High Energy Phys.} {\bfseries 18} (1998) 1} [\href{https://arxiv.org/abs/hep-ph/9709356}{{\ttfamily hep-ph/9709356}}].

\bibitem{PhysRevLett.40.279}
F.~Wilczek, \emph{Problem of strong $p$ and $t$ invariance in the presence of instantons}, \href{https://doi.org/10.1103/PhysRevLett.40.279}{\emph{Phys. Rev. Lett.} {\bfseries 40} (1978) 279}.

\bibitem{PhysRevD.53.2236}
J.-w.~Lee and I.-g.~Koh, \emph{Galactic halos as boson stars}, \href{https://doi.org/10.1103/PhysRevD.53.2236}{\emph{Phys. Rev. D} {\bfseries 53} (1996) 2236}.

\bibitem{Di_Valentino_2021}
E.D.~Valentino, O.~Mena, S.~Pan, L.~Visinelli, W.~Yang, A.~Melchiorri et~al., \emph{In the realm of the hubble tension{\textemdash}a review of solutions}, \href{https://doi.org/10.1088/1361-6382/ac086d}{\emph{Classical and Quantum Gravity} {\bfseries 38} (2021) 153001}.

\bibitem{Joudaki_2016}
S.~Joudaki et~al., \emph{{CFHTLenS} revisited: assessing concordance with planck including astrophysical systematics}, \href{https://doi.org/10.1093/mnras/stw2665}{\emph{Monthly Notices of the Royal Astronomical Society} {\bfseries 465} (2016) 2033}.

\bibitem{Hildebrandt_2016}
H.~Hildebrandt et~al., \emph{{KiDS}-450: cosmological parameter constraints from tomographic weak gravitational lensing}, \href{https://doi.org/10.1093/mnras/stw2805}{\emph{Monthly Notices of the Royal Astronomical Society} {\bfseries 465} (2016) 1454}.

\bibitem{Riess_2019}
A.G.~Riess, S.~Casertano, W.~Yuan, L.M.~Macri and D.~Scolnic, \emph{Large magellanic cloud cepheid standards provide a 1{\%} foundation for the determination of the hubble constant and stronger evidence for physics beyond $\ensuremath{\Lambda}${CDM}}, \href{https://doi.org/10.3847/1538-4357/ab1422}{\emph{The Astrophysical Journal} {\bfseries 876} (2019) 85}.

\bibitem{Ade2016pap16}
P.A.R.~Ade et~al., \emph{Planck 2015 results}, \href{https://doi.org/10.1051/0004-6361/201525833}{\emph{Astronomy \& Astrophysics} {\bfseries 594} (2016) A24}.

\bibitem{Addison_2016}
G.E.~Addison, Y.~Huang, D.J.~Watts, C.L.~Bennett, M.~Halpern, G.~Hinshaw et~al., \emph{Quantifying discordance in the 2015 planck cmb spectrum}, \href{https://doi.org/10.3847/0004-637x/818/2/132}{\emph{The Astrophysical Journal} {\bfseries 818} (2016) 132}.

\bibitem{Kitching_2016}
T.D.~Kitching, L.~Verde, A.F.~Heavens and R.~Jimenez, \emph{Discrepancies between {CFHTLenS} cosmic shear and planck: new physics or systematic effects?}, \href{https://doi.org/10.1093/mnras/stw707}{\emph{Monthly Notices of the Royal Astronomical Society} {\bfseries 459} (2016) 971}.

\bibitem{Couchot_2017}
F.~Couchot, S.~Henrot-Versill{\'{e} }, O.~Perdereau, S.~Plaszczynski, B.R.~d'Orfeuil, M.~Spinelli et~al., \emph{Relieving tensions related to the lensing of the cosmic microwave background temperature power spectra}, \href{https://doi.org/10.1051/0004-6361/201527740}{\emph{Astronomy \& Astrophysics} {\bfseries 597} (2017) A126}.

\bibitem{Milgrim1983}
M.~Milgrom, \emph{{A Modification of the Newtonian dynamics as a possible alternative to the hidden mass hypothesis}}, \href{https://doi.org/10.1086/161130}{\emph{Astrophys.J.} {\bfseries 270} (1983) 365}.

\bibitem{Milgrim1983a}
M.~Milgrom, \emph{{A Modification of the Newtonian dynamics: Implications for galaxies}}, \href{https://doi.org/10.1086/161131}{\emph{Astrophys.J.} {\bfseries 270} (1983) 371}.

\bibitem{Milgrim1983b}
M.~Milgrom, \emph{{A modification of the Newtonian dynamics: implications for galaxy systems}}, \href{https://doi.org/10.1086/161132}{\emph{Astrophys.J.} {\bfseries 270} (1983) 384}.

\bibitem{Milgrom2011}
M.~Milgrom, \emph{{MD or DM? Modified dynamics at low accelerations vs dark matter}}, {\emph{PoS} {\bfseries HRMS2010} (2010) 033} [\href{https://arxiv.org/abs/1101.5122}{{\ttfamily 1101.5122}}].

\bibitem{Bekenstein1984}
J.~Bekenstein and M.~Milgrom, \emph{{Does the missing mass problem signal the breakdown of Newtonian gravity?}}, \href{https://doi.org/10.1086/162570}{\emph{Astrophys.J.} {\bfseries 286} (1984) 7}.

\bibitem{Bekenstein2009}
J.D.~Bekenstein, \emph{{Relativistic MOND as an alternative to the dark matter paradigm}}, \href{https://doi.org/10.1016/j.nuclphysa.2009.05.122}{\emph{Nucl.Phys.} {\bfseries A827} (2009) 555C} [\href{https://arxiv.org/abs/0901.1524}{{\ttfamily 0901.1524}}].

\bibitem{Milgrom1986}
M.~Milgrom, \emph{{Solutions for the modified Newtonian dynamics field equation}}, \href{https://doi.org/10.1086/164021}{\emph{Astrophys.J.} {\bfseries 302} (1986) 617}.

\bibitem{Moffat2005}
J.~Moffat, \emph{{Scalar-tensor-vector gravity theory}}, \href{https://doi.org/10.1088/1475-7516/2006/03/004}{\emph{JCAP} {\bfseries 0603} (2006) 004} [\href{https://arxiv.org/abs/gr-qc/0506021}{{\ttfamily gr-qc/0506021}}].

\bibitem{Brownstein2005}
J.~Brownstein and J.~Moffat, \emph{{Galaxy rotation curves without non-baryonic dark matter}}, \href{https://doi.org/10.1086/498208}{\emph{Astrophys.J.} {\bfseries 636} (2006) 721} [\href{https://arxiv.org/abs/astro-ph/0506370}{{\ttfamily astro-ph/0506370}}].

\bibitem{Brownstein2005a}
J.~Brownstein and J.~Moffat, \emph{{Galaxy cluster masses without non-baryonic dark matter}}, \href{https://doi.org/10.1111/j.1365-2966.2006.09996.x}{\emph{Mon.Not.Roy.Astron.Soc.} {\bfseries 367} (2006) 527} [\href{https://arxiv.org/abs/astro-ph/0507222}{{\ttfamily astro-ph/0507222}}].

\bibitem{Moffat2005a}
J.~Moffat and V.~Toth, \emph{{Modified Gravity: Cosmology without dark matter or Einstein's cosmological constant}},  \href{https://arxiv.org/abs/0710.0364}{{\ttfamily 0710.0364}}.

\bibitem{Bekenstein2005}
J.D.~Bekenstein, \emph{{Relativistic gravitation theory for the MOND paradigm}}, \href{https://doi.org/10.1103/PhysRevD.70.083509, 10.1103/PhysRevD.70.083509 10.1103/PhysRevD.71.069901, 10.1103/PhysRevD.71.069901}{\emph{Phys.Rev.} {\bfseries D70} (2004) 083509} [\href{https://arxiv.org/abs/astro-ph/0403694}{{\ttfamily astro-ph/0403694}}].

\bibitem{Dam1970}
H.~van Dam and M.~Veltman, \emph{{Massive and massless Yang-Mills and gravitational fields}}, \href{https://doi.org/10.1016/0550-3213(70)90416-5}{\emph{Nucl.Phys.} {\bfseries B22} (1970) 397}.

\bibitem{Zakharov1970}
V.~Zakharov, \emph{{Linearized gravitation theory and the graviton mass}}, {\emph{JETP Lett.} {\bfseries 12} (1970) 312}.

\bibitem{Babichev2010}
E.~Babichev, C.~Deffayet and R.~Ziour, \emph{{The Recovery of General Relativity in massive gravity via the Vainshtein mechanism}}, \href{https://doi.org/10.1103/PhysRevD.82.104008}{\emph{Phys.Rev.} {\bfseries D82} (2010) 104008} [\href{https://arxiv.org/abs/1007.4506}{{\ttfamily 1007.4506}}].

\bibitem{Babichev2013}
E.~{Babichev} and M.~{Crisostomi}, \emph{{Restoring general relativity in massive bigravity theory}}, \href{https://doi.org/10.1103/PhysRevD.88.084002}{\emph{Phys.Rev.} {\bfseries D88} (2013) 084002} [\href{https://arxiv.org/abs/1307.3640}{{\ttfamily 1307.3640}}].

\bibitem{Overduin1998}
J.~Overduin and P.~Wesson, \emph{{Kaluza-Klein gravity}}, \href{https://doi.org/10.1016/S0370-1573(96)00046-4}{\emph{Phys.Rept.} {\bfseries 283} (1997) 303} [\href{https://arxiv.org/abs/gr-qc/9805018}{{\ttfamily gr-qc/9805018}}].

\bibitem{Ponce1993}
J.~{Ponce de Leon} and P.S.~{Wesson}, \emph{{Exact solutions and the effective equation of state in Kaluza-Klein theory}}, \href{https://doi.org/10.1063/1.530028}{\emph{Journal of Mathematical Physics} {\bfseries 34} (1993) 4080}.

\bibitem{Wesson1992}
P.S.~{Wesson} and J.P.~{de Leon}, \emph{{Kaluza-Klein equations, Einstein's equations, and an effective energy-momentum tensor.}}, \href{https://doi.org/10.1063/1.529834}{\emph{Journal of Mathematical Physics} {\bfseries 33} (1992) 3883}.

\bibitem{de2010schwarzschild}
J.P.~de~Leon, \emph{Schwarzschild-like exteriors for stars in kaluza-klein gravity}, {\emph{arXiv preprint arXiv:1003.3151} (2010) }.

\bibitem{moraes2016cosmic}
P.~Moraes, \emph{Cosmic acceleration from varying masses in five dimensions}, {\emph{International Journal of Modern Physics D} {\bfseries 25} (2016) 1650009}.

\bibitem{Einstein}
A.~Einstein, \emph{Meaning of relativity}, Princeton Univ. Press, Princeton, NJ (1923), \href{https://doi.org/ISBN: 9780691164083}{ISBN: 9780691164083}.

\bibitem{sciama1953origin}
D.W.~Sciama, \emph{On the origin of inertia}, {\emph{Monthly Notices of the Royal Astronomical Society} {\bfseries 113} (1953) 34}.

\bibitem{Brans1961}
C.~{Brans} and R.H.~{Dicke}, \emph{{Mach's Principle and a Relativistic Theory of Gravitation}}, \href{https://doi.org/10.1103/PhysRev.124.925}{\emph{Physical Review} {\bfseries 124} (1961) 925}.

\bibitem{fujii2003scalar}
Y.~Fujii and K.~Maeda, \emph{{The scalar-tensor theory of gravitation}}, Cambridge Monographs on Mathematical Physics, Cambridge University Press, Cambridge (7, 2007), \href{https://doi.org/10.1017/CBO9780511535093}{10.1017/CBO9780511535093}.

\bibitem{faraoni2004cosmology}
V.~Faraoni, \emph{Cosmology in Scalar-Tensor Gravity}, vol.~139 of \emph{Fundamental Theories of Physics}, Springer, Dordrecht, 1~ed. (2004), \href{https://doi.org/10.1007/978-1-4020-1989-0}{10.1007/978-1-4020-1989-0}.

\bibitem{Hoyle1964}
F.~{Hoyle} and J.V.~{Narlikar}, \emph{{A New Theory of Gravitation}}, \href{https://doi.org/10.1098/rspa.1964.0227}{\emph{Royal Society of London Proceedings Series A} {\bfseries 282} (1964) 191}.

\bibitem{hoyle1964c}
F.~Hoyle and J.~Narlikar, \emph{The c-field as a direct particle field}, \href{https://doi.org/10.1098/rspa.1964.0225}{\emph{Proceedings of the Royal Society of London. Series A. Mathematical and Physical Sciences} {\bfseries 282} (1964) 178}.

\bibitem{hoyle1964time}
F.~Hoyle and J.V.~Narlikar, \emph{Time symmetric electrodynamics and the arrow of time in cosmology}, \href{https://doi.org/10.1098/rspa.1964.0002}{\emph{Proceedings of the Royal Society of London. Series A. Mathematical and Physical Sciences} {\bfseries 277} (1964) 1}.

\bibitem{hoyle1964avoidance}
F.~Hoyle and J.V.~Narlikar, \emph{On the avoidance of singularities in c-field cosmology}, \href{https://doi.org/10.1098/rspa.1964.0076}{\emph{Proceedings of the Royal Society of London. Series A. Mathematical and Physical Sciences} {\bfseries 278} (1964) 465}.

\bibitem{hoyle1964gravitational}
F.~Hoyle and J.~Narlikar, \emph{On the gravitational influence of direct particle fields}, \href{https://doi.org/10.1098/rspa.1964.0226}{\emph{Proceedings of the Royal Society of London. Series A. Mathematical and Physical Sciences} {\bfseries 282} (1964) 184}.

\bibitem{PhysRevD.104.044001}
R.~Roy, A.B.~Abdikamalov, D.~Ayzenberg, C.~Bambi, S.~Riaz and A.~Tripathi, \emph{Testing the weak-equivalence principle near black holes}, \href{https://doi.org/10.1103/PhysRevD.104.044001}{\emph{Phys. Rev. D} {\bfseries 104} (2021) 044001}.

\bibitem{Wagner:2012ui}
T.A.~Wagner, S.~Schlamminger, J.H.~Gundlach and E.G.~Adelberger, \emph{{Torsion-balance tests of the weak equivalence principle}}, \href{https://doi.org/10.1088/0264-9381/29/18/184002}{\emph{Class. Quant. Grav.} {\bfseries 29} (2012) 184002} [\href{https://arxiv.org/abs/1207.2442}{{\ttfamily 1207.2442}}].

\bibitem{rosi2017quantum}
G.~Rosi, G.~D’Amico, L.~Cacciapuoti, F.~Sorrentino, M.~Prevedelli, M.~Zych et~al., \emph{Quantum test of the equivalence principle for atoms in coherent superposition of internal energy states}, {\emph{Nature communications} {\bfseries 8} (2017) 15529}.

\bibitem{10.1093/mnrasl/slaa143}
S.-C.~Yang, W.-B.~Han and G.~Wang, \emph{Tests of weak equivalence principle with the gravitational wave signals in the ligo–virgo catalogue gwtc-1}, \href{https://doi.org/10.1093/mnrasl/slaa143}{\emph{Monthly Notices of the Royal Astronomical Society: Letters} {\bfseries 499} (2020) L53} [\href{https://arxiv.org/abs/https://academic.oup.com/mnrasl/article-pdf/499/1/L53/54638219/mnrasl\_499\_1\_l53.pdf}{{\ttfamily https://academic.oup.com/mnrasl/article-pdf/499/1/L53/54638219/mnrasl\_499\_1\_l53.pdf}}].

\bibitem{PhysRevLett.129.121102}
{\scshape MICROSCOPE Collaboration} collaboration, \emph{$microscope$ mission: Final results of the test of the equivalence principle}, \href{https://doi.org/10.1103/PhysRevLett.129.121102}{\emph{Phys. Rev. Lett.} {\bfseries 129} (2022) 121102}.

\bibitem{huber2000precision}
F.~Huber, R.~Lewis, E.~Messerschmid and G.~Smith, \emph{Precision tests of einstein's weak equivalence principle for antimatter}, {\emph{Advances in Space Research} {\bfseries 25} (2000) 1245}.

\bibitem{Yang:2019tzi}
S.-C.~Yang, W.-B.~Han and G.~Wang, \emph{{Tests of weak equivalence principle with the gravitational wave signals in the LIGO\textendash{}Virgo catalogue GWTC-1}}, \href{https://doi.org/10.1093/mnrasl/slaa143}{\emph{Mon. Not. Roy. Astron. Soc.} {\bfseries 499} (2020) L53} [\href{https://arxiv.org/abs/1912.10758}{{\ttfamily 1912.10758}}].

\bibitem{Voisin:2020lqi}
G.~Voisin, I.~Cognard, P.C.C.~Freire, N.~Wex, L.~Guillemot, G.~Desvignes et~al., \emph{{An improved test of the strong equivalence principle with the pulsar in a triple star system}}, \href{https://doi.org/10.1051/0004-6361/202038104}{\emph{Astron. Astrophys.} {\bfseries 638} (2020) A24} [\href{https://arxiv.org/abs/2005.01388}{{\ttfamily 2005.01388}}].

\bibitem{Jammer}
M.~Jammer, \emph{Concepts of Mass in Contemporary Physics and Philosophy}, Princeton University Press, Princeton, NJ (2000).

\bibitem{Hoyle1980}
J.V.N.~F.~Hoyle, \emph{The physics-astronomy frontier}, W.H.Freeman \& Co. Ltd., San Francisco (1981).

\bibitem{Weinberg:1972kfs}
S.~Weinberg, \emph{{Gravitation and Cosmology}: {Principles and Applications of the General Theory of Relativity}}, John Wiley and Sons, New York (1972).

\bibitem{Einstein:1916vd}
A.~Einstein, \emph{{The foundation of the general theory of relativity.}}, \href{https://doi.org/10.1002/andp.19163540702}{\emph{Annalen Phys.} {\bfseries 49} (1916) 769}.

\bibitem{bondi1952cosmology}
H.~Bondi, \emph{Cosmology}, Cambridge monographs on physics, University Press, Cambridge (1952).

\bibitem{sciama1964physical}
D.W.~Sciama, \emph{The physical structure of general relativity}, \href{https://doi.org/10.1103/RevModPhys.36.463}{\emph{Reviews of Modern Physics} {\bfseries 36} (1964) 463}.

\bibitem{berman2008machian}
M.S.~Berman, \emph{On the machian origin of inertia}, \href{https://doi.org/10.1007/s10509-008-9915-3}{\emph{Astrophysics and Space Science} {\bfseries 318} (2008) 269}.

\bibitem{Licata_2016}
I.~Licata, C.~Corda and E.~Benedetto, \emph{A machian request for the equivalence principle in extended gravity and nongeodesic motion}, \href{https://doi.org/10.1134/s0202289316010102}{\emph{Gravitation and Cosmology} {\bfseries 22} (2016) 48}.

\bibitem{mashhoon2003gravitoelectromagnetism}
B.~Mashhoon, \emph{Gravitoelectromagnetism: a brief review}, {\emph{arXiv preprint gr-qc/0311030} (2003) }.

\bibitem{mashhoon2001gravitoelectromagnetism1}
B.~Mashhoon, \emph{{Gravitoelectromagnetism}},  in \emph{{Spanish Relativity Meeting on Reference Frames and Gravitomagnetism (EREs2000)}}, 9, 2000, \href{https://doi.org/10.1142/9789812810021_0009}{DOI} [\href{https://arxiv.org/abs/gr-qc/0011014}{{\ttfamily gr-qc/0011014}}].

\bibitem{mashhoon2001gravitomagnetism}
B.~Mashhoon, F.~Gronwald and H.I.~Lichtenegger, \emph{Gravitomagnetism and the clock effect},  in \emph{Gyros, Clocks, Interferometers...: Testing Relativistic Graviy in Space}, pp.~83--108, Springer, 2001.

\bibitem{Verbiest2008}
J.P.W.~{Verbiest}, M.~{Bailes}, W.~{van Straten}, G.B.~{Hobbs}, R.T.~{Edwards}, R.N.~{Manchester} et~al., \emph{{Precision Timing of PSR J0437-4715: An Accurate Pulsar Distance, a High Pulsar Mass, and a Limit on the Variation of Newton's Gravitational Constant}}, \href{https://doi.org/10.1086/529576}{\emph{ApJ} {\bfseries 679} (2008) 675} [\href{https://arxiv.org/abs/0801.2589}{{\ttfamily 0801.2589}}].

\bibitem{taylor1993particle}
J.~Taylor, \emph{Particle astrophysics, ivth rentr{\'e}es de blois, ed}, {\emph{G. Fontaine, \& J. Tran Thanh Van (Gif-sur-Yvette, France: {\'E}ditions Fronti{\`e}res)} {\bfseries 367} (1993) }.

\bibitem{Nordtvedt1990}
K.~Nordtvedt, \emph{G\ifmmode \dot{}\else \.{}\fi{}/g and a cosmological acceleration of gravitationally compact bodies}, \href{https://doi.org/10.1103/PhysRevLett.65.953}{\emph{Phys. Rev. Lett.} {\bfseries 65} (1990) 953}.

\bibitem{kaspi1994high}
V.~Kaspi, J.~Taylor and M.~Ryba, \emph{High-precision timing of millisecond pulsars. 3: Long-term monitoring of psrs b1855+ 09 and b1937+ 21}, {\emph{The Astrophysical Journal} {\bfseries 428} (1994) 713}.

\bibitem{Garc_a_Berro_2011}
E.~Garc{\'{\i} }a-Berro, P.~Lor{\'{e}}n-Aguilar, S.~Torres, L.G.~Althaus and J.~Isern, \emph{An upper limit to the secular variation of the gravitational constant from white dwarf stars}, \href{https://doi.org/10.1088/1475-7516/2011/05/021}{\emph{Journal of Cosmology and Astroparticle Physics} {\bfseries 2011} (2011) 021}.

\bibitem{garcia1995rate}
E.~Garc{\'\i}a-Berro, M.~Hernanz, J.~Isern and R.~Mochkovitch, \emph{The rate of change of the gravitational constant and the cooling of white dwarfs}, {\emph{Monthly Notices of the Royal Astronomical Society} {\bfseries 277} (1995) 801}.

\bibitem{corsico2013independent}
A.H.~C{\'o}rsico, L.G.~Althaus, E.~Garc{\'\i}a-Berro and A.D.~Romero, \emph{An independent constraint on the secular rate of variation of the gravitational constant from pulsating white dwarfs}, {\emph{Journal of Cosmology and Astroparticle Physics} {\bfseries 2013} (2013) 032}.

\bibitem{Althaus_2011}
L.G.~Althaus, A.H.~C{\'{o} }rsico, S.~Torres, P.~Lor{\'{e}}n-Aguilar, J.~Isern and E.~Garc{\'{\i}}a-Berro, \emph{The evolution of white dwarfs with a varying gravitational constant}, \href{https://doi.org/10.1051/0004-6361/201015849}{\emph{Astronomy and Astrophysics} {\bfseries 527} (2011) A72}.

\bibitem{PhysRevLett.93.261101}
J.G.~Williams, S.G.~Turyshev and D.H.~Boggs, \emph{Progress in lunar laser ranging tests of relativistic gravity}, \href{https://doi.org/10.1103/PhysRevLett.93.261101}{\emph{Phys. Rev. Lett.} {\bfseries 93} (2004) 261101}.

\bibitem{krasinsky2004secular}
G.A.~Krasinsky and V.A.~Brumberg, \emph{Secular increase of astronomical unit from analysis of the major planet motions, and its interpretation}, {\emph{Celestial mechanics and dynamical astronomy} {\bfseries 90} (2004) 267}.

\bibitem{Psaltis:2005ai}
D.~Psaltis, \emph{{Constraining Brans-Dicke gravity with millisecond pulsars in ultracompact binaries}},  \href{https://arxiv.org/abs/astro-ph/0501234}{{\ttfamily astro-ph/0501234}}.

\bibitem{Avilez:2013dxa}
A.~Avilez and C.~Skordis, \emph{{Cosmological constraints on Brans-Dicke theory}}, \href{https://doi.org/10.1103/PhysRevLett.113.011101}{\emph{Phys. Rev. Lett.} {\bfseries 113} (2014) 011101} [\href{https://arxiv.org/abs/1303.4330}{{\ttfamily 1303.4330}}].

\bibitem{Tan:2023fyl}
J.~Tan and B.~Wang, \emph{{Constraints on Brans-Dicke gravity from neutron star-black hole merger events using higher harmonics}}, \href{https://doi.org/10.1103/PhysRevD.109.084036}{\emph{Phys. Rev. D} {\bfseries 109} (2024) 084036} [\href{https://arxiv.org/abs/2312.07017}{{\ttfamily 2312.07017}}].

\bibitem{Amirhashchi:2019jpf}
H.~Amirhashchi and A.K.~Yadav, \emph{{Constraining an exact Brans\textendash{}Dicke gravity theory with recent observations}}, \href{https://doi.org/10.1016/j.dark.2020.100711}{\emph{Phys. Dark Univ.} {\bfseries 30} (2020) 100711} [\href{https://arxiv.org/abs/1908.04735}{{\ttfamily 1908.04735}}].

\bibitem{Billyard:1998kg}
A.~Billyard, A.~Coley and J.~Ibanez, \emph{{On the asymptotic behavior of cosmological models in scalar tensor theories of gravity}}, \href{https://doi.org/10.1103/PhysRevD.59.023507}{\emph{Phys. Rev. D} {\bfseries 59} (1999) 023507} [\href{https://arxiv.org/abs/gr-qc/9807055}{{\ttfamily gr-qc/9807055}}].

\bibitem{barrow1997behavior}
J.D.~Barrow and P.~Parsons, \emph{Behavior of cosmological models with varying g}, \href{https://doi.org/10.1103/PhysRevD.55.1906}{\emph{Physical Review D} {\bfseries 55} (1997) 1906}.

\bibitem{mimoso1995anisotropic}
J.P.~Mimoso and D.~Wands, \emph{Anisotropic scalar-tensor cosmologies}, \href{https://doi.org/10.1103/PhysRevD.52.5612}{\emph{Physical Review D} {\bfseries 52} (1995) 5612}.

\bibitem{barrow1994perfect}
J.D.~Barrow and J.P.~Mimoso, \emph{Perfect fluid scalar-tensor cosmologies}, {\emph{Physical Review D} {\bfseries 50} (1994) 3746}.

\bibitem{Faraoni:1999yp}
V.~Faraoni, \emph{{Illusions of general relativity in Brans-Dicke gravity}}, \href{https://doi.org/10.1103/PhysRevD.59.084021}{\emph{Phys. Rev. D} {\bfseries 59} (1999) 084021} [\href{https://arxiv.org/abs/gr-qc/9902083}{{\ttfamily gr-qc/9902083}}].

\bibitem{Bondi:1996md}
H.~Bondi and J.~Samuel, \emph{{The Lense-Thirring effect and Mach's principle}}, \href{https://doi.org/10.1016/S0375-9601(97)00117-5}{\emph{Phys. Lett. A} {\bfseries 228} (1997) 121} [\href{https://arxiv.org/abs/gr-qc/9607009}{{\ttfamily gr-qc/9607009}}].

\bibitem{delCastillo:2020wka}
G.F.T.~del Castillo, \emph{{An introduction to the Kaluza-Klein formulation}}, \href{https://doi.org/10.31349/RevMexFisE.17.27}{\emph{Rev. Mex. Fis. E} {\bfseries 17} (2020) 27}.

\bibitem{friedman2013covariant}
Y.~Friedman and T.~Scarr, \emph{Covariant uniform acceleration},  in \emph{Journal of Physics: Conference Series}, vol.~437, p.~012009, IOP Publishing, 2013.

\bibitem{Moffat2009}
J.~Moffat and V.~Toth, \emph{{Fundamental parameter-free solutions in modified gravity}}, \href{https://doi.org/10.1088/0264-9381/26/8/085002}{\emph{Class.Quant.Grav.} {\bfseries 26} (2009) 085002} [\href{https://arxiv.org/abs/0712.1796}{{\ttfamily 0712.1796}}].

\bibitem{sanders1984anti}
R.~Sanders, \emph{Anti-gravity and galaxy rotation curves}, {\emph{Astronomy and Astrophysics (ISSN 0004-6361), vol. 136, no. 2, July 1984, p. L21-L23.} {\bfseries 136} (1984) L21}.

\bibitem{sanders2002modified}
R.H.~Sanders and S.S.~McGaugh, \emph{Modified newtonian dynamics as an alternative to dark matter}, {\emph{Annual Review of Astronomy and Astrophysics} {\bfseries 40} (2002) 263}.

\bibitem{tully1977new}
R.B.~Tully and J.R.~Fisher, \emph{A new method of determining distances to galaxies}, {\emph{Astronomy and Astrophysics, vol. 54, no. 3, Feb. 1977, p. 661-673.} {\bfseries 54} (1977) 661}.

\bibitem{faber1976velocity}
S.~Faber and R.E.~Jackson, \emph{Velocity dispersions and mass-to-light ratios for elliptical galaxies}, \href{https://doi.org/10.1086/154215}{\emph{The Astrophysical Journal} {\bfseries 204} (1976) 668}.

\bibitem{mcgaugh2004mass}
S.S.~McGaugh, \emph{The mass discrepancy-acceleration relation: disk mass and the dark matter distribution}, {\emph{The Astrophysical Journal} {\bfseries 609} (2004) 652}.

\bibitem{tully1997ursa}
R.B.~Tully and M.A.~Verheijen, \emph{The ursa major cluster of galaxies. ii. bimodality of the distribution of central surface brightnesses}, \href{https://doi.org/10.1086/304318}{\emph{The Astrophysical Journal} {\bfseries 484} (1997) 145}.

\bibitem{sanders1994faber}
R.~Sanders, \emph{A faber-jackson relation for clusters of galaxies: Implications for modified dynamics}, {\emph{Astronomy and Astrophysics} {\bfseries 284} (1994) L31}.

\bibitem{das2023aaspects}
S.~Das, \emph{Aspects of machian gravity (ii): Testing theory against rotation curves of 175 sparc galaxies},  2023.

\bibitem{das2023aspects5}
S.~Das, \emph{Aspects of machian gravity (iii): Testing theory against galaxy cluster mass}, {\emph{arXiv preprint arXiv:2312.06312} (2023) }.

\bibitem{King1966}
I.R.~{King}, \emph{{The structure of star clusters. IV. Photoelectric surface photometry in nine globular clusters}}, \href{https://doi.org/10.1086/109918}{\emph{Astronomical Journal} {\bfseries 71} (1966) 276}.

\bibitem{Cavaliere1976}
A.~{Cavaliere} and R.~{Fusco-Femiano}, \emph{{X-rays from hot plasma in clusters of galaxies}}, {\emph{Astronomy and Astrophysics} {\bfseries 49} (1976) 137}.

\bibitem{reiprich2003cosmological}
T.H.~Reiprich, \emph{Cosmological implications and physical properties of an x-ray flux-limited sample of galaxy clusters},  2003.

\bibitem{SPARC}
F.~{Lelli}, S.S.~{McGaugh} and J.M.~{Schombert}, \emph{{SPARC: Mass Models for 175 Disk Galaxies with Spitzer Photometry and Accurate Rotation Curves}}, \href{https://doi.org/10.3847/0004-6256/152/6/157}{\emph{The Astrophysical Journal} {\bfseries 152} (2016) 157} [\href{https://arxiv.org/abs/1606.09251}{{\ttfamily 1606.09251}}].

\bibitem{coquereaux1990theory}
R.~Coquereaux and G.~Esposito-Farese, \emph{The theory of kaluza-klein-jordan-thiry revisited},  in \emph{Annales de l'IHP Physique th{\'e}orique}, vol.~52, pp.~113--150, 1990.

\bibitem{de1988cosmological}
J.P.~De~Leon, \emph{Cosmological models in a kaluza-klein theory with variable rest mass}, \href{https://doi.org/10.1007/BF00758909}{\emph{General relativity and gravitation} {\bfseries 20} (1988) 539}.

\bibitem{wesson2005equivalence}
P.S.~Wesson, \emph{The equivalence principle as a probe for higher dimensions}, \href{https://doi.org/10.1142/S0218271805007991}{\emph{International journal of modern physics D} {\bfseries 14} (2005) 2315}.

\bibitem{moraes2015cosmological}
P.H.~Moraes, \emph{Cosmological solutions from induced matter model applied to 5d f (r, t) f (r, t) gravity and the shrinking of the extra coordinate}, {\emph{The European Physical Journal C} {\bfseries 75} (2015) 1}.

\bibitem{halpern2000behavior}
P.~Halpern, \emph{Behavior of kasner cosmologies with induced matter}, {\emph{Physical Review D} {\bfseries 63} (2000) 024009}.

\bibitem{Williams2015}
L.L.~Williams, \emph{Field equations and lagrangian for the kaluza metric evaluated with tensor algebra software}, \href{https://doi.org/https://doi.org/10.1155/2015/901870}{\emph{Journal of Gravity} {\bfseries 2015} (2015) 901870} [\href{https://arxiv.org/abs/https://onlinelibrary.wiley.com/doi/pdf/10.1155/2015/901870}{{\ttfamily https://onlinelibrary.wiley.com/doi/pdf/10.1155/2015/901870}}].

\bibitem{hoyle1963mach}
F.~Hoyle and J.V.~Narlikar, \emph{Mach’s principle and the creation of matter}, {\emph{Proceedings of the Royal Society of London. Series A. Mathematical and Physical Sciences} {\bfseries 273} (1963) 1}.

\bibitem{godel1949example}
K.~G{\"o}del, \emph{An example of a new type of cosmological solutions of einstein's field equations of gravitation}, {\emph{Reviews of modern physics} {\bfseries 21} (1949) 447}.

\end{thebibliography}\endgroup

\end{document}